\documentclass[11pt, oneside, tikz]{article}   	

\usepackage[utf8]{inputenc}
\usepackage[parfill]{parskip}    	
\usepackage{graphicx}				
\usepackage{multicol}
\usepackage{vwcol}
\usepackage[english,italian]{babel}
			
\usepackage{amssymb}
\usepackage{mathtools}
\usepackage{upgreek}
\usepackage{multido}
\usepackage{relsize,scalefnt}

\RequirePackage{amsmath}%
\usepackage{soul}

\usepackage{import}

\usepackage{setspace}
 
\newcommand{\Repeat}{\multido{\i=1+1}}

\usepackage{scrextend}
\usepackage[toc,page]{appendix}
\usepackage[usenames, dvipsnames]{xcolor}
\usepackage{tikz} 
\usepackage{pgfplots}
\usepackage{xlop} 
\usepackage{listings}
\usepackage{algpseudocode}
\usepackage{algorithm}
\usepackage[breaklinks]{hyperref}
\hypersetup{
    colorlinks,
    citecolor=black,
    filecolor=black,
    linkcolor=black,
    urlcolor=black,
    breaklinks=true
}
\usepackage{cancel}
\usepackage{enumitem}
\usepackage{titlecaps}

\usepackage[left=1in,right=1in,top=1in,bottom=1in]{geometry}
\usepackage{polynom}

\usepackage{url}
\usepackage{breakurl}

\pgfplotsset{filter discard warning=false}
\newcommand{\Modp}[1] {\ \ (\mathrm{mod}\ #1)}

\iffalse
	\newcommand{\todo}[1] {\textit{\color{cyan}TODO: #1}}
	\newcommand{\todone}[1] {{\color{gray} \st{\textit{TODO: #1}} }}
	\newcommand{\todonot}[1] {\st{\textit{TODO: #1}}}
\else
	\newcommand{\todo}[1] {}
	\newcommand{\todone}[1] {}
	\newcommand{\todonot}[1] {}
\fi

\newcommand{\note}[1] {\textit{Note: #1}}
\newcommand{\term}[1] {\textit{#1}}

\newcommand{\gexp}[2] {g^{#1^#2}}

\newcommand{\gset}[3] {\left\lbrace \gexp{#1}{#2} \right\rbrace_{#3}}
\newcommand{\gsetsid} {\gset{s}{i}{i \in [d]}}
\newcommand{\gsetasid} {\gset{\alpha s}{i}{i \in \{0, \ldots, d\}}}
\newcommand{\maligned}[1] {\begin{aligned}#1\end{aligned}\mkern-9mu}
\newcommand{\flalignskip}[1] {
	{\setlength{\abovedisplayskip}{5pt}
	\setlength{\belowdisplayskip}{3pt}
	\begin{flalign*}
		#1
	\end{flalign*}
	}
}

\newcommand{\coef}[2]{c_{\textsc{#1},#2}}

\newcommand{\prng}[1] {e\left( #1 \right)}
\newcommand{\psf}[1]{_{\mathsf{#1}}}

\newcommand{\zkSNARKraw}[0]{zk-SNARK}
\newcommand{\zkSNARK}{\term{\zkSNARKraw}}

\newcommand{\zkSNARKOP}{\term{\zkSNARK{}OP}}
\newcommand{\zeroKnowledge}{\term{zero-knowledge}}
\newcommand{\ZeroKnowledge}{\term{Zero-knowledge}}
\newcommand{\zeroKnowledgeProof}{\term{zero-knowledge proof}}

\newcommand{\ZeroKnowledgeProofs}{\term{Zero-knowledge proofs}}
\newcommand{\succinctness}{\term{succinctness}}

\newcommand{\checkPolynomialRestriction}{polynomial restriction}
\newcommand{\checkCofactors}{polynomial cofactors}

\newcommand{\checkOperation}{valid operation}
\newcommand{\checkOperations}{\checkOperation{s}}

\newcommand{\checkVarPolynomials}{variable polynomials restriction}

\newcommand{\checkConsistency}{variable values consistency}

\newcommand{\smallmspace}{
	\setlength{\abovedisplayskip}{3pt}
	\setlength{\belowdisplayskip}{0pt}
}

\newcommand{\plotgraph}[4]{
\begin{tikzpicture}
	\begin{axis}[
		width=13.5em,height=17em,
		grid=both,
		xmax=3.7,xmin=-0.5,ymin=-0.5,ymax=4.99,
		domain=-0.5:5,y domain=-0.5:5,
		xticklabel={},yticklabel={},minor tick num=0,
	    axis lines = middle,
	    xlabel=$x$,ylabel={#1},label style = {at={(ticklabel cs:1.13)}},
	    yticklabel=\empty,
		#4	    	
	]
		#2
	\end{axis}
	
	#3
\end{tikzpicture}
}

\newcommand{\plotgrid}[3]{
	\plotgraph{#1}{#2}{#3}{}
}

\newcommand{\plotdot}[3]{
	\node[label={150:{#3}},circle,red,inner sep=1.3pt,fill] at (axis cs:#1,#2) {};
}

\newcommand{\plotdotangle}[4]{
	\node[label={#3:{#4}},circle,red,inner sep=1.3pt,fill] at (axis cs:#1,#2) {};
}

\newcommand{\plotpoly}[2]{
	\plot[thick, #2] plot[samples=100, smooth] expression{#1};
}

\newcommand{\plotpolylabeled}[4]{
	\plot[thick, #2] plot[samples=100, smooth] expression{#1} node [pos=0.9, above left, #4] {#3};
}

\newcommand{\polyl}{-(2*x^2)/3 + x + 5/3}
\newcommand{\polyr}{(-x^2 + 5*x)/2}
\newcommand{\polyo}{3*x^2/20 - 29*x/20 + 53 /10}

\newcommand{\zkProtocol}[3] {
	\smallmspace
	\begin{itemize}
		\smallmspace
		\item Setup
		\begin{itemize}
			#1
		\end{itemize}
		\item Proving
		\begin{itemize}
			#2
		\end{itemize}
		\item Verification
		\begin{itemize}
			#3
		\end{itemize}
	\end{itemize}
}

\newtheorem{dummytheorem}{Dummy-Theorem}[section]

\newtheorem{remark}[dummytheorem]{Remark}

\title{Why and How \zkSNARKraw\ Works: Definitive Explanation}
\author{Maksym Petkus\\\small{maksym@petkus.info}}
\date{}							

\usepackage[style=alphabetic,sorting=none,maxbibnames=99,dateabbrev=false,urldate=iso8601,backref=false,backrefstyle=none,backend=bibtex]{biblatex} 
\begin{document}

\maketitle

\setlength{\abovedisplayskip}{3pt}
\setlength{\belowdisplayskip}{2pt}
\setlength{\abovedisplayshortskip}{0pt}
\setlength{\belowdisplayshortskip}{0pt}

\begin{otherlanguage}{english}

\begin{abstract}
	Despite the existence of multiple great resources on \zkSNARK\ construction, from original papers\footcite{cryptoeprint:2011:443, cryptoeprint:2013:279} to explainers \footcite{zksnarksInNutshell:2016, qapProgramsZeroToHero:2016, zkSnarksUnderTheHood:2017, snarkEplain:2017}, due to the sheer number of moving parts the subject remains a black box for many. While some pieces of the puzzle are given one can not see the full picture without the missing ones.

	Hence the focus of this work is to shed light onto the topic with a straightforward and clean approach based on examples and answering many whys along the way so that more individuals can appreciate the state of the art technology, its innovators and ultimately the beauty of math.
	
	Paper's contribution is a simplistic exposition with a sufficient and gradually increasing level of complexity, necessary to understand \zkSNARK\ without any prerequisite knowledge of the subject, cryptography or advanced math. The primary goal is not only to explain how it works but why it works and how it came to be this way.

	\vspace{45ex}
\end{abstract}

\end{otherlanguage}

\pagebreak
\renewcommand*\contentsname{Contents}
\tableofcontents
\pagebreak

\section{Preface}

While initially planned as short, the work now spans several dozens of pages, nevertheless it requires very little pre-requisite knowledge, and one can freely skip familiar parts.

Do not worry if you are not acquainted with some of the used math symbols, there will be just a few, and they will be introduced gradually, one at a time.

\section{Introduction}

\ZeroKnowledge\ \term{succinct non-interactive arguments of knowledge} (\zkSNARK) is the truly ingenious method of proving that something is true without revealing any other information, however, why it is useful in the first place?

\ZeroKnowledgeProofs\ are advantageous in a myriad of application, including:

\begin{itemize}
	\item Proving statement on private data:
	\begin{itemize}
		\item Person $A$ has more than $X$ in his bank account
		\item In the last year, a bank did not transact with an entity $Y$
		\item Matching DNA without revealing full DNA
		\item One has a credit score higher than $Z$
	\end{itemize}
	
	\item Anonymous authorization:
	\begin{itemize}
		\item Proving that requester $R$ has right to access web-site's restricted area without revealing its identity (e.g., login, password)
		\item Prove that one is from the list of allowed countries/states without revealing from which one exactly
		\item Prove that one owns a monthly pass to a subway/metro without revealing card's id
	\end{itemize}	
	
	\item Anonymous payments:	
	\begin{itemize}
		\item Payment with full detachment from any kind of identity\footcite{cryptoeprint:2014:349}
		\item Paying taxes without revealing one's earnings
	\end{itemize}
	
	\item Outsourcing computation:
	\begin{itemize}
		\item Outsource an expensive computation and validate that the result is correct without redoing the execution; it opens up a category of trustless computing
		\item Changing a blockchain model from everyone computes the same to one party computes and everyone verifies
	\end{itemize}
\end{itemize}

As great as it sounds on the surface the underlying method is a ``marvel'' of mathematics and cryptography and is being researched  for the 4th decade since its introduction in 1985 in the principal work ``The Knowledge Complexity of Interactive Proof-systems" \cite{Goldwasser:1985:KCI:22145.22178} with subsequent introduction of the non-interactive proofs \cite{Blum:1988:NZA:62212.62222} which are especially essential in the context of blockchains.

In any \zeroKnowledgeProof\ system, there is a \term{prover} who wants to convince a \term{verifier} that some \term{statement} is true without revealing any other information, e.g., \term{verifier} learns that the prover has more than $X$ in his bank account but nothing else (i.e., the actual amount is not disclosed). A protocol should satisfy three properties:
\begin{itemize}
	\item Completeness --- if the \term{statement} is true then a \term{prover} can convince a \term{verifier}
	\item Soundness --- a cheating \term{prover} can not convince a \term{verifier} of a false \term{statement}
	\item Zero-knowledge --- the interaction only reveals if a \term{statement} is true and nothing else
\end{itemize}

The \zkSNARK\ term itself was introduced in \cite{cryptoeprint:2011:443}, building on \cite{groth2010short} with following  Pinocchio protocol \cite{cryptoeprint:2012:215, cryptoeprint:2013:279} making it applicable for general computing.

\section{The Medium of a Proof}

Let us start simple and try to prove something without worrying about the zero-knowledge, non-interactivity, its form,	 and applicability.

Imagine that we have an array of bits of length $10$, and we want to prove to a verifier (e.g., program) that all those bits are set to $1$, i.e., we \term{know} an array such that every element equals to $1$.
$$b = [\ \framebox{?} \Repeat{10}{,\ \framebox{?}}\ ]$$

Hello
Verifier can only check (i.e., read) one element at a time. In order to verify the statement one can proceed by reading elements in some arbitrary order and checking if it is truly equal to 1 and if so the confidence in that statement after the first check is $\frac{1}{10} = 10\%$, or statement is invalidated altogether if the bit equals to 0. A verifier must proceed to the next round until he reaches sufficient confidence. In some cases, one may trust a prover and require only 50\% confidence which means that 5 checks must be executed, in other cases where 95\% confidence is needed all cells must be checked. It is clear that the downside of such a proving protocol is that one must do the number of checks proportionate to the number of elements, which is non-practical if we consider arrays of millions of elements.

Let us consider polynomials, which can be visualized as a curve on a graph, shaped by a mathematical equation:

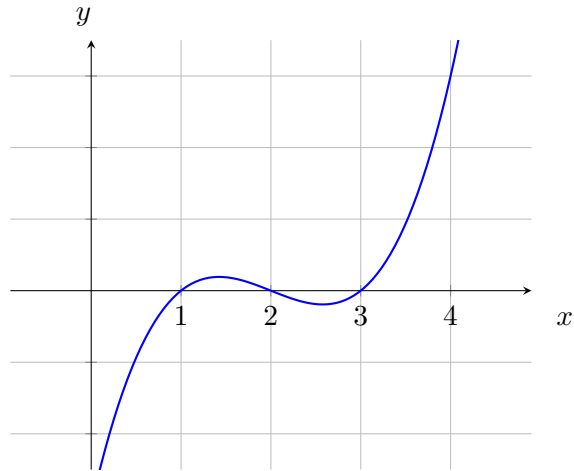
\begin{center}
	\begin{tikzpicture}	     
		\begin{axis}[
			grid=major,
			ymin=-5,ymax=7,xmax=4.9,xmin=-0.9,
			xticklabel={},yticklabel=\empty,minor tick num=0,
	        axis lines = middle,
	        xlabel=$x$,ylabel=$y$,label style = {at={(ticklabel cs:1.1)}}
	    	]
			\plot[thick, blue] plot[samples=100, smooth] expression{x^3 -6*x^2 + 11*x - 6};
		\end{axis}    
	\end{tikzpicture}
\end{center}

The above curve corresponds to the polynomial: $f(x) = x^3 -6x^2 + 11x - 6$. The degree of a polynomial is determined by its greatest exponent of $x$, which in this case is 3.

Polynomials have an advantageous property, namely, if we have two non-equal polynomials of degree at most $d$, they can intersect at no more than $d$ points. For example, let us modify the original polynomial slightly $x^3 -6x^2 + \mathbf{10}x - \mathbf{5}$ and visualize it in green:

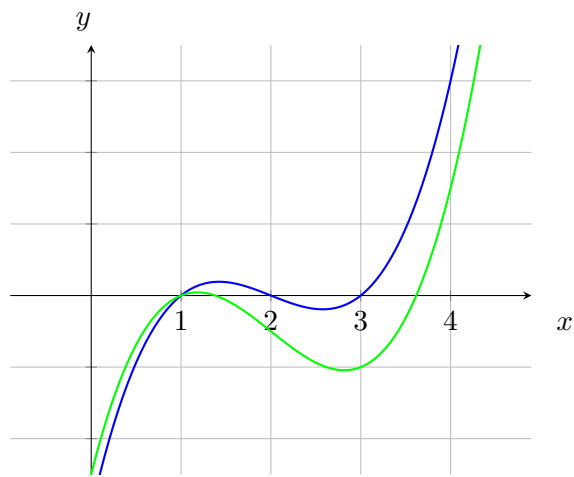
\begin{center}
	\begin{tikzpicture}
		\begin{axis}[
			grid=major,
			ymin=-5,ymax=7,xmax=4.9,xmin=-0.9,
			xticklabel={},yticklabel=\empty,minor tick num=0,
		        axis lines = middle,
		        xlabel=$x$,ylabel=$y$,label style = {at={(ticklabel cs:1.1)}}
		    	]
			\plot[thick, blue] plot[samples=100, smooth] expression{x^3 -6*x^2 + 11*x - 6};
			
			\plot[thick, green] plot[samples=100, smooth] expression{x^3 -6*x^2 + 10*x - 5};
		\end{axis}
	\end{tikzpicture}
\end{center}

Such a tiny change produces a dramatically different result. In fact, it is impossible to find two non-equal polynomials, which share a consecutive chunk of a curve\footnote{Excluding a single point chunk case}.

This property flows from the method of finding shared points.
If we want to find intersections of two polynomials, we need to equate them. For example, to find where a polynomial crosses an $x$-axis (i.e., $f(x) = 0$), we equate $x^3 -6x^2 + 11x - 6 = 0$, and solutions to such an equation will be those shared points: $x = 1$, $x = 2$ and $x = 3$, also you can clearly see that this is true on the previous graph, where the blue curve crosses the $x$-axis line.

Likewise, we can equate our original and modified version of polynomials to find their intersections.
\begin{align*}
	x^3 -6x^2 + 11x - 6 &= x^3 -6x^2 + \mathbf{10}x - \mathbf{5} \\
	x - 1 &= 0
\end{align*}

The resulting polynomial is of degree 1 with an obvious solution $x = 1$. Hence only one intersection:

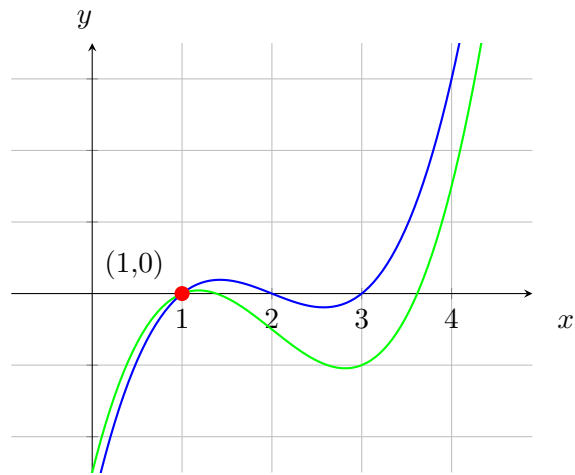
\begin{center}
	\begin{tikzpicture}
		\begin{axis}[
			grid=major,
			ymin=-5,ymax=7,xmax=4.9,xmin=-0.9,
			xticklabel={},yticklabel=\empty,minor tick num=0,
		        axis lines = middle,
		        xlabel=$x$,ylabel=$y$,label style = {at={(ticklabel cs:1.1)}}
		    	]
			\plot[thick, blue] plot[samples=100, smooth] expression{x^3 -6*x^2 + 11*x - 6};
			
			\plot[thick, green] plot[samples=100, smooth] expression{x^3 -6*x^2 + 10*x - 5};
			\node[label={150:{(1,0)}},circle,red,inner sep=2pt,fill] at (axis cs:1,0) {};
		\end{axis}
	\end{tikzpicture}
\end{center}

The result of any such equation for arbitrary degree $d$ polynomials is always another polynomial of degree at most $d$, since there is no multiplication to produce higher degrees. Example: $5x^3 + 7x^2 - x + 2 = 3x^3 - x^2 + 2x - 5$, which simplifies to $2x^3 + 8x^2 -3x + 7 = 0$. And the Fundamental Theorem of Algebra tells us that a degree $d$ polynomial can have at most $d$ solutions\footnote{More on this in section \ref{section:factorization}}, and therefore at most $d$ shared points.

Hence we can conclude that evaluation\footnote{More on polynomial evaluation: \cite{polyEvaluation:2013}} of any polynomial at an arbitrary point is akin to the representation of its unique identity. Let us evaluate our example polynomials at $x = 10$.
$$ x^3 - 6x^2 + 11x - 6 = 504 $$
$$ x^3 - 6x^2 + 10x - 5 = 495 $$

In fact out of all choices of $x$ to evaluate, only at most 3 choices will have equal evaluations in those polynomials and all others will differ.

That is why if a prover claims to \term{know} some polynomial (no matter how large its degree is) that the verifier also knows, they can follow a simple protocol to verify the statement:
\begin{itemize}
	\item Verifier chooses a random value for $x$ and evaluates his polynomial locally
	\item Verifier gives $x$ to the prover and asks to evaluate the polynomial in question
	\item Prover evaluates his polynomial at $x$ and gives the result to the verifier
	\item Verifier checks if the local result is equal to the prover's result, and if so then the statement is proven with a high confidence
\end{itemize}

If we, for example, consider an integer range of $x$ from 1 to $10^{77}$, the number of points where evaluations are different is $10^{77} - d$. Henceforth the probability that $x$ accidentally ``hits" any of the $d$ shared points is equal to $\displaystyle \frac{d}{10^{77}}$, which is considered negligible.

\note{the new protocol requires only one round and gives overwhelming confidence (almost $100\%$ assuming $d$ is sufficiently smaller than the upper bound of the range) in the statement compared to the inefficient bit check protocol.}

That is why polynomials are at the very core of \zkSNARK, although it is likely that other proof mediums exist as well.

\section{Non-Interactive Zero-Knowledge of a Polynomial}

\subsection{Proving Knowledge of a Polynomial}

We start with a problem of proving the knowledge of a polynomial and make our way to a generic approach. We will discover many other properties of polynomials along the way.
  
The discussion so far has focused on a weak notion of a proof, where parties have to trust each other because there are no measures yet to enforce the rules of the protocol. For example, the prover is not required to know a polynomial, and he can use any other means available to him to come up with a correct result. Moreover, if the amplitude of the verifier's polynomial evaluations is not large, let us say 10, the verifier can guess a number, and there is a non-negligible probability that it will be accepted. We have to address such weakness of the protocol, but first what does it means to know a polynomial? A polynomial can be expressed in the form (where $n$ is the degree of the polynomial):
$$c_n x^n + ... + c_1 x^1 + c_0 x^0$$
If one stated that he or she knows a polynomial of degree 1 (i.e., $c_1 x^1 + c_0$), that means that what one really \emph{knows} is the coefficients $c_0, c_1$. Moreover, coefficients can have any value, including $0$.

Let us say that the prover claims to know a degree 3 polynomial, such that $x = 1$ and $x = 2$ are two of all possible solutions. One of such valid polynomials is $x^3 - 3x^2 + 2x = 0$. For $x = 1$: $1 - 3 + 2 = 0$. For $x = 2$: $8 - 12 + 4 = 0$.

Let us first look more closely at the anatomy of the solution.

\subsection{Factorization} \label{section:factorization}

\newcommand{\polyExFactor}{x^3 - 3x^2 + 2x}

The Fundamental Theorem of Algebra states that any polynomial can be factored into linear polynomials (i.e., a degree 1 polynomials representing a line), as long it is solvable. Consequently, we can represent any valid polynomial as a product of its factors:
$$(x-a_0)(x-a_1)...(x-a_n) = 0$$ 
Also, if any of these factors is zero then the whole equation is zero, henceforth all the $a$-s are the only solutions.

In fact, our example can be factored into the following polynomial:
$$\polyExFactor = (x - 0)(x - 1)(x - 2)$$

And the solutions are (values of $x$): $0, 1, 2$, you can check this easily on either form of the polynomial, but the factorized form has all the solutions (also called roots) on the surface.

Getting back to the prover's claim that he knows a polynomial of degree 3 with the roots 1 and 2, this means that his polynomial has the form: 
$$ (x - 1)(x - 2) \cdot \ldots $$
In other words $(x - 1)$ and $(x - 2)$ are the cofactors of the polynomial in question. Hence if the prover wants to prove that indeed his polynomial has those roots without disclosing the polynomial itself, he needs to prove that his polynomial $p(x)$ is the multiplication of those cofactors $t(x) = (x - 1)(x - 2)$, called \term{target polynomial}, and some arbitrary polynomial $h(x)$ (equals to $x-0$ in our example), i.e.:
\begin{equation*}
	p(x) = t(x) \cdot h(x)
\end{equation*}

In other words, there exists some polynomial $h(x)$ which makes $t(x)$ equal to $p(x)$, therefore $p(x)$ contains $t(x)$, consequently $p(x)$ has all roots of $t(x)$, the very thing to be proven.

A natural way to find $h(x)$ is through the division $h(x) = \frac{p(x)}{t(x)}$. If the prover cannot find such $h(x)$ that means that $p(x)$ does not have the necessary cofactors $t(x)$, in which case the polynomials division will have a remainder.

In our example if we divide $p(x) = \polyExFactor$ by the $t(x) = (x-1)(x-2) = x^2 -3x + 2$:

\begin{center}
	\polylongdiv{x^3 - 3x^2 + 2x}{(x-1)(x-2)}
\end{center}

\note{the denominator is to the left, the result is to the top right, and the remainder is to the bottom\footnote{Polynomial division explanation with examples is available at \cite{dividingByAPoly:2014}}.}

We have got the result $h(x) = x$ without remainder.

\note{for simplicity, onwards we will use polynomial's letter variable to denote its evaluation, e.g., $p = p(r)$}

Using our polynomial identity check protocol we can compare polynomials $p(x)$ and $t(x) \cdot\nobreak h(x)$:
\begin{itemize}
	\item Verifier samples a random value $r$, calculates $t = t(r)$ (i.e., evaluates) and gives $r$ to the prover
	\item Prover calculates $h(x) = \frac{p(x)}{t(x)}$ and evaluates $p(r)$ and $h(r)$; the resulting values $p, h$ are provided to the verifier
	\item Verifier then checks that $p = t \cdot h$, if so those polynomials are equal, meaning that $p(x)$ has $t(x)$ as a cofactor.
\end{itemize}

To put this into practice, let us execute this protocol for our example:
\begin{eqnarray*}
	p(x) = x^3 - 3x^2 + 2x \\
	t(x) = (x-1)(x-2)
\end{eqnarray*}

\newcommand{\constr}{23}
\begin{itemize}
	\item Verifier samples a random value $\constr$, calculates $t = t(\constr) = (\constr - 1)(\constr - 2) = 462$ 
			and gives $\constr$ to the prover
	\item Prover calculates $h(x) = \frac{p(x)}{t(x)} = x$,
			evaluates $p = p(\constr) = 10626$ and $h = h(\constr) = 23$ and provides $p, h$ to the verifier
	\item Verifier then checks that $p = t \cdot h$: $ 10626 = 462 \cdot 23 $, which is true, and therefore the statement is proven
\end{itemize}

On the contrary, if the prover uses a different $p'(x)$ which does not have the necessary cofactors, for example $p'(x) = {\color{red}2}x^3 - 3x^2 + 2x$, then:

\begin{center}
	$h(x) \quad=$\polylongdiv{2x^3 - 3x^2 + 2x}{(x-1)(x-2)}
\end{center}

We will get $2x + 3$ with the remainder $7x - 6$, i.e.: $p(x) = t(x) \times (2x + 3) + 7x - 6$. This means that the prover will have to divide the remainder by the $t(r)$ in order to evaluate $h(x) = 2x + 3 + \frac{7x - 6}{t(x)}$. Therefore because of the random selection of $x$ by the verifier, there is a low\footnote{But still non-negligible} probability that the evaluation of the remainder $7x-6$ will be evenly divisible by the evaluation of $t(x)$, henceforth if verifier will additionally check that $p$ and $h$ must be integers, such proofs will be rejected. 
However, the check requires the polynomial coefficients to be integers too, creating a significant limitation to the protocol.

That is the reason to introduce cryptographic primitives which make such division impossible, even if the raw evaluations happen to be divisible.

\note{although the author's chief objective is simplicity, including the set of math symbols in use, it would be detrimental for further sections to omit the ubiquitous symbol prime:\ \ {$'$}\ . Its essential purpose is to signify some transformation or derivation of the original variable or function, e.g., if we want to multiply $v$ by $2$ and assign it to a separate variable, we could use prime: $v' = 2 \cdot v$.}

\begin{remark} \label{remark:prePairingIssues}
	Now we can check a polynomial for specific properties without learning the polynomial itself, so this already gives us some form of \zeroKnowledge\ and \succinctness. Nonetheless, there are multiple issues with this construction:
	\begin{itemize}
		\item Prover may not know the claimed polynomial $p(x)$ at all. He can calculate evaluation $t = t(r)$, select a random number $h$ and set $p = t \cdot h$, which will be accepted by the verifier as valid, since equation holds.
		\item Because prover knows the random point $x = r$, he can construct any polynomial which has one shared point at $r$ with $t(r) \cdot h(r)$.
		\item In the original statement, prover claims to know a polynomial of a particular degree, in the current protocol there is no enforcement of degree. Hence prover can cheat by using a polynomial of higher degree which also satisfies the cofactors check.
	\end{itemize}
\end{remark}

We will address all of the issues in the following sections.

\subsection{Obscure Evaluation}

Two first issues of remark \ref{remark:prePairingIssues} are possible because values are presented at raw, prover knows $r$ and $t(r)$. It would be ideal if those values would be given as a black box, so one cannot temper with the protocol, but still able to compute operations on those obscure values. Something similar to the hash function, such that when computed it is hard to go back to the original input.

\subsubsection{Homomorphic Encryption}

That is exactly what homomorphic encryption is designed for. Namely, it allows to encrypt a value and be able to apply arithmetic operations on such encryption. There are multiple ways to achieve homomorphic properties of encryption, and we will briefly introduce a simple one.

The general idea is that we choose a base\footnote{There are certain properties that base number needs to have} natural number $g$ (say 5) and to encrypt a value we exponentiate $g$ to the power of that value. For example, if we want to encrypt the number 3:
$$5^3 = 125$$

Where 125 is the encryption of $3$. If we want to multiply this encrypted number by $2$, we raise it to the exponent of $2$:
$$ 125^2 = 15625 = \left(5^3\right)^2 = 5^{2 \times 3} = 5^6 $$

We were able to multiply an unknown value by 2 and keep it encrypted. We can also add two encrypted values through multiplication, for example, 3 + 2:
$$ 5^3 \cdot 5^2 = 5^{3 + 2} = 5^5 = 3125 $$

Similarly, we can subtract encrypted numbers through division, for example, $5 - 3$:
$$\frac{5^5}{5^3} = 5^5 \cdot 5^{-3} = 5^{5 - 3} = 5 ^ 2 = 25$$

However, since the base $5$ is public, it is quite easy to go back to the secret number, dividing encrypted by 5 until the result is 1. The number of steps is the secret number.

\subsubsection{Modular Arithmetic}

That is where the modular arithmetic comes into play. The idea of modular arithmetic is following: instead of having an infinite set of numbers we declare that we select only first $n$ natural numbers, i.e., $0, 1, \ldots, n-1$, to work with, and if any given integer falls out of this range, we ``wrap" it around. For example, let us choose six first numbers. To illustrate this, consider a circle with six ticks of equal units; this is our range\footnote{Usually referred to as finite field}.

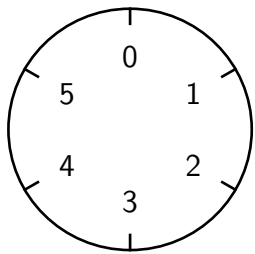
\newcommand{\mrad}{1.6}
\newcommand{\mulen}{3.14 * 2 * \mrad / 6}

\begin{tikzpicture}[line cap=rect,line width=1pt]
	\filldraw [fill=white] (0,0) circle [radius=\mrad cm];
	\foreach \angle [count=\xi from 0] in {90,30,...,-240}
	{
	  \draw[line width=1pt] (\angle:\mrad - 0.2) -- (\angle:\mrad);
	  \node[font=\large] at (\angle:\mrad - 0.64) {\textsf{\xi}};
	}
\end{tikzpicture}

Now let us see where the number eight will land. As an analogy, we can think of it as a rope, the length of which is eight units:

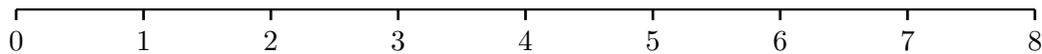
\begin{tikzpicture}
	\draw[thick] (0,0) -- (8 * \mulen,0) ;
	\foreach \x in {0,1,...,8}
   		\draw[line width=1pt] (\x*\mulen,0pt) -- (\x*\mulen,-4pt) node[anchor=north] {$\x$};
\end{tikzpicture}

If we attach the rope to the beginning of the circle

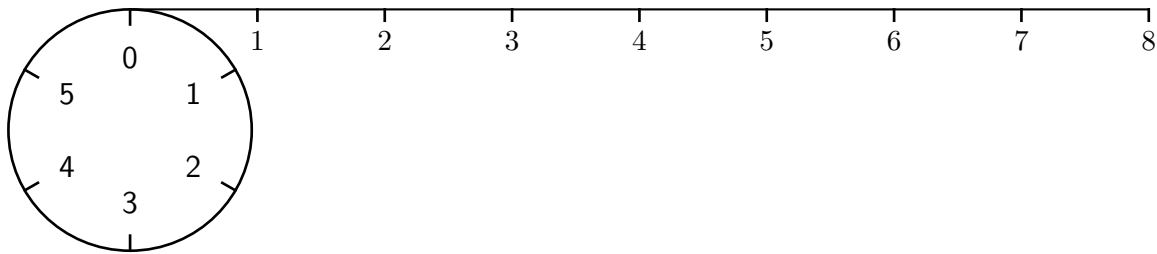
\begin{tikzpicture}[line cap=rect,line width=1pt]
	\filldraw [fill=white] (0,0) circle [radius=\mrad cm];
	\foreach \angle [count=\xi from 0] in {90,30,...,-240}
	{
	  \draw[line width=1pt] (\angle:\mrad - 0.2) -- (\angle:\mrad);
	  \node[font=\large] at (\angle:\mrad - 0.64) {\textsf{\xi}};
	}
	
	\draw[thick] (0, \mrad) -- (8 * \mulen, \mrad) ;
	\foreach \x in {1,...,8}
   		\draw[line width=1pt] (\x*\mulen, \mrad) -- (\x*\mulen, \mrad-0.15) node[anchor=north] {$\x$};
\end{tikzpicture}

and start wrapping the rope around it, after one rotation we still have a portion of the rope left:

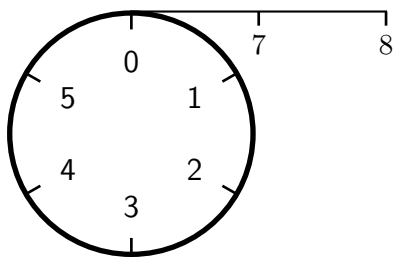
\begin{tikzpicture}[line cap=rect,line width=1pt]
	\filldraw [fill=white, line width=2pt] (0,0) circle [radius=\mrad cm];
	
	\foreach \angle [count=\xi from 0] in {90,30,...,-240}
	{
	  \draw[line width=1pt] (\angle:\mrad - 0.2) -- (\angle:\mrad);
	  \node[font=\large] at (\angle:\mrad - 0.64) {\textsf{\xi}};
	}
	
	\draw[thick] (0, \mrad + 0.02) -- (2 * \mulen, \mrad + 0.02) ;
	\foreach \xi [count=\x from 1] in {7,...,8}
   		\draw[line width=1pt] (\x*\mulen, \mrad) -- (\x*\mulen, \mrad-0.15) node[anchor=north] {\xi};
\end{tikzpicture}

Therefore if we continue the process, the rope will end right at the tick \#2.

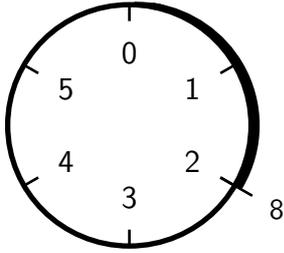
\begin{tikzpicture}[line cap=rect,line width=1pt]
	\filldraw [fill=white, line width=2pt] (0,0) circle [radius=\mrad cm];
	
	\draw[line width=2pt] ([shift=(80:\mrad)]0.1,0.003) arc (80:90:\mrad);
	\draw[line width=2.5pt] ([shift=(-30:\mrad)]0.07,0) arc (-30:80:\mrad);	
	
	\foreach \angle [count=\xi from 0] in {90,30,...,-240}
	{
	  \draw[line width=1pt] (\angle:\mrad - 0.2) -- (\angle:\mrad);
	  \node[font=\large] at (\angle:\mrad - 0.64) {\textsf{\xi}};
	}
	
	\draw[line width=1pt] (-30:\mrad - 0.2) -- (-30:\mrad + 0.25);
	\node[] at (-30:\mrad + 0.64) {\textsf{8}};
\end{tikzpicture}

It is the result of the modulo operation. No matter how long the rope is it will always stop at one of the circle's ticks. Therefore the modulo operation will keep it in certain bounds (in this case from 0 to 5). The 15-units rope will stop at 3, i.e., 6 + 6 + 3 (two full circles with 3-units leftover). The negative numbers work the same way, and the only difference is that we wrap it in the opposite direction, for $-8$ the result will be 4.

Moreover, we can perform arithmetic operations, and the result will always be in the scope of $n$ numbers. We will use the notation ``$\mathrm{mod}\ n$" for now on to denote the range of numbers. For example:
\begin{eqnarray*}
	3 \times 5 = 3 \Modp{6}\\
	5 + 2 = 1 \Modp{6}
\end{eqnarray*}

Furthermore, the most important property is that the order of operations does not matter, e.g., we can perform all operations first and then apply modulo or apply modulo after every operation. For example $(2 \times 4 - 1) \times 3 = 3 \Modp{6}$ is equivalent to:
\begin{eqnarray*}
	2 \times 4 = 2 \Modp{6} \\	
	2 - 1 = 1 \Modp{6}  \\
	1 \times 3 = 3 \Modp{6}
\end{eqnarray*}

So why on earth is that helpful? It turns out that if we use modulo arithmetic, having a result of operation it is non-trivial to go back to the original numbers because many different combinations will have the same result:
\begin{eqnarray*}
	5 \times 4 = 2 \Modp{6} \\
	4 \times 2 = 2 \Modp{6} \\
	2 \times 1 = 2 \Modp{6} \\
	\ldots
\end{eqnarray*}
Without the modular arithmetic, the size of the result gives a clue to its solution. This piece of information is hidden otherwise, while common arithmetic properties are preserved.

\subsubsection{Strong Homomorphic Encryption}

If we go back to the homomorphic encryption and use modular arithmetic, for example with modulo 7, we will get:
\begin{eqnarray*}
	5^1 = 5 \Modp{7} \\
	5^2 = 4 \Modp{7} \\
	5^3 = 6 \Modp{7} \\
	\ldots
\end{eqnarray*}
And different exponents will have the same result:
\begin{eqnarray*}
	5^5 = 3 \Modp{7} \\
	5^{11} = 3 \Modp{7} \\
	5^{17} = 3 \Modp{7} \\
	\ldots
\end{eqnarray*}
This is where it gets \emph{hard} to find the exponent. In fact, if modulo is sufficiently large, it becomes infeasible to do so, and a good portion of the modern-day cryptography is based on the ``hardness" of this problem.

All the homomorphic properties of the scheme are preserved in the modular realm:
\begin{alignat*}{2}
	\mathrm{encryption:}	&\quad	& 5^3& = 6 \Modp{7}\\
	\mathrm{multiplication:}	&\quad\quad	 & 6^2 = {(5^3)}^2 = 5^6& = 1 \Modp{7}\\
	\mathrm{addition:}&\quad 	& 5^3 \cdot 5^2 = 5^5& = 3 \Modp{7}
\end{alignat*}
\note{modular division is a bit more complicated and out of the scope.}

Let us explicitly state the encryption function: $E(v) = g^v \Modp{n}$, where $v$ is the value we want to encrypt.

\begin{remark}\label{remark:helimits}
There are limitations to this homomorphic encryption scheme while we can multiply an encrypted value by an unencrypted value, we cannot multiply (and divide) two encrypted values, as well as we cannot exponentiate an encrypted value. While unfortunate from the first impression, these properties will turn out to be the cornerstone of \zkSNARK. The limitations are addressed in section \ref{section:pairings}
\end{remark}

\subsubsection{Encrypted Polynomial}

Armed with such tools, we can now evaluate a polynomial with an encrypted random value of $x$ and modify the \zeroKnowledge\ protocol accordingly.

Let us see how we can evaluate a polynomial $p(x) = x^3 - 3x^2 + 2x$. As we have established previously to know a polynomial is to know its coefficients, in this case those are: 1, -3, 2. 
Because homomorphic encryption does not allows to exponentiate an encrypted value, we've must been given encrypted values of powers of $x$ from 1 to 3: $E(x), E(x^2), E(x^3)$, so that we can evaluate the encrypted polynomial as follows:
\begin{gather*}
	{E\left(x^3\right)}^1 \cdot {E\left(x^2\right)}^{-3} \cdot {E\left(x\right)}^{2} = \\
	{\left(g^{x^3}\right)}^1 \cdot {\left(g^{x^2}\right)}^{-3} \cdot {\Big( g^{x} \Big)}^2 =\\
	g^{1x^3} \cdot g^{-3x^2} \cdot g^{2x} =\\
	g^{x^3 - 3x^2 + 2x}
\end{gather*}
As the result of such operations, we have an encrypted evaluation of our polynomial at some unknown to us $x$. This is quite a powerful mechanism, and because of the homomorphic property, the encrypted evaluations of the same polynomials are always the same in encrypted space.

We can now update the previous version of the protocol, for a polynomial of degree $d$:
\begin{itemize}
	\item Verifier 
		\begin{itemize}
			\item samples a random value $s$, i.e., secret
			\item calculates encryptions of $s$ for all powers $i$ in $0, 1, ..., d$, i.e.: $E(s^i) = g^{s^i}$
			\item evaluates unencrypted \term{target polynomial} with $s$:  $t(s)$
			\item encrypted powers of $s$ are provided to the prover: $E(s^0), E(s^1), ..., E(s^d)$
		\end{itemize} 
	\item Prover 
	\begin{itemize}
		\item calculates polynomial $h(x) = \frac{p(x)}{t(x)}$
		\item using encrypted powers $g^{s^0}, g^{s^1}, \ldots, g^{s^d}$ and coefficients $c_0, c_1, \ldots, c_n$ evaluates
		\flalignskip{
			& E\left( p(s) \right) = g^{p(s)} = \left(g^{s^d}\right)^{c_d} \cdots \left(g^{s^1}\right)^{c_1} \cdot \left(g^{s^0}\right)^{c_0} \text{\ \ and similarly\ \ } E\left( h(s) \right) =  g^{h(s)} &
		}
		\item the resulting $g^p$ and $g^h$ are provided to the verifier
	\end{itemize}
	\item Verifier
	\begin{itemize}
		\item The last step for the verifier is to checks that $p = t(s) \cdot h$ in encrypted space:
		\flalignskip{
			& g^p = \left(g^h\right)^{t(s)} \quad \Rightarrow \quad
			g^p = g^{t(s) \cdot h} &
		}
	\end{itemize}
\end{itemize}
\note{because the prover does not know anything about $s$, it makes it hard to come up with non-legitimate but still matching evaluations.}

While in such protocol the prover's agility is limited he still can use any other means to forge a proof without actually using the provided encryptions of powers of $s$, for example, if the prover claims to have a satisfactory polynomial using only 2 powers $s^3$ and $s^1$, that is not possible to verify in the current protocol.

\subsection{Restricting a Polynomial} \label{restricting polynomial}

The knowledge of a polynomial is the knowledge of its coefficients $c_0, c_1, \ldots, c_i$ and the way we ``assign" those coefficients in the protocol is through exponentiation of the corresponding encrypted powers of the secret value $s$ (i.e., $E\left(s^i\right)^{c_i} = g^{c_i \cdot s^i}$). We do already restrict a prover in the selection of encrypted powers of $s$, but such restriction is not enforced, e.g., one could use any possible means to find some arbitrary values $z_p$ and $z_h$ which satisfy equation $z_p = \left(z_h\right)^{t(s)}$ and provide them to the verifier instead of $g^p$ and $g^h$. For example, for some random $r$ $z_h = g^r$ and $z_p = \left(g^{t(s)}\right)^r$, where $g^{t(s)}$ can be computed from the provided encrypted powers of $s$. That is why verifier needs the proof that only supplied encryptions of powers of $s$ were used to calculate $g^p$ and $g^h$ and nothing else.

Let us consider an elementary example of a degree 1 polynomial with one variable and one coefficient $f(x) = c \cdot x$ and correspondingly the encryption of the $s$ is provided $E(s) = g^s$. What we are looking for is to make sure that only encryption of $s$, i.e., $g^s$, was homomorphically ``multiplied" by some arbitrary coefficient $c$ and nothing else. So the result must always be of the form $\left(g^s\right)^c$ for some arbitrary $c$. 

A way to do this is to require to perform the same operation on another \textit{shifted} encrypted value alongside with the original one, acting as an arithmetic analog of ``checksum", ensuring that the result is exponentiation of the original value.

This is achieved through the Knowledge-of-Exponent Assumption (or KEA), introduced in \cite{damgaard1991towards}, more precisely:

\begin{itemize}
	\item Alice has a value $a$, that she wants Bob to exponentiate to any power\footnote{Where $a$ is a generator of a finite field group used}, the single requirement is that only this $a$ can be exponentiated and nothing else, to ensure this she:
	\begin{itemize}
		\item chooses a random $\alpha$
		\item calculates $a'  = a^\alpha \Modp{n}$
		\item provides the tuple $(a, a')$ to Bob and asks to perform same arbitrary exponentiation of each value and reply with the resulting tuple $(b, b')$ where the exponent ``$\alpha$-shift" remains the same, i.e., $b^\alpha = b' \Modp{n}$
	\end{itemize}
	\item because Bob cannot extract $\alpha$ from the tuple $(a, a')$ other then through a brute-force\footnote{The proof is provided in the original paper} which is infeasible, it is conjectured that the only way Bob can produce a valid response is through the procedure:
	\begin{itemize}
	\item chose some value $c$
	\item calculate $b = (a)^c \Modp{n}$ \quad and \quad $b' = {(a')}^c \Modp{n}$
	\item reply with $(b, b')$
	\end{itemize}
	\item having the response and $\alpha$, Alice checks the equality: 
		\begin{flalign*}
			(b)^\alpha &= b' &\\
			\left(a^c\right)^\alpha &= {(a')}^c &\\
			a^{c \cdot \alpha} &= {\left(a^\alpha\right)}^c &
		\end{flalign*}
	\item conclusions:
	\begin{itemize}
	\item Bob has applied the same exponent (i.e., $c$) to both values of the tuple
	\item Bob could only use the original Alice's tuple to maintain the $\alpha$ relationship
	\item Bob \textit{knows} the applied exponent $c$, because the only way to produce valid $(b, b')$ is to use the same exponent
	\item Alice has not learned $c$ for the same reason Bob cannot learn $\alpha$\footnote{Although the $c$ is \textit{encrypted} its range of possible values might not be sufficient to preserve \zeroKnowledge\ property which will be addressed in the section \ref{section:polyzk}.
	}
	\end{itemize}
\end{itemize}

Ultimately such protocol provides a proof to Alice that Bob indeed exponentiated $a$ by some value known to him, and he could not do any other operation, e.g., multiplication, addition, since this would erase the $\alpha$-shift relationship. 

In the homomorphic encryption context, exponentiation is the multiplication of the encrypted value. We can apply the same construction in the case with the simple one-coefficient polynomial $f(x) = c \cdot x$:
\begin{itemize}
	\item Verifier chooses random $s, \alpha$ and provides evaluation for $x = s$ for power 1 and its ``shift": $\left(g^s, g^{\alpha \cdot s}\right)$
	\item Prover applies the coefficient $c$: $\left( \left(g^s\right)^c, \left(g^{\alpha \cdot s}\right)^c \right) = \left(g^{c \cdot s}, g^{\alpha \cdot c \cdot s}\right)$
	\item Verifier checks: $\left(g^{c \cdot s}\right)^\alpha = g^{\alpha \cdot c \cdot s}$
\end{itemize}

Such construction \textit{restricts} the prover to use only the encrypted $s$ provided, therefore prover could have assigned coefficient $c$ only to the polynomial provided by the verifier. We can now scale such one-term polynomial\footnote{Monomial} approach to a multi-term polynomial because the coefficient assignment of each term is calculated separately and then homomorphically ``added" together (this approach was introduced by Jens Groth in \cite{groth2010short}). So if the prover is given encrypted exponentiations of $s$ alongside with their \textit{shifted} values he can evaluate original and shifted polynomial, where the same check must hold.
In particular, for a degree $d$ polynomial:

\begin{itemize}
	\item Verifier provides encrypted powers $g^{s^0}, g^{s^1}, \ldots, g^{s^d}$ and their shifts $g^{\alpha s^0}, g^{\alpha s^1}, \ldots, g^{\alpha s^d}$
	\item Prover:
	\begin{itemize}[leftmargin=8pt]
		\item evaluates encrypted polynomial with provided powers of $s$: \\
	$g^{p(s)} = 
	\left(g^{s^0}\right)^{c_0} \cdot \left(g^{s^1}\right)^{c_1} \cdot \ldots \cdot \left(g^{s^d}\right)^{c_d} = 
	g^{c_0 s^0 + c_1 s^1 + \ldots + c_d s^d}$

		\item evaluates encrypted ``shifted" polynomial with the corresponding $\alpha$-shifts of the powers of $s$: \\
	$ g^{\alpha p(s)} = 
	\left(g^{\alpha s^0}\right)^{c_0} \cdot \left(g^{\alpha s^1}\right)^{c_1} \cdot \ldots \cdot \left(g^{\alpha s^d}\right)^{c_d} = 
	g^{c_0 \alpha s^0 + c_1 \alpha s^1 + \ldots + c_d \alpha s^d} = 
	g^{\alpha (c_0 s^0 + c_1 s^1 + \ldots + c_d s^d)}$
	
		\item provides the result as $g^p, g^{p'}$ to the verifier
	\end{itemize}
	
	\item Verifier checks: ${\left(g^p\right)}^\alpha = g^{p'}$
\end{itemize}

\todone{? provide explanation that if $g^{\alpha s^0}$ is supplied that would allow prover to use other values? While it actually requires use of the provided generator. Conclusion: it is legal with the framework.}

For our previous example polynomial $p(x) = x^3 - 3x^2 + 2x$ this would be:
\begin{itemize}
	\item Verifier provides $E(s^3), E(s^2), E(s)$ and their shifts $E(\alpha s^3), E(\alpha s^2), E(\alpha s)$
	\item Prover evaluates:\\ 
	$g^p = g^{p(s)} = 
	\left(g^{s^3}\right)^1 \cdot \left(g^{s^2}\right)^{-3} \cdot \Big(g^s\Big)^2 = 
	g^{s^3} \cdot g^{-3s^2} \cdot g^{2s} = g^{s^3 - 3s^2 + 2s} $ \\
	$g^{p'} = g^{\alpha p(s)} = 
	\left(g^{\alpha s^3}\right)^1 \cdot \left(g^{\alpha s^2}\right)^{-3} \cdot \Big(g^{\alpha s}\Big)^2 = 
	g^{\alpha s^3} \cdot g^{-3 \alpha s^2} \cdot g^{2 \alpha s} = g^{\alpha(s^3 - 3 s^2 + 2s)} $
	\item Verifier checks ${\left(g^p\right)}^\alpha = g^{p'}$:
	\begin{flalign*}
		{\left(g^{s^3 - 3s^2 + 2s}\right)}^\alpha &= g^{\alpha(s^3 - 3 s^2 + 2s)} &&\\
		g^{\alpha(s^3 - 3 s^2 + 2s)} &= g^{\alpha(s^3 - 3 s^2 + 2s)} &&
	\end{flalign*}
\end{itemize}

Now we can be sure that the prover did not use anything else other than the provided by verifier polynomial, since there is no other way to preserve the $\alpha$-shift. 
Also if a verifier would want to ensure exclusion of some power(s) of $s$ in a prover's polynomial, e.g., $j$, he will not provide encryption $g^{s^j}$ and its \textit{shift} $g^{\alpha s^j}$.

Compared to what we have started with, we now have a robust protocol. However there is still a significant drawback to the \zeroKnowledge\ property, regardless of encryption: while theoretically polynomial coefficients $c_i$ can have a vast range of values, in reality, it might be quite limited (6 in the previous example), which means that the verifier could brute-force limited range of coefficients combinations until the result is equal to the prover's answer. For instance if we consider the range of $100$ values for each coefficient, the degree 2 polynomial would total to 1 million of distinct combinations, which considering  brute-force would require less than 1 million iterations. Moreover, the secure protocol should be secure even in cases where there is only one coefficient, and its value $\text{is}\ 1$.

\todone{note: $c^0$ alone opens door for the proof forgery}

\subsection{Zero-Knowledge} \label{section:polyzk}

Because verifier can extract knowledge about the unknown polynomial $p(x)$ only from the data sent by the prover, let us consider those provided values (the proof): $g^p, g^{p'}, g^h$. 
They participate in the following checks:
\begin{align*}
	g^p &= \left(g^h\right)^{t(s)} & &\text{(polynomial\  $p(x)$ has roots of $t(x)$)}\\	
	\left(g^p\right)^\alpha &= g^{p'} & &\text{(polynomial of a correct form is used)}
\end{align*}

The question is how do we alter the proof such that the checks still hold, but no knowledge can be extracted? One answer can be derived from the previous section: we can ``shift'' those values by some random number $\delta$ (delta), e.g., $\left(g^{p}\right)^\delta$. Now, in order to extract the knowledge, one first needs to find $\delta$ which is considered infeasible. Moreover, such randomization is statistically indistinguishable from random.

To maintain relationships let us examine the verifier's checks. One of the prover's values is on each side of the equations. Therefore if we ``shift" each of them with the same $\delta$ the equations must remain balanced.

Concretely, prover samples a random $\delta$  and exponentiates his proof values with it $\left(g^{p(s)}\right)^\delta$, $\left(g^{h(s)}\right)^\delta$, $\left(g^{\alpha p(s)}\right)^\delta$ and provides to the verifier for verification: 
\begin{align*}
	\left(g^p\right)^\delta &= \left(\left(g^h\right)^\delta\right)^{t(s)} \\
	\left(\left(g^p\right)^\delta\right)^\alpha &= \left(g^{p'}\right)^\delta
\end{align*}

After consolidation we can observe that the check still holds:
\begin{align*}
	g^{\delta \cdot p} &= g^{\delta \cdot t(s) h} \\
	g^{\delta \cdot \alpha p} &= g^{\delta \cdot p'} 
\end{align*}

\note{how easily the zero-knowledge is woven into the construction, this is often referred to as ``free'' zero-knowledge.}

\subsection{Non-Interactivity}

Till this point, we had an \emph{interactive \zeroKnowledge}\ scheme. Why is that the case? Because the proof is only valid for the original verifier, nobody else (other verifiers) can trust the same proof since:
\begin{itemize}
	\item the verifier could collude with the prover and disclose those secret parameters $s, \alpha$ which allows to fake the proof, as mentioned in remark \ref{remark:prePairingIssues}
	\item the verifier can generate fake proofs himself for the same reason
	\item verifier have to store $\alpha$ and $t(s)$ until all relevant proofs are verified, which allows an extra attack surface with possible leakage of secret parameters
\end{itemize}

Therefore a separate interaction with every verifier is required in order for a statement (knowledge of polynomial in this case) to be proven.

While interactive proof system has its use cases, for example when a prover wants to convince only a dedicated verifier 
(called designated verifier\footnote{More on designated verifier in \cite{jakobsson1996designated}}) such that the proof cannot be re-used to prove same statement to others,
 it is quite inefficient when one needs to convince many parties simultaneously (e.g., in distributed systems such as blockchain) or permanently. Prover would be required to stay online at all times and perform the same computation for every verifier.
 
Hence, we need the secret parameters to be reusable, public, trustworthy and infeasible to abuse. 

Let us first consider how would we secure the secrets ($t(s), \alpha$) after they are produced. We can encrypt them the same way verifier encrypts powers of $s$ before sending  to the prover. However as mentioned in the remark \ref{remark:helimits}, the homomorphic encryption we use does not support the multiplication of two encrypted values, which is necessary for both verification checks to multiply encryptions of $t(s)$ and $h$ as well as $p$ and $\alpha$.
This is where cryptographic pairings fit in.

\subsubsection{Multiplication of Encrypted Values}\label{section:pairings}

Cryptographic pairings (bilinear map) is a mathematical construction, denoted as a function $e(g^*, g^*)$, which given two encrypted inputs (e.g., $g^a, g^b$) from one set of numbers allows to map them deterministically to their multiplied representation in a different output set of numbers, i.e., $e(g^a, g^b) = e(g,g)^{ab}$:

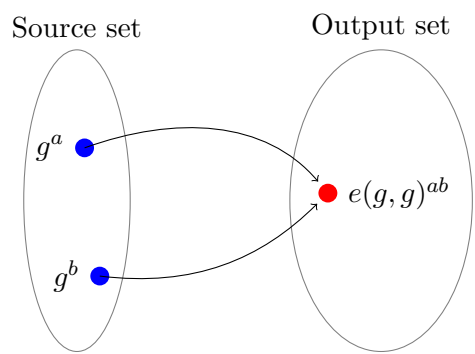
\begin{center}
	\begin{tikzpicture}
		\coordinate (A) at (0.1cm,0.7cm);
		\coordinate (B) at (0.3cm,-1cm);
		\coordinate (T) at (3.3cm,0.1cm);
	
		\node[] at (0,2.3cm) {Source set};
		\draw[gray] (0,0) ellipse (0.7cm and 2cm);
		
		\node[label={180:{$g^a$}},circle,blue,inner sep=2.5pt,fill] at (A) {};
		\node[label={180:{$g^b$}},circle,blue,inner sep=2.5pt,fill] at (B) {};	
		
		\node[] at (4cm,2.3cm) {Output set};
		\draw[gray] (4cm,0) ellipse (1.2cm and 2cm);
		
		\node[label={0:{$e(g, g)^{ab}$}},circle,red,inner sep=2.5pt,fill] at (T) {};	
			
		\draw[->, shorten >= 0.2cm] (A) to[out=20,in=130] (T);
		\draw[->, shorten >= 0.2cm] (B) to[out=-7,in=225] (T);
		
	\end{tikzpicture}
\end{center}

\todone{graph showing source group and a target group set with arrows suggesting mapping, where the final destination is in different field}.

Because the source and output number sets\footnote{Usually referred to as a group.} are different the result of the pairing is not usable as an input for another pairing operation. We can look at the output set (also called ``target set") as  being from a ``different universe." Therefore we cannot multiply the result by another encrypted value and suggested by the name itself we can only multiply two encrypted values at a time. 

In some sense, it resembles a hash function, which maps all possible input values to an element in the set of possible output values and it is not trivially reversible.

\note{from first glance, such limitation must only impede a dependent functionality, ironically in the \zkSNARK\ case it is a paramount property on which security of the scheme holds, see remark \ref{pairingSecurity}.}

\todone{graph of big input set and smaller output set with arrow?}

A rudimentary (and technically incorrect) mathematical analogy for pairing function $e(g^*, g^*)$ would be to state that there is a way to ``swap" each input's base and exponent, such that base $g$ is modified in the process of transformation into exponent, e.g., $g^a \rightarrow a^{\mathbf{g}}$. Both ``swapped" inputs are then multiplied together, such that raw $a$ and $b$ values get multiplied under the same exponent, e.g.:
$$ e(g^a, g^b) = a^{\mathbf{g}} \cdot b^{\mathbf{g}} = \left( ab \right)^{\mathbf{g}} $$

Therefore because the base gets altered during the ``swap" using the result $\left( ab \right)^{\mathbf{g}}$ in another pairing (e.g., $\prng{\left( ab \right)^{\mathbf{g}}, g^c }$) would not produce desired encrypted multiplication $abc$.

The core properties of pairings can be expressed in the equations:
\begin{eqnarray*}
	e(g^a, g^b) = e(g^b, g^a) = e(g^{ab}, g^1) = e(g^1, g^{ab}) = {e(g^1, g^a)}^b = {e(g^1, g^1)}^{ab} = \ldots
\end{eqnarray*}

Technically the result of a pairing is an encrypted product of raw values under a different generator $\mathbf{g}$ of the target set, i.e., $e(g^a, g^b) = \mathbf{g}^{ab}$. Therefore it has properties of the homomorphic encryption, e.g., we can add the encrypted products of multiple pairings together:
\begin{eqnarray*}
	e(g^a, g^b) \cdot e(g^c, g^d) = \mathbf{g}^{ab} \cdot \mathbf{g}^{cd} = \mathbf{g}^{ab + cd} = e(g, g)^{ab + cd}
\end{eqnarray*}

\note{cryptographic pairing is leveraging elliptic curves to achieve these properties, therefore from now on notation\ \ $g^n$\ \ will represent a generator point on a curve added to itself $n$ times instead of a multiplicative group generator which we have used in previous sections.}

The survey \cite{cryptoeprint:2004:064} provides a starting point for exploration of the cryptographic pairings.

\subsubsection{Trusted Party Setup}

Having cryptographic pairings, we are now ready to set up secure public and reusable parameters. Let us assume that we trust a single honest party to generate secrets $s$ and $\alpha$. As soon as $\alpha$ and all necessary powers of $s$ with corresponding $\alpha$-shifts are encrypted ($g^\alpha, g^{s^i}, g^{\alpha s^i}$ for $i$ in $0, 1, \ldots, d$), the raw values must be deleted.

These parameters are usually referred to as \textit{common reference string} or CRS. After CRS is generated any prover and any verifier can use it in order to conduct non-interactive zero-knowledge proof protocol.
While non-crucial, the optimized version of CRS will include encrypted evaluation of the \term{target polynomial} $g^{t(s)}$. 

Moreover CRS is divided into two groups (for $i$ in $0, 1, \ldots, d$):
\begin{itemize}
	\item Proving key\footnote{Also called \textit{evaluation key}}: $(g^{s^i}, g^{\alpha s^i})$
	\item Verification key: $(g^{t(s)}, g^{\alpha})$
\end{itemize}

Being able to multiply encrypted values the verifier can check the polynomials in the last step of the protocol:
\begin{itemize}
	\item Having verification key verifier processes received encrypted polynomial evaluations $g^p, g^h, g^{p'}$ from the prover:
	\begin{itemize}
	\item checks that\ \ $p = t \cdot h$\ \ in encrypted space:
	
	$\prng{g^p, g^1} = \prng{g^t, g^h} \quad$ which is equivalent to
	$\quad\prng{g, g}^{p} = \prng{g, g}^{t \cdot h}$
	
	\item checks polynomial restriction: \\
	$\prng{g^p, g^\alpha} = \prng{g^{p'}, g}$
	\end{itemize}
\end{itemize}

\subsubsection{Trusting One out of Many}

While the trusted setup is efficient, it is not effective since multiple users of CRS will have to trust that one deleted $\alpha$ and $s$, since currently there is no way to prove that\footnote{Proof of ignorance is an area of active research \cite{cryptoeprint:2018:896}}. Hence it is necessary to minimize or eliminate that trust. Otherwise, a dishonest party would be able to produce fake proofs without being detected.

One way to achieve that is by generating a \textit{composite} CRS by multiple parties employing mathematical tools introduced in previous sections, such that neither of those parties knows the secret. Here is an approach, let us consider three participants Alice, Bob and Carol with corresponding indices A, B and C, for $i$ in $1, 2, \ldots, d$:
\begin{itemize}
	\item Alice samples her random $s_A$ and $\alpha_A$ and publishes her CRS:\\ 
	$\left( g^{s^i_A}, g^{\alpha_A}, g^{\alpha_A s^i_A} \right)$
	\item Bob samples his $s_B$ and $\alpha_B$ and augments Alice's encrypted CRS through homomorphic multiplication: \\
	$\left( \left(g^{s^i_A}\right)^{s^i_B}, \left(g^{\alpha_A}\right)^{\alpha_B}, \left(g^{\alpha_A s^i_A}\right)^{\alpha_B s^i_B} \right) = 
	\left(
		g^{{(s_A s_B)}^i}, 
		g^{\alpha_A \alpha_B}, 
		g^{\alpha_A \alpha_B {(s_A  s_B)}^i}
	\right)$
	
	and publishes the resulting two-party Alice-Bob CRS: \\
	$ \left(
		g^{s\psf{AB}^i}, 
		g^{\alpha\psf{AB}}, 
		g^{\alpha\psf{AB}\, s\psf{AB}^i}
	\right) $
	\item So does Carol with her $s_C$ and $\alpha_C$: \\
	$\left( 
		\left( g^{s\psf{AB}^i} \right)^{s^i_C}, 
		\Big( g^{\alpha\psf{AB}} \Big)^{\alpha_C}, 
		\left( g^{\alpha\psf{AB}\, s\psf{AB}^i} \right)^{\alpha_C s^i_C} 
	\right) = 
	\left( 
		g^{\left(s_A s_B s_C \right)^i}, 
		g^{\alpha_A \alpha_B \alpha_C}, 
		g^{\alpha_A \alpha_B \alpha_C {(s_A  s_B s_C)}^i} 
	\right) $ 
	
	and publishes Alice-Bob-Carol CRS: \\
	$\left( 
		g^{s\psf{ABC}^i}, 
		g^{\alpha\psf{ABC}}, 
		g^{\alpha\psf{ABC}\, s\psf{ABC}^i} 
	\right)$
\end{itemize}

As the result of such protocol, we have composite $s^i = s^i_A s^i_B s^i_C$, and $\alpha = \alpha_A \alpha_B \alpha_C$ and no participant learns secret parameters of other participants unless they are colluding. In fact, in order to learn $s$ and $\alpha$, one must collude with every other participant. Therefore even if one out of all is honest, it will be infeasible to produce fake proofs.

\note{this process can be repeated for as many participants as necessary.}

The question one might have is how to verify that participant have been consistent with every value of CRS, because an adversary can sample multiple different $s_1, s_2, \ldots$ and $\alpha_1, \alpha_2, \ldots $, and use those randomly for different powers of $s$ (or provide random numbers as an augmented common reference string), rendering CRS invalid and unusable.

Luckily, because we can multiply encrypted values using pairings, we are able to perform consistency check, starting with the first parameter and ensuring that every next is derived from it. Every published CRS by participants can be checked as follows:
\begin{itemize}
	\item We take power 1 of $s$ as canonical value and check every other power for consistency with it: \\
	$\left. e\left(g^{s^i}, g\right) = e\left(g^{s^1}, g^{s^{i-1}}\right) \right|_{i \in \{2, \ldots, d\}}$ \\
	for example:
	\begin{itemize}
	\item Power 2: $e\left(g^{s^2}, g\right) = e\left(g^{s^1}, g^{s^{1}}\right) \Rightarrow {e(g,g)}^{s^2} = {e(g,g)}^{s^{1+1}}$
	\item Power 3: $e\left(g^{s^3}, g\right) = e\left(g^{s^1}, g^{s^{2}}\right) \Rightarrow {e(g,g)}^{s^3} = {e(g,g)}^{s^{1+2}}$
	, etc.
	\end{itemize}
	\item We now check if the $\alpha$-shift of values in the previous step is correct:\\
	$\left. e\left(g^{s^i}, g^{\alpha}\right) = e\left(g^{\alpha s^i}, g\right)\right|_{i \in [d]}$ \\
	for example:
	\begin{itemize}
		\item Power 3: $e\left(g^{s^3}, g^{\alpha}\right) = e\left(g^{\alpha s^3}, g\right) \Rightarrow {e(g,g)}^{s^3 \cdot \alpha} = {e(g,g)}^{\alpha s^3} $, etc.
	\end{itemize}
	
	\vspace*{0.5em}
	\item[] where $i \in \{2, \ldots, d\}$ is a shortened form of ``$i$ is in $2, 3, \ldots, d$" and $[d]$ is a shortened form of $1, 2, \ldots, d$, which is the more convenient notation for the next sections
\end{itemize}

\todone{describe that every additional participant who augments CRS has to publish encrypted version of his secrets so as to prove that they are truly augmenting previous version and not construction from scratch}

Notice that while we verify that every participant is consistent with their secret parameters, the requirement to use previously published CRS is not enforced for every next party (Bob and Carol in our example). Hence if an adversary is the last in the chain he can ignore the previous CRS and construct valid parameters from scratch, as if he was the first in the chain, therefore being the only one who knows secret $s$ and $\alpha$.

We can address this by additionally requiring every participant except the first one to encrypt and publish his secret parameters, for example, Bob also publishes:
	\[ \left. \left(
		 g^{s^i_B}, g^{\alpha_B}, g^{\alpha_B s^i_B}
	\right) \right|_{i \in [d]} \]
	
This allows to validate that Bob's CRS is a proper multiple of Alice's parameters, for $i$ in $1, 2, \ldots, d$:
\begin{itemize}
	\item $\prng{g^{s^i\psf{AB}}, g} = \prng{g^{s^i_A}, g^{s^i_B}}$
	\item $\prng{g^{\alpha\psf{AB}}, g} = \prng{g^{\alpha_A}, g^{\alpha_B}}$
	\item $\prng{g^{\alpha\psf{AB}\, s^i\psf{AB}}, g} = \prng{g^{\alpha_A s^i_A}, g^{\alpha_B s^i_B}}$
\end{itemize}

Similarly Carol will have to prove that her CRS is a proper multiple of Alice-Bob's CRS.

This is a robust CRS setup scheme which does not rely entirely on any single party. In fact, it is sufficient if only one party is honest and deletes and never shares its secret parameters, even if all other parties have colluded. So the more there are unrelated  participants in CRS setup\footnote{Sometimes called ceremony \cite{ceremony:2016}} the faintest the possibility of fake proofs, the probability becomes negligible if competing parties are participating. The scheme allows involving other untrusted parties who are in doubt about the legibility of the setup because verification step ensures they are not sabotaging (which also includes usage of weak $\alpha$ and $s$) the final common reference string.

\subsection{Succinct Non-Interactive Argument of Knowledge of Polynomial}\label{section:zkSNARKOP}

We are now ready to consolidate the evolved \zkSNARKOP\ protocol. Being formal, for brevity, we will be using curly brackets to denote a set of elements populated by the subscript next to it, for example $\left\{ s^i \right\}_{i \in [d]}$ denotes a set $s^1, s^2, \ldots, s^d$.

Having agreed upon \term{target polynomial} $t(x)$ and degree $d$ of the prover's polynomial:

\zkProtocol{
	\item sample random values $s, \alpha$
	\item calculate encryptions $g^\alpha$ and
		$\gsetsid, \gsetasid$
	\item proving key: $\left( \gsetsid, \gsetasid \right)$
	\item verification key: 
		$ \left( g^\alpha, g^{t(s)} \right) $	
}{
	\item assign coefficients $\left\{ c_i \right\}_{i \in \{0,\ldots, d\}}$ (i.e., knowledge), $p(x) = c_d x^d + \cdots + c_1 x^1 + c_0 x^0$
	\item calculate polynomial $h(x) = \frac{p(x)}{t(x)}$
	\item evaluate encrypted polynomials $g^{ p(s)}$ and $g^{ h(s)}$ using $\gsetsid$ 
	\item evaluate encrypted shifted polynomial $g^{ \alpha p(s)}$ using $\gsetasid$
	\item sample random $\delta$
	\item set the randomized proof $\pi = \left(g^{\delta p(s)}, g^{\delta h(s)}, g^{\delta \alpha p(s)} \right)$
}{
	\item parse proof $\pi$ as $\left(g^{p}, g^{h}, g^{p'} \right)$
	\item check \checkPolynomialRestriction\ \ \ $\prng{g^{p'}, g} = \prng{g^{p}, g^\alpha}$
	\item check \checkCofactors \ \ $e\left(g^p, g\right) = e\left(g^{t(s)}, g^h\right)$
}

\begin{remark}\label{pairingSecurity}
	If it would be possible to reuse result of pairing for another multiplication such protocol would be completely insecure because the prover can assign $g^{p'} = \prng{g^{p}, g^\alpha}$ which would then pass the ``\checkPolynomialRestriction" check:
	$$ e\left(\prng{g^{p}, g^\alpha}, g\right) = e\left(g^{p}, g^\alpha\right) $$
\end{remark}

\subsubsection{Conclusions}
We came to the zero-knowledge succinct non-interactive arguments of knowledge protocol for the knowledge of a polynomial problem, which is a niche use-case. While one can claim that a prover can easily construct such polynomial $p(x)$ just by multiplying $t(x)$ by another bounded polynomial to make it pass the test, the construction is still useful.

\todone{the security of the protocol is not proven}

Verifier knows that the prover has a valid polynomial but not which particular one. We could add additional proofs of other properties of the polynomial such as: divides by multiple polynomials, is a square of a polynomial. There could be a service which accepts, stores and rewards all the attested polynomials, or there is a need in an encrypted evaluation of unknown polynomials of a necessary form. However, having universal scheme would allow for a myriad of applications.

\section{General-Purpose Zero-Knowledge Proofs}

We have paved our way with a simple yet sufficient example involving most of the \zkSNARK\ machinery, and it is now possible to advance the scheme to execute zero-knowledge programs.

\subsection{Computation}

Let us consider a simple program in pseudocode:

\begin{algorithm}[H] 
\caption{Operation depends on an input}
\label{alg:DependsOnInput}
\begin{algorithmic} 
    \Function{\textnormal{calc}}{w, a, b}
    		\If{w}
    			\State \Return{a $\times$ b}
    		\Else
    			\State \Return{a + b}
    		\EndIf
    \EndFunction
\end{algorithmic}
\end{algorithm}

From a high-level view, it is quite unrelated to polynomials, which we have the protocol for. Therefore we need to find a way to convert a program into the polynomial form. The first step then is to translate the program into the language of math, which is relatively easy, the same statement can be expressed as following (assuming $w$ is either 0 or 1):
$$f(w, a, b) = w (a \times b) + (1 - w) (a + b) $$

Executing calc(1, 4, 2) and evaluating $f(1, 4, 2)$ will yield the same result: 8. Conversely calc(0, 4, 2) and $f(0, 4, 2)$ would both be resolved to 6. We can express any kind of finite program in such a way. 

What we need to prove then (in this example), is that for the input $(1, 4, 2)$ of expression $f(w, a, b)$ the output is $8$, in other words, we check the equality:
$$w (a \times b) + (1 - w) (a + b) = 8$$

\subsection{Single Operation}
We now have a general computation expressed in a mathematical language, but we still need to translate it into the realm of polynomials. Let us have a closer look at what computation is in a nutshell. Any computation at it is core consists of elemental operations of the form:
$$ \mathsf{left\ operand}\quad \mathbf{operator}\quad \mathsf{right\ operand}\quad =\quad \mathsf{output} $$

Two operands (i.e., values) are being operated upon by an operator (e.g., $+, -, \times, \div$). For example for operands 2 and 3 and operator ``multiplication" these will resolve to $2 \times 3 = 6$.  Because any complex computation (or a program) is just a series of operations, firstly we need to find out how single such operation can be represented by a polynomial.

\subsubsection{Arithmetic Properties of Polynomials}

Let us see how polynomials are related to arithmetic operations.
If you take two polynomials $f(x)$ and $g(x)$ and try, for example, to multiply them $h(x) = f(x) \times g(x)$, the result of evaluation of $h(x)$ at any $x = r$ will be the multiplication of results of evaluations of $f(r)$ and $g(r)$. Let us consider two following polynomials:
$f(x) = 2x^2 - 9x + 10 $ and $g(x) = -4x^2 + 15 x -9$. Visualized in the form of graph:

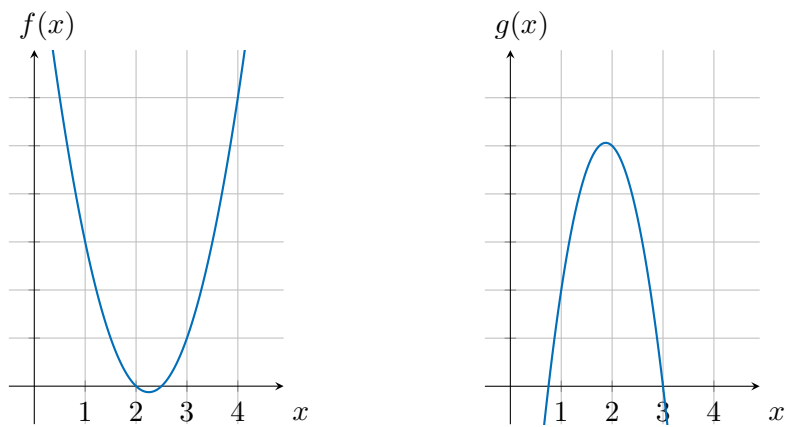
\newcommand{\polyF}{(x*(2*x - 9) + 10)}
\newcommand{\polyG}{(x*(-4*x + 15) - 9)}

\newcommand{\plotArithEx}[4]{
	\plotgraph{#1}{
		\plotpoly{(#2)}{NavyBlue};
		#3
	}{#4}
	{ymin=-0.79, ymax=6.99, xmax=4.9, xtick={1,...,4}, ytick={1,...,12}}
}

\begin{center}
	\plotArithEx{$f(x)$}{\polyF}{}{}
	\qquad\qquad
	\plotArithEx{$g(x)$}{\polyG}{}{}
\end{center}

For $x = 1$ these will evaluate to: $f(1) = 2 -9 + 10 = 3$,\ \ $g(1) = -4 + 15 -9 = 2$.

Let us multiply the polynomials: $h(x) = f(x) \times g(x) =-8 x^4 + 66 x^3 - 193 x^2 + 231 x -90$.
Visually multiplication can be seen as:

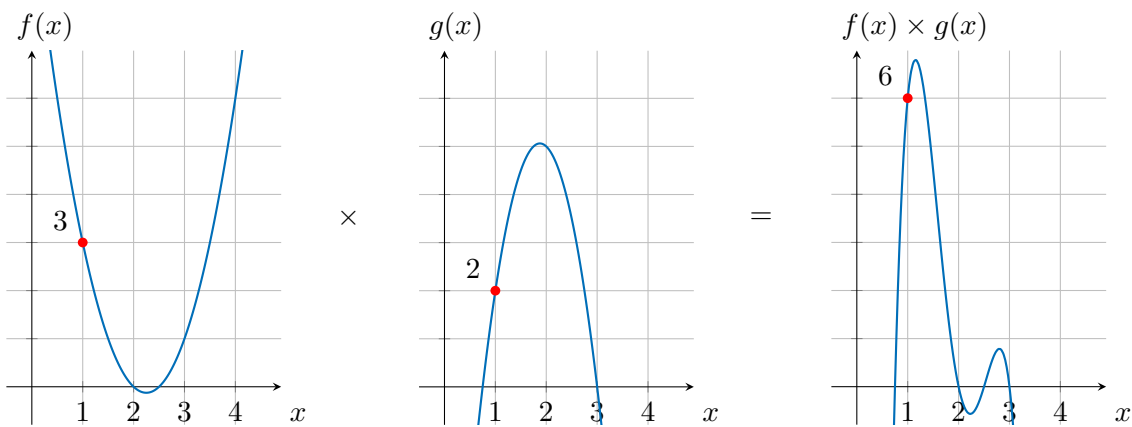
\begin{center}
	\plotArithEx{$f(x)$}
		{\polyF}
		{\plotdot{1}{3}{3}}
		{\node at (4.5,2.75) {$\times$}; }
	\plotArithEx{$g(x)$}
		{\polyG}
		{\plotdot{1}{2}{$2$}}
		{\node at (4.5,2.75) {$=$}; }
	\plotArithEx{$f(x) \times g(x)$}
		{\polyG * \polyF}
		{\plotdot{1}{6}{6}}
		{}
\end{center}

If we examine evaluations at $x = 1$ on the resulting polynomial $f(x) \times g(x)$ we will get: $h(1) = -8 + 66 -193 + 231 - 90 = 6$, hence the values at $x = 1$ of $f(x)$ and $g(x)$ has multiplied, and respectively at every other $x$.

Likewise if we add $f(x)$ and $g(x)$ we will get $-2x^2 + 6x + 1$ which 
evaluates to $5$ at $x = 1$.

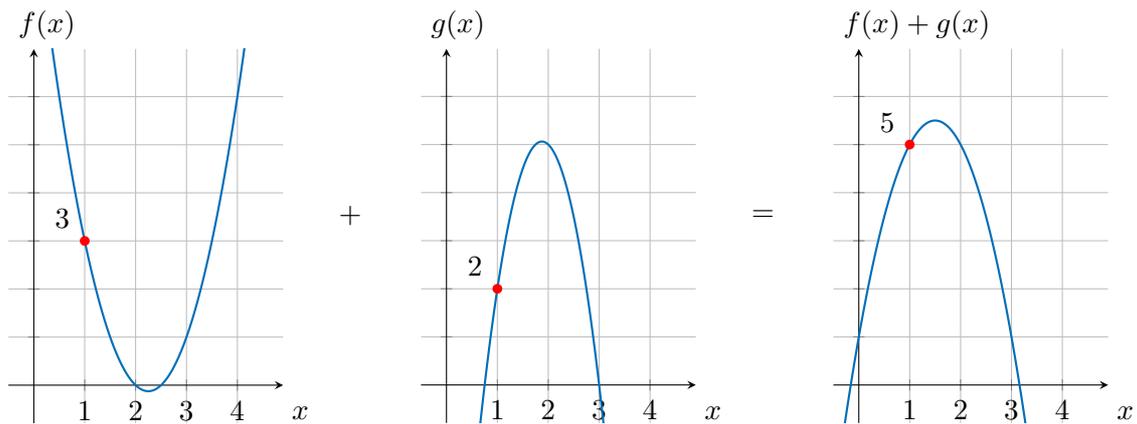
\begin{center}
	\plotArithEx{$f(x)$}
		{\polyF}
		{\plotdot{1}{3}{3}}
		{\node at (4.5,2.75) {$+$};}
	\plotArithEx{$g(x)$}
		{\polyG}
		{\plotdot{1}{2}{$2$}}
		{\node at (4.5,2.75) {$=$};}
	\plotArithEx{$f(x) + g(x)$}
		{\polyG + \polyF}
		{\plotdot{1}{5}{5}}
		{}
\end{center}

\note{evaluations at other $x$-s were also added together, e.g., examine $x = 2, x = 3$.}

If we can represent operand values as polynomials (and we indeed can as outlined) then through the arithmetic properties, we will be able to get the result of an operation imposed by an operand.

\subsection{Enforcing Operation}
If a prover claims to have the result of multiplication of two numbers how does verifier checks that?
To prove the correctness of a single operation, we must enforce the correctness of the output (result) for the operands provided.
If we look again at the form of operation:
\[ \mathsf{left\ operand}\quad \mathbf{operator}\quad \mathsf{right\ operand}\quad =\quad \mathsf{output} \]

The same can be represented as an \term{operation polynomial}:
$$ l(x) \ \mathbf{operator} \ r(x) = o(x) $$
where for some chosen $a$:
\begin{itemize}
	\item $l(x)$ - at $a$ represents (evaluates to) the value of the left operand 
	\item $r(x)$ - at $a$ represents the value of the right operand
	\item $o(x)$ - at $a$ represents the result (output) of the operation
\end{itemize}

Therefore if the operands and the output are represented correctly for the operation by those polynomials, then the evaluation of\ \ $l(a) \ \mathbf{operator} \ r(a) = o(a)$\ \ should hold. And moving \term{output polynomial} $o(x)$ to the left side of the equation\ \ $l(a) \ \mathbf{operator} \ r(a) - o(a) = 0$\ \ is surfacing the fact that the \term{operation polynomial}\ \  $l(x) \ \mathbf{operator} \ r(x) - o(x) = 0$ \ \ has to evaluate to 0 at $a$, if the value represented by the \term{output polynomial} $o(x)$ is the correct result produced by the $\mathbf{operator}$ on the values represented by \term{operand polynomials} $l(x)$ and $r(x)$. Henceforth \term{operation polynomial} must have the root $a$ if it is valid, and consequently, it must contain cofactor $(x - a)$ as we have established previously (see \hyperref[section:factorization]{factorization, section \ref*{section:factorization}}), which is the \term{target polynomial} we prove against, i.e., $t(x) = x - a$.

For example, let us consider operation: \[3 \times 2 = 6\]

It can be represented by simple polynomials $l(x) = 3x$, $r(x) = 2x$, $o(x) = 6x$, which evaluate to the corresponding values for $a = 1$, i.e., $l(1) = 3;\ r(1) = 2;\ o(1) = 6$.

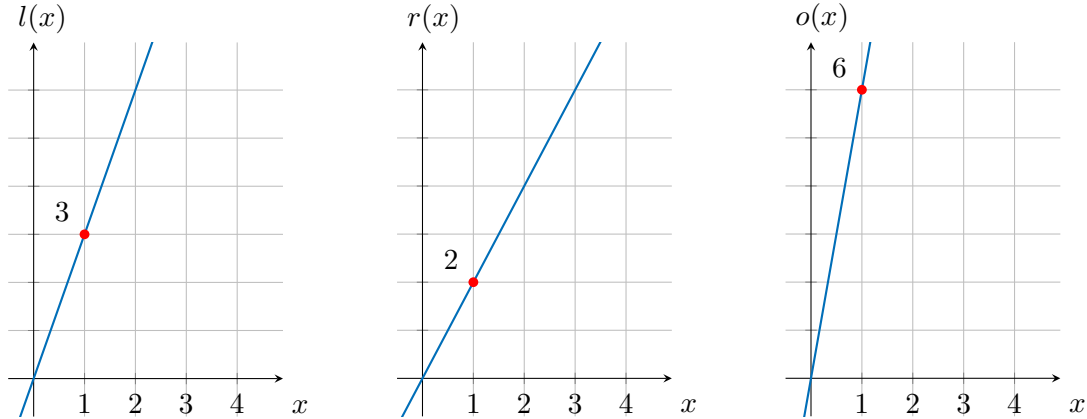
\begin{center}
	\plotArithEx{$l(x)$}
		{3*x}
		{\plotdot{1}{3}{3}}
		{}
	\quad
	\plotArithEx{$r(x)$}
		{2*x}
		{\plotdot{1}{2}{$2$}}
		{}
	\quad
	\plotArithEx{$o(x)$}
		{6*x}
		{\plotdot{1}{6}{6}}
		{}
\end{center}

\note{The value of\ \ $a$ can be arbitrary.}

The operation polynomial then will be:
\begin{gather*}
	l(x) \times r(x) = o(x) \\
	3x \times 2x = 6x \\
	6x^2 - 6x = 0
\end{gather*}
Which is visualised as:

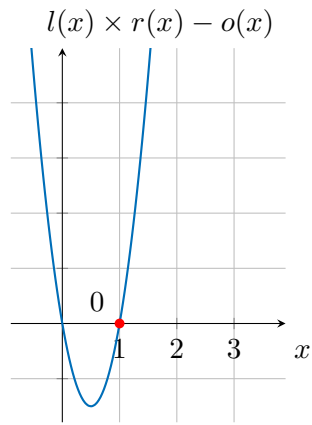
\begin{center}
	\plotgraph{$l(x) \times r(x) - o(x)$}{
		\plotpoly{6*x^2 - 6*x}{NavyBlue};
		\plotdot{1}{0}{0}
	}{}
	{ymin=-1.79, ymax=4.99, xmin=-0.9, xmax=3.9, domain=-1:5, xtick={1,...,4}, ytick={-1,...,7}}
\end{center}

It is noticeable that the operation polynomial has $(x-1)$ as a co-factor:
$$ 6x^2 - 6x = 6x (x - 1) $$

Therefore if the prover provides such polynomials $l(x), r(x), o(x)$ instead of former $p(x)$ then the verifier will accept it as valid, since it is divisible by $t(x)$. On the contrary if the prover tries to cheat and substitutes output value with 4, e.g., $o(x) = 4x$, then the \term{operation polynomial} will be $6x^2 - 4x = 0$:

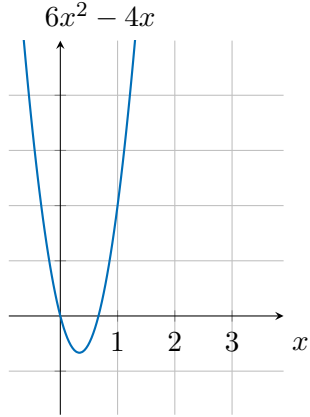
\begin{center}
	\plotgraph{$6x^2 - 4x$}{
		\plotpoly{6*x^2 - 4*x}{NavyBlue};
	}{}
	{ymin=-1.79, ymax=4.99, xmin=-0.9, xmax=3.9, domain=-1:5, xtick={1,...,4}, ytick={-1,...,7}}
\end{center}

Which is not have a solution $x=1$, henceforth $l(x) \times r(x) - o(x)$ is not divisible by $t(x)$ without remainder: \\
$h(x) = $\polylongdiv{6x^2 - 4x}{(x-1)} $\quad\quad \Rightarrow \quad 
h(x) = 6x + 2 + \frac{2}{x-1}$

Hence such \textit{inconsistent operation} will not be accepted by the verifier\footnote{As described in \hyperref[section:factorization]{section \ref*{section:factorization}}}.

\subsection{Proof of Operation} \label{sec:proofOfOperation}

Let us modify our latest protocol to support a single multiplication operation proof. Recall that previously we had proof of knowledge of polynomial $p(x)$, but now we deal with three $l(x), r(x), o(x)$. While we could define $p(x) = l(x) \times r(x) - o(x)$ there are two counterargument. Firstly, in our protocol, the multiplication of encrypted values (i.e., $l(s) \times r(s)$) is not possible in the proving stage, since pairings can only be used once and it is required for the ``\checkPolynomialRestriction " check. Secondly, this would leave an opportunity for the prover to modify the structure of polynomial at will but still maintain a valid cofactor $t(x)$, for example $p(x) = l(x)$ or $p(x) = l(x) - r(x)$ or even $p(x) = l(x) \times r(x) + o(x)$, as long as $p(x)$ has root $a$. Such modification effectively means that the proof is about a different statement, which is certainly not desired.

That is why the evaluations of polynomials $l(s)$, $r(s)$, $o(s)$ have to be provided separately by the prover. This means that the \emph{knowledge of polynomial} must be adjusted. In essence what a verifier needs to check in encrypted space is that $l(s) \times r(s) - o(s) = t(s) h(s)$. While a verifier can perform multiplication using cryptographic pairings, the subtraction ($-o(x)$) is an expensive operation\footnote{Would require to find inverse of $g^{o(s)}$} that is why we move $o(x)$ to the right side of the equation: $l(x) r(x) = t(x) h(x) + o(x)$. In encrypted space verifier's check translates to:
\begin{align*}
	e\left(g^{l(s)}, g^{r(s)}\right) &= e\left(g^{t(s)}, g^{h(s)}\right) \cdot e\left(g^{o(s)}, g\right) \\
	e(g, g)^{l(s) r(s)} &= e(g,g)^{t(s)h(s)} \cdot e(g,g)^{o(s)} \\
	e(g, g)^{l(s) r(s)} &= e(g,g)^{t(s)h(s) + o(s)}
\end{align*}

\note{recall that the result of cryptographic pairings supports encrypted addition through multiplication, see section \ref{section:pairings}}.
		
While the \emph{setup} stage stays unchanged, here is the updated protocol:
\begin{itemize}
	\item Proving
	\begin{itemize}
		\item assign corresponding coefficients to the $l(x)$, $r(x)$, $o(x)$
		\item calculate polynomial $h(x) = \frac{l(x) \times r(x) - o(x)}{t(x)}$
		\item evaluate encrypted polynomials $g^{l(s)}$, $g^{r(s)}$, $g^{o(s)}$ and $g^{h(s)}$ using $\gsetsid$ 
		\item evaluate encrypted shifted polynomials $g^{\alpha l(s)}$, $g^{\alpha r(s)}$, $g^{\alpha o(s)}$ using $\gsetasid$
		\item set proof $\pi = \left( g^{l(s)}, g^{r(s)}, g^{o(s)}, g^{h(s)}, g^{\alpha l(s)}, g^{\alpha r(s)}, g^{\alpha o(s)} \right)$
	\end{itemize}
\item Verification
	\begin{itemize}
		\item parse proof $\pi$ as $\left( g^l, g^r, g^o, g^h, g^{l'}, g^{r'}, g^{o'} \right)$
		\item \checkPolynomialRestriction{s} check:\\
		$e(g^{l'}, g) = e(g^{l}, g^\alpha)$ \\
		$e(g^{r'}, g) = e(g^{r}, g^\alpha)$ \\
		$e(g^{o'}, g) = e(g^{o}, g^\alpha)$
		\item \checkOperation\ check: $ e\left(g^{l}, g^{r}\right) = e\left(g^{t(s)}, g^{h}\right) \cdot e\left(g^{o}, g\right) $
	\end{itemize}
\end{itemize}

Such protocol allows to prove that the result of multiplication of two values is computed correctly. 

One might notice that in the updated protocol we had to let go of the \zeroKnowledge{} component. The reason for this is to make the transition simpler. We will get back to it in a later section.

\subsection{Multiple Operations}

\newcommand{\lcolor}{ForestGreen}
\newcommand{\rcolor}{blue}
\newcommand{\ocolor}{red}
\newcommand{\lo}[1]{{\color{\lcolor} #1}}
\newcommand{\ro}[1]{{\color{\rcolor} #1}}
\newcommand{\oo}[1]{{\color{\ocolor} #1}}

\newcommand{\operspace}[4]{
\lo{#1} #4 {\times} #4 \ro{#2} #4 =& #4 \oo{#3}
}

\newcommand{\oper}[3]{
	\operspace{#1}{#2}{#3}{\quad}
}

\newcommand{\opershort}[3]{
	\operspace{#1}{#2}{#3}{\enspace}
}
We can prove a single operation, but how do we scale to prove multiple operations (which is our ultimate goal)? Let us try to add just one another operation. Consider the need to compute the product: $a \times b \times c$. In the elemental operation model this would mean two operations:
\begin{align*}
	\oper{a}{b}{r_1} \\
	\oper{r_1}{c}{r_2}
\end{align*}

As discussed previously we can represent one such operation by making operand polynomials evaluate to a corresponding value at some arbitrary $x$, for example $1$. Having this the properties of polynomials does not restrict us in representing other values at different $x$, for example $2$, e.g.:

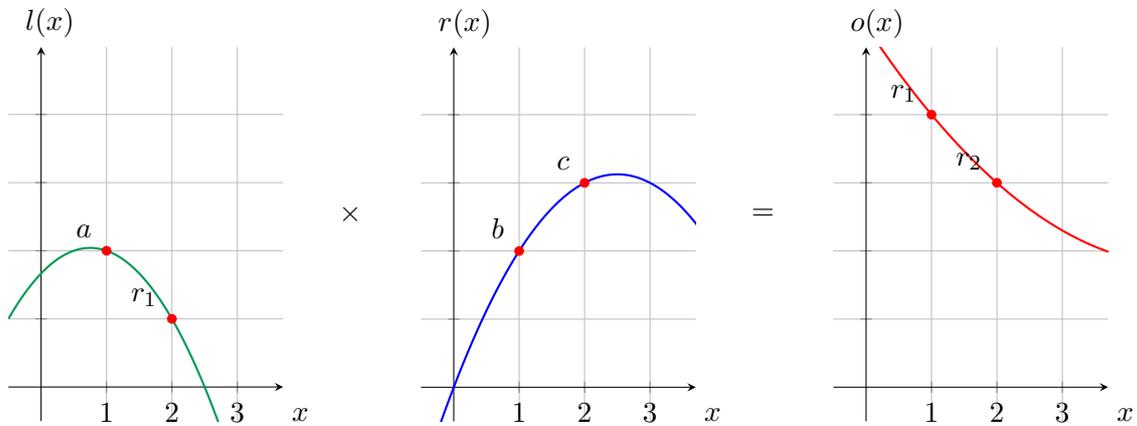
\begin{center}
	\plotgrid{$l(x)$}{
		\plotpoly{\polyl}{\lcolor};
		\plotdot{1}{2}{$a$}
		\plotdot{2}{1}{$r_1$}
	}{
		\node at (4.5,2.75) {$\times$};
	}
	\plotgrid{$r(x)$}{
		\plotpoly{\polyr}{\rcolor};
		\plotdot{1}{2}{$b$}
		\plotdot{2}{3}{$c$}
	}{
		\node at (4.5,2.75) {$=$};
	}
	\plotgrid{$o(x)$}{
		\plotpoly{\polyo}{\ocolor};
		\plotdot{1}{4}{$r_1$}
		\plotdot{2}{3}{$r_2$}
	}{}
\end{center}

Such independence allows us to \emph{execute} two operations at once without ``mixing" them together, i.e., no interfering. The result of such polynomial arithmetic will be:
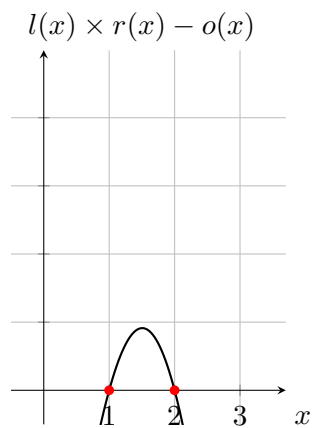
\begin{center}
	\plotgrid{$l(x) \times r(x) - o(x)$}{
		\plotpoly{(\polyl) * (\polyr) - (\polyo)}{black};
		\plotdot{1}{0}{}
		\plotdot{2}{0}{}
	}{}
\end{center}

Where it is visible that the operation polynomial has roots $x=1$ and $x=2$. Therefore both operations are \emph{executed} correctly.

Let us have a look at example of 3 multiplications $2 \times 1 \times 3 \times 2$, which can be executed as follows:
\begin{align*}
	\oper{2}{1}{2} \\
	\oper{2}{3}{6} \\
	\oper{6}{2}{12}
\end{align*}

We need to represent those as operand polynomials, such that for operations represented by $x \in \{1, 2, 3\}$ the $l(x)$ pass correspondingly through $\lo{2, 2}$ and $\lo{6}$, i.e., through points $(1,\lo{2}), (2, \lo{2}), (3, \lo{6})$, and similarly $r(x) \ni (1, \ro{1}), (2, \ro{3}), (3, \ro{2})$ and $o(x) \ni (1,\oo{2}), (2, \oo{6}), (3, \oo{12})$. 

However, how do we find such polynomials which passes through those points? For any case where we have more than one point, a particular mathematical method has to be used.

\subsubsection{Polynomial Interpolation}

In order to construct \term{operand} and \term{output polynomials} we need a method which given a set of points produces a \emph{curved} polynomial in such a way that it passes through all those points, it is called \term{interpolation}
There are different ways available:
\begin{itemize}
	\item Set of equations with unknowns
	\item Newton polynomial	
	\item Neville's algorithm
	\item Lagrange polynomials
	\item Fast Fourier transform
\end{itemize}

Let us use the former for example. The idea of such method is that there exists a unique polynomial of degree at most $n$ with yet \term{unknown coefficients} which pass through given $n+1$ points such that for each point $\left\{ \left(x_i, y_i\right) \right\}_{i \in [n+1]}$ the polynomial evaluated at $x_i$ should be equal to $y_i$. In our case for three points it will be polynomial of degree 2 of the form:
\[ ax^2 + bx + c = y \]
Let us \term{equalize} the evaluated polynomial for each point of the \term{left operand polynomial} (green) and solve the system of equations by expressing each coefficient in terms of others:

\begin{gather*}
	\begin{cases}
		l(1) = 2 \\
		l(2) = 2 \\
		l(3) = 6
	\end{cases}
	\Rightarrow	
	\begin{cases}
	a(1)^2 + b \cdot 1 + c = 2 \\
	a(2)^2 + b \cdot 2 + c = 2 \\
	a(3)^2 + b \cdot 3 + c = 6
	\end{cases}
	\Rightarrow	
	\begin{cases}
	a + b + c = 2 \\
	4a + 2b + c = 2 \\
	9a + 3b + c = 6 
	\end{cases}
	\Rightarrow
	\begin{cases}
	a = 2 - b - c \\
	2b = 2 - 4(2 - b - c) - c \\
	c = 6 - 9(2 - b - c) - 3b
	\end{cases}
	\\ \Rightarrow 
	\begin{cases}
	a = 2 - b - c \\
	b = \frac{6 - 3c}{2} \\
	c = -12 + 6b + 9c
	\end{cases}
	\Rightarrow
	\begin{cases}
	a = 2 - b - c \\
	b = \frac{6 - 3c}{2} \\
	c = -12 + 6(\frac{6 - 3c}{2}) + 9c
	\end{cases}	
	\Rightarrow
	\begin{cases}
	a = 2 \\
	b = -6 \\
	c = 6
	\end{cases}
\end{gather*}

Therefore the \term{left operand polynomial} is:
$$ \lo{l(x)} = 2x^2 - 6x + 6 $$
Which corresponds to the following graph:

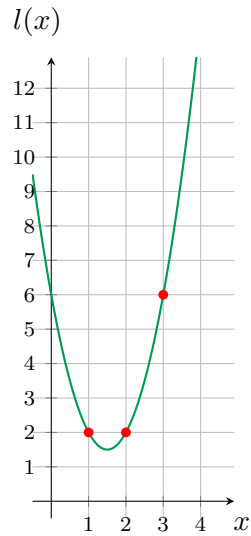
\newcommand{\polyexl}{2*x^2 - 6*x + 6}
\newcommand{\polyexr}{(-3*x^2 + 13*x - 8)/2}
\newcommand{\polyexo}{x^2 + x}

\begin{center}
	\pgfplotsset{every tick label/.append style={font=\scriptsize}}
	\plotgraph{$l(x)$}{
		\plotpoly{\polyexl}{\lcolor};
		\plotdot{1}{2}{}
		\plotdot{2}{2}{}
		\plotdot{3}{6}{}
	}{}{ymax=12.9, xmax=4.9, height=20em, width=11em, xtick={1,...,4}, ytick={1,...,12},yticklabel={}}
\end{center}

We can find $r(x)$ and $o(x)$ in the same way:
\begin{equation*}
	\ro{r(x)} = \frac{-3x^2 + 13x - 8}{2}; \quad \oo{o(x)} = x^2 + x
\end{equation*}

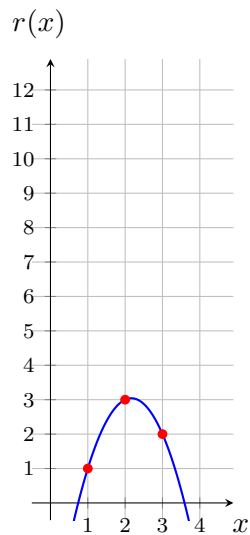
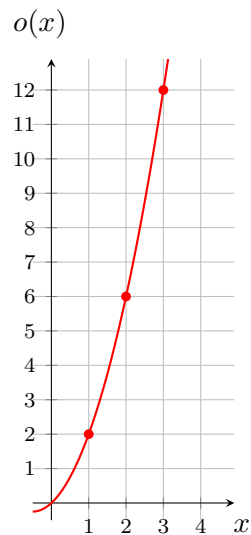
\begin{center}
	\pgfplotsset{every tick label/.append style={font=\scriptsize}}
	\plotgraph{$r(x)$}{
		\plotpoly{\polyexr}{\rcolor};
		\plotdot{1}{1}{}
		\plotdot{2}{3}{}
		\plotdot{3}{2}{}
	}{}{ymax=12.9, xmax=4.9, height=20em, width=11em, xtick={1,...,4}, ytick={1,...,12},yticklabel={}}
	\qquad\qquad\qquad
	\plotgraph{$o(x)$}{
		\plotpoly{\polyexo}{\ocolor};
		\plotdot{1}{2}{}
		\plotdot{2}{6}{}
		\plotdot{3}{12}{}
	}{}{ymax=12.9, xmax=4.9, height=20em, width=11em, xtick={1,...,4}, ytick={1,...,12},yticklabel={}}
\end{center}

\subsubsection{Multi-Operation Polynomials}

Now we have operand polynomials which represent three operations, let us see step-by-step how the correctness of each operation is verified. Recall that a verifier is looking for equality $l(x) \times r(x) - o(x) = t(x)h(x)$. In this case, because the operations are represented at points $x \in \{1,2,3\}$ the target polynomial has to evaluate to $0$ at those $x$-s, in other words, the roots of the $t(x)$ must be 1, 2 and 3, which in elementary form is:

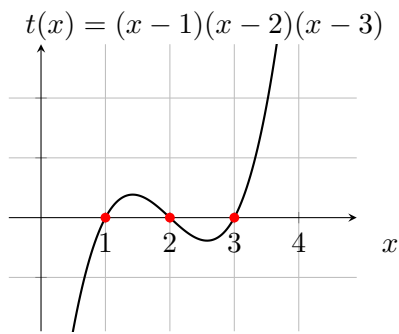
\newcommand{\polyext}{(x - 1)*(x - 2)*(x - 3)}
\begin{center}
	\plotgraph{$t(x) = (x - 1)(x - 2)(x - 3)$}{
		\plotpoly{\polyext}{};
		\plotdot{1}{0}{}
		\plotdot{2}{0}{}
		\plotdot{3}{0}{}
	}{}{ymin=-1.9, ymax=2.9, xmax=4.9, height=14em, width=16em, xtick={1,...,4}, ytick={-1,...,4}, label style = {at={(ticklabel cs:1.15)}}}
\end{center}

Firstly, $l(x)$ and $r(x)$ are multiplied which results in:

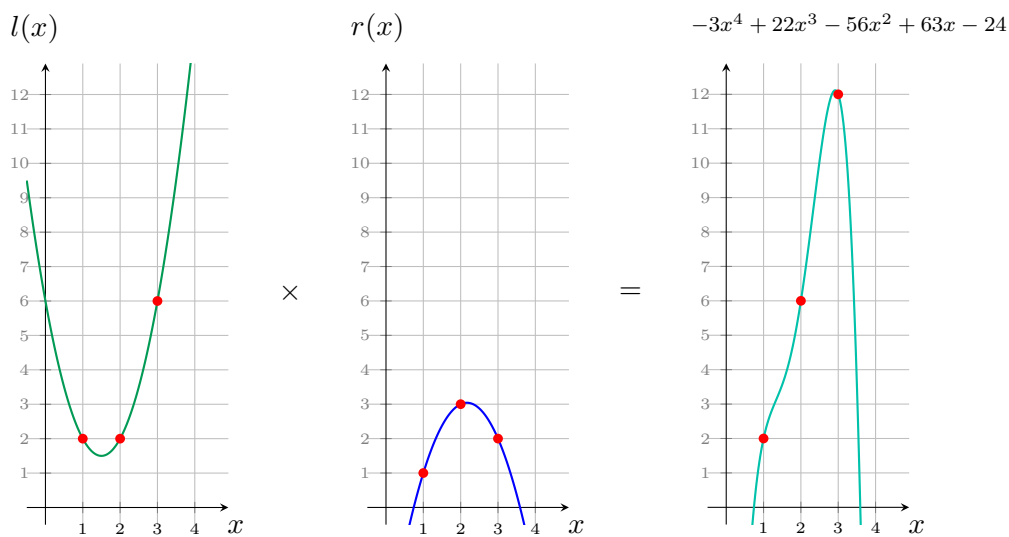
\definecolor{LRcolor}{RGB}{0, 193, 170}
\begin{center}
\pgfplotsset{every tick label/.append style={font=\tiny}}
\pgfplotsset{every y tick label/.append style={color=gray}}
	\plotgraph{$l(x)$}{
		\plotpoly{\polyexl}{\lcolor};
		\plotdot{1}{2}{}
		\plotdot{2}{2}{}
		\plotdot{3}{6}{}
	}{
		\node at (9em,8em) {$\times$};	
	}{ymax=12.9, xmax=4.9, height=20em, width=11em, xtick={1,...,4}, ytick={1,...,12}, yticklabel={}}
	\plotgraph{$r(x)$}{
		\plotpoly{\polyexr}{\rcolor};
		\plotdot{1}{1}{}
		\plotdot{2}{3}{}
		\plotdot{3}{2}{}
	}{
		\node at (9em,8em) {$=$};		
	}{ymax=12.9, xmax=4.9, height=20em, width=11em, xtick={1,...,4}, ytick={1,...,12}, yticklabel={}}
	\plotgraph{\scriptsize{$-3x^4 + 22x^3 - 56x^2 + 63x - 24$}}{
		\plotpoly{(\polyexl) * (\polyexr)}{LRcolor};
		\plotdot{1}{2}{}
		\plotdot{2}{6}{}
		\plotdot{3}{12}{}
	}{}{ymax=12.9, xmax=4.9, height=20em, width=11em, xtick={1,...,4}, ytick={1,...,12}, restrict y to domain=-10:14, yticklabel={}}
\end{center}

Secondly, the $o(x)$ is subtracted from the result of $l(x) \times r(x)$:

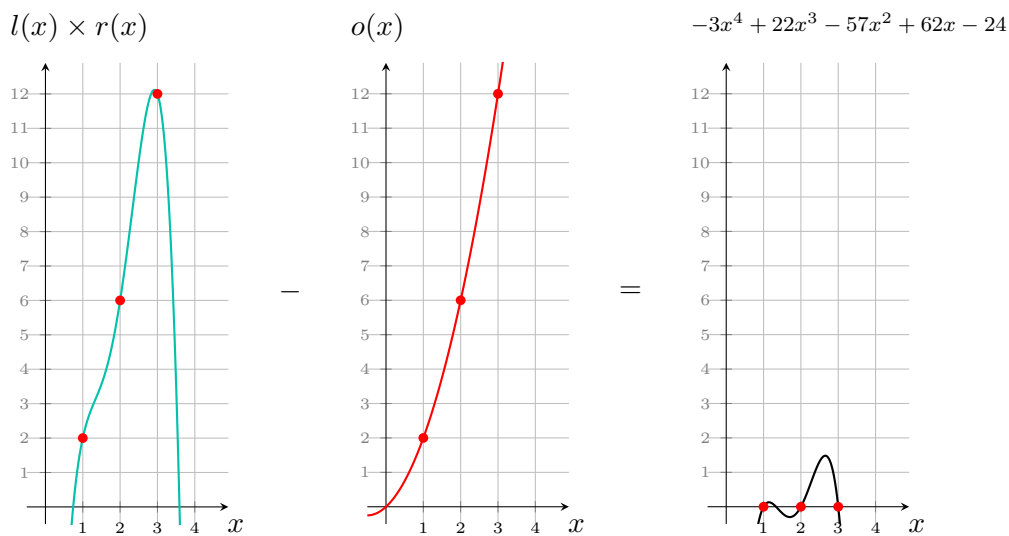
\begin{center}
	\pgfplotsset{every tick label/.append style={font=\tiny}}
	\pgfplotsset{every y tick label/.append style={color=gray}}
	\plotgraph{$l(x) \times r(x)$}{
		\plotpoly{(\polyexl) * (\polyexr)}{LRcolor};
		\plotdot{1}{2}{}
		\plotdot{2}{6}{}
		\plotdot{3}{12}{}
	}{
		\node at (9em,8em) {$-$};	
	}{ymax=12.9, xmax=4.9, height=20em, width=11em, xtick={1,...,4}, ytick={1,...,12}, restrict y to domain=-10:14, yticklabel={}}
	\plotgraph{$o(x)$}{
		\plotpoly{\polyexo}{\ocolor};
		\plotdot{1}{2}{}
		\plotdot{2}{6}{}
		\plotdot{3}{12}{}
	}{
		\node at (9em,8em) {$=$};	
	}{ymax=12.9, xmax=4.9, height=20em, width=11em, xtick={1,...,4}, ytick={1,...,12}, restrict y to domain=-10:14, yticklabel={}}
	\plotgraph{\scriptsize{$-3x^4 + 22x^3 - 57x^2 + 62x - 24$}}{
		\plotpoly{(\polyexl)*(\polyexr)-(\polyexo)}{};
		\plotdot{1}{0}{}
		\plotdot{2}{0}{}
		\plotdot{3}{0}{}
	}{}
	{ymax=12.9, xmax=4.9, height=20em, width=11em, xtick={1,...,4}, ytick={1,...,12}, restrict y to domain=-10:14, yticklabel={}}
\end{center}

Where it is already visible that every operands multiplication corresponds to a correct result.
For the last step a \term{prover} needs to present a valid cofactor: 
$$h(x) = \frac{l(x) \times r(x) - o(x)}{t(x)} = \frac{-3x^4 + 22x^3 - 57x^2 + 62x - 24}{(x - 1)(x - 2)(x - 3)}$$

Using long division we get:

$h(x) =$ \polylongdiv{-3x^4 + 22x^3 - 57x^2 + 62x - 24}{(x-1)(x-2)(x-3)}

\newcommand{\polyexh}{-3*x + 4}
With $h(x) = -3x + 4$ a \term{verifier} can compute $t(x)h(x)$:
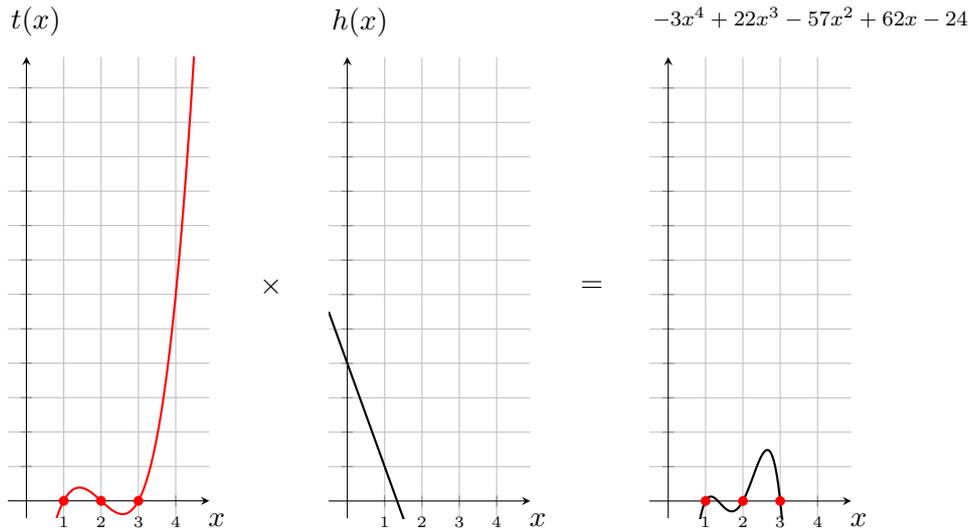
\begin{center}
	\pgfplotsset{every tick label/.append style={font=\tiny}}
	\pgfplotsset{every y tick label/.append style={color=gray}}
	\plotgraph{$t(x)$}{
		\plotpoly{\polyext}{\ocolor};
		\plotdot{1}{0}{}
		\plotdot{2}{0}{}
		\plotdot{3}{0}{}
	}{
		\node at (9em,8em) {$\times$};	
	}{ymax=12.9, xmax=4.9, height=20em, width=11em, xtick={1,...,4}, ytick={1,...,12}, restrict y to domain=-10:14}
	\plotgraph{$h(x)$}{
		\plotpoly{-3*x + 4}{};
	}{
		\node at (9em,8em) {$=$};	
	}
	{ymax=12.9, xmax=4.9, height=20em, width=11em, xtick={1,...,4}, ytick={1,...,12}, restrict y to domain=-10:14}
	\plotgraph{\scriptsize{$-3x^4 + 22x^3 - 57x^2 + 62x - 24$}}{
		\plotpoly{(\polyext)*(\polyexh)}{};
		\plotdot{1}{0}{}
		\plotdot{2}{0}{}
		\plotdot{3}{0}{}
	}{}
	{ymax=12.9, xmax=4.9, height=20em, width=11em, xtick={1,...,4}, ytick={1,...,12}, restrict y to domain=-10:14}
\end{center}

It is now evident that $l(x) \times r(x) - o(x) = t(x) h(x)$ which is what had to be proven.

\subsection{Variable Polynomials}

With such an approach, we can prove many operations at once (e.g., millions and more), but there is a critical downside to it.

If the ``program," execution for which is being proved, uses the same \term{variable}, either as an operand or as output, in different operations, for example:
\begin{gather*}
	\begin{aligned}
	\oper{a}{b}{r_1} \\
	\oper{a}{c}{r_2}
	\end{aligned}
\end{gather*}

The $\lo{a}$ will have to be represented in the \term{left operand polynomial} for both operations as:

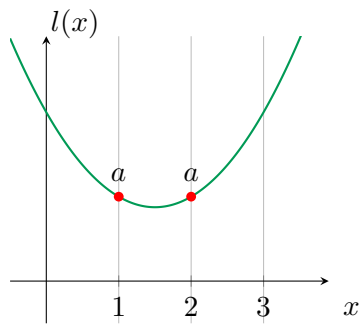
\begin{center}
	\plotgraph{$l(x)$}{
		\plotpoly{(x-1)*(x-2)/2+1}{\lcolor};
		\plotdotangle{1}{1}{90}{$a$}
		\plotdotangle{2}{1}{90}{$a$}
	}{}{ymax=2.9, xmax=3.9, height=14em, width=15em, ytick=\empty}
\end{center}

Nevertheless, because our protocol allows prover to set any coefficients to a polynomial, he is not restricted from setting different values of $a$ for different operations (i.e., represented by some $x$), e.g.:

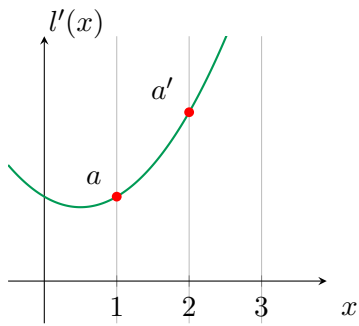
\begin{center}
	\plotgraph{$l'(x)$}{
		\plotpoly{(x+1-1)*(x+1-2)/2+1}{\lcolor};
		\plotdot{1}{1}{$a$}
		\plotdot{2}{2}{$a'$}
	}{}{ymax=2.9, xmax=3.9, height=14em, width=15em, ytick=\empty}
\end{center}

This freedom breaks consistency and allows prover to prove the execution of some other program which is not what verifier is interested in. Therefore we must ensure that any variable can only have a single value across every operation it is used in.

\note{variable in this context differs from the regular computer science definition in a sense that it is immutable and is only assigned once per execution.}

\subsubsection{Single-Variable Operand Polynomial}\label{section:single-var-oper-poly}

Let us consider a simple case (as with the current example) where we have only one variable (e.g., $a$) used in all left operands represented by the \term{left operand polynomial} $l(x)$. We have to find out if it is possible to ensure that this polynomial represents  the same values of $a$ for every operation. The reason why a prover can set different values is that he has control over each coefficient for every exponentiation of $x$. Therefore if those coefficients were constant, that would solve the variability problem.

May us have a closer look at polynomials containing equal values. For example examine two polynomials representing equal values for the two operations correspondingly (i.e., at $x=1$ and $x=2$), where the first polynomial contains value $1$ and the second contains value $2$:

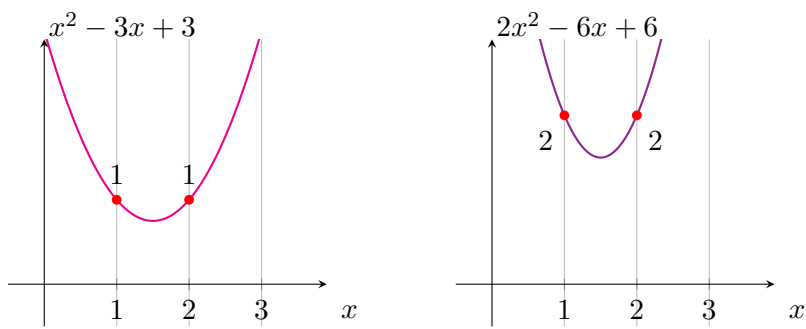
\begin{center}
	\plotgraph{$x^2 -3x + 3$}{
		\plotpoly{(x-1)*(x-2)+1}{Magenta};
		\plotdotangle{1}{1}{90}{$1$}
		\plotdotangle{2}{1}{90}{$1$}
	}{}{ymax=2.9, xmax=3.9, height=14em, width=15em, ytick=\empty}
	\qquad
	\plotgraph{$2x^2 -6x + 6$}{
		\plotpoly{2*(x-1)*(x-2)+2}{Plum};
		\plotdotangle{1}{2}{260}{$2$}
		\plotdotangle{2}{2}{280}{$2$}
	}{}{ymax=2.9, xmax=3.9, height=14em, width=15em, ytick=\empty}
\end{center}

Notice that the corresponding coefficients are proportional in each polynomial, such that coefficients in the second are twice as large as in the first, i.e.:
\begin{equation*}
	2x^2 -6x + 6 = 2 \times (x^2 -3x + 3)
\end{equation*}

Therefore when we want to change all the values simultaneously in a polynomial we need to change its proportion, this is due to arithmetic properties of polynomials, if we multiply a polynomial by a number, evaluations at every possible $x$ will also be multiplied (i.e., scaled). To verify, try to multiply the first polynomial by 3 or any other number. 

Consequently, if a verifier needs to enforce the prover to set the same value in all operations, then it should only be possible to modify the proportion and not the individual coefficients. 

So how coefficients proportion can be preserved? We can start by considering what is provided as proof for the \term{left operand polynomial}. It is an encrypted evaluation of $l(x)$ at some secret $s$: $g^{l(s)}$, i.e., it is an encrypted number. We already know from section \ref{restricting polynomial} how to restrict a verifier to use only the provided exponents of $s$ through an $\alpha$-shift, such that homomorphic multiplication is the single operation available.

Similarly to restricting a single exponent, the verifier can restrict the whole polynomial at once. Instead of providing separate encryptions $g^{s^1}, g^{s^2}, \ldots, g^{s^d}$ and their $\alpha$-shifts $g^{\alpha s^1}, g^{\alpha s^2}, \ldots, g^{\alpha s^d}$ the protocol proceeds:

\zkProtocol{
	\item construct the respective \term{operand polynomial} $l(x)$ with corresponding coefficients
	\item sample random $\alpha$ and $s$
	\item set proving key with encrypted $l(s)$ and it is ``shifted" pair: $\left( g^{l(s)}, g^{\alpha l(s)} \right)$
	\item set verification key: $\left( g^\alpha \right)$
} {
	\item having operand's value $v$
	\begin{itemize}
		\item multiply \term{operand polynomial}: $\left( g^{l(s)} \right)^v$
		\item multiply shifted \term{operand polynomial}: $\left( g^{\alpha l(s)} \right)^v$
	\end{itemize}
	\item provide \term{operand polynomial} multiplication proof: $\left( g^{v\, l(s)}, g^{v\, \alpha l(s)} \right)$
} {
	\item parse the proof as $\left( g^{l}, g^{l'} \right)$
	\item verify proportion: $\prng{g^{l'}, g} = \prng{g^{l}, g^\alpha}$ 
}

Prover needs to respond with the same $\alpha$-shift and because he cannot recover $\alpha$ from the proving key the only way to maintain the shift is to multiply both encryptions $g^{l(s)}$ and $g^{\alpha l(s)}$ by the same value. Therefore prover cannot modify individual coefficients of $l(x)$, for example if $l(x) = ax^2 + bx + c$ he can only multiply the whole polynomial at once by some value $v$: $v(ax^2 + bx + c)= vax^2 + vbx + vc$. Multiplication by another polynomial is not available since \term{pairings}, and $\alpha$-shifts of individual exponents of $s$ are not available. Prover cannot add or subtract either since $ g^{\alpha(l(x) + a'x^2 + c')} \neq g^{\alpha l(x)} \cdot g^{a' x^2} \cdot g^{c'}$ (this, again, requires the knowledge of unencrypted $\alpha$).

We now have the protocol, but how \term{operand polynomial} $l(x)$ should be constructed? Since any integer can be derived by multiplying $1$, the polynomial should evaluate to $1$ for every corresponding operation, e.g.:
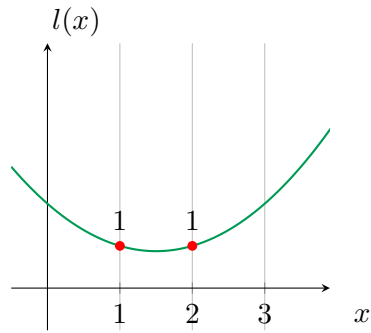
\begin{center}
	\plotgraph{$l(x)$}{
		\plotpoly{((x-1)*(x-2)/2+1)/2}{\lcolor};
		\plotdotangle{1}{0.5}{90}{$1$}
		\plotdotangle{2}{0.5}{90}{$1$}
	}{}{ymax=2.9, xmax=3.9, height=14em, width=15em, ytick=\empty, label style = {at={(ticklabel cs:1.16)}}}
\end{center}

This allows a prover to \term{assign} the value of $a$:
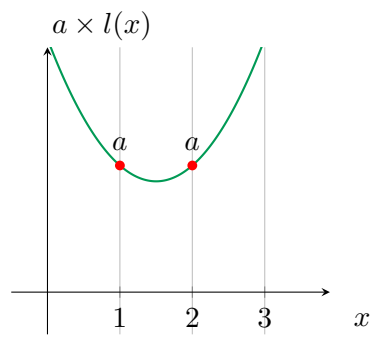
\begin{center}
	\plotgraph{$a \times l(x)$}{
		\plotpoly{3*((x-1)*(x-2)/2+1)/2}{\lcolor};
		\plotdotangle{1}{1.5}{90}{$a$}
		\plotdotangle{2}{1.5}{90}{$a$}
	}{}{ymax=2.9, xmax=3.9, height=14em, width=15em, ytick=\empty, label style = {at={(ticklabel cs:1.16)}}}
\end{center}

\begin{remark} \label{remark:polyModification}
	Since verification key contains $g^\alpha$	it is possible to add (or subtract) an arbitrary value $v'$ to the polynomial, i.e.:
	\begin{align*}
		g^{v l(s)} \cdot g^{v'} &= g^{v l(s) + v'} \\
		g^{\alpha v l(s)} \cdot \left(g^{\alpha}\right)^{v'} &= g^{\alpha (v l(s) + v')} \\
		\prng{g^{\alpha (v l(s) + v')}, g} &= \prng{g^{v l(s) + v'}, g^\alpha}
	\end{align*}
	
	Therefore it is possible to modify the polynomial beyond what is intended by the verifier and prove a different statement. We will address this shortcoming in section \ref{subsection:poly-non-malleability}.
\end{remark}

\subsubsection{Multi-Variable Operand Polynomial}

We are now able to singularly set value only if all left operands use the same variable. What if we add another one $d$:

\begin{center}
	\parbox[c]{10em}{
	 	\begin{align*}
			\oper{a}{b}{r_1} \\
			\oper{a}{c}{r_2} \\
			\oper{d}{c}{r_3} \\
		\end{align*}
	}
	\qquad\qquad\quad
	\parbox[c]{10em}{
		\plotgraph{$l(x)$}{
			\plotpoly{(x-1)*(x-2)/2+1}{\lcolor};
			\plotdotangle{1}{1}{90}{$a$}
			\plotdotangle{2}{1}{90}{$a$}
			\plotdot{3}{2}{$d$}
		}{}{ymax=2.9, xmax=3.9, height=14em, width=15em, ytick=\empty}
	}
\end{center}

If we have used the same approach we would not be able to set the value separately for each variable, and every distinct variable will be multiplied altogether. Hence such restricted polynomial can support only one \term{variable}. If we examine properties of polynomials, we will see that adding polynomials together adds distinct evaluations of those polynomials. Therefore we can separate the \term{operand polynomial} $l(x)$ into \term{operand variable polynomials} $l_a(x)$ and $l_d(x)$ (note the subscripts) such that \term{variables} $a$ and $d$ are \term{assigned} and restricted separately similarly to the previous section and then added together to represent variables of all left operands. Because we add \term{operand variable polynomials} together, we need to ensure that only one of all the \term{variables} is represented for each operation by the operand polynomial. 

Using the arithmetic properties we can construct each \term{operand variable polynomial} such that if \term{variable} is used as an operand in the corresponding operation then it evaluates to $1$, otherwise to $0$. Consecutively $0$ multiplied by any value will remain zero and when added together it will be ignored. For our example $l_a(x)$ must conform to evaluations $l_a(1) = 1$, $l_a(2) = 1$ and $l_a(3) = 0$ and $l_d(x)$ is zero at 1 and 2 but $1$ at $x=3$:
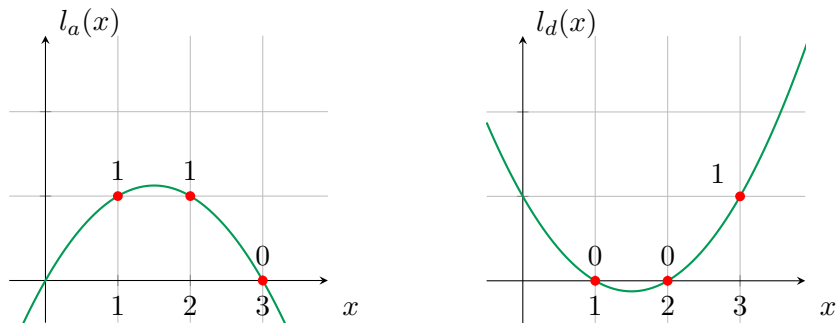
\begin{center}
	\plotgraph{$\ \ \ l_a(x)$}{
			\plotpoly{(-x^2 + 3*x)/2}{\lcolor};
			\plotdotangle{1}{1}{90}{$1$}
			\plotdotangle{2}{1}{90}{$1$}
			\plotdotangle{3}{0}{90}{$0$}
	}{}{ymax=2.9, xmax=3.9, height=14em, width=15em, ytick={}}
	\qquad\quad
	\plotgraph{$\ \ \ l_d(x)$}{
			\plotpoly{(x^2 - 3*x + 2)/2}{\lcolor};
			\plotdotangle{1}{0}{90}{$0$}
			\plotdotangle{2}{0}{90}{$0$}
			\plotdot{3}{1}{$1$}
	}{}{ymax=2.9, xmax=3.9, height=14em, width=15em, ytick={}}
\end{center}

Consequently we can set the value of each variable separately and just add them together to get the operand polynomial, for example if $a=3$ and $d=2$:

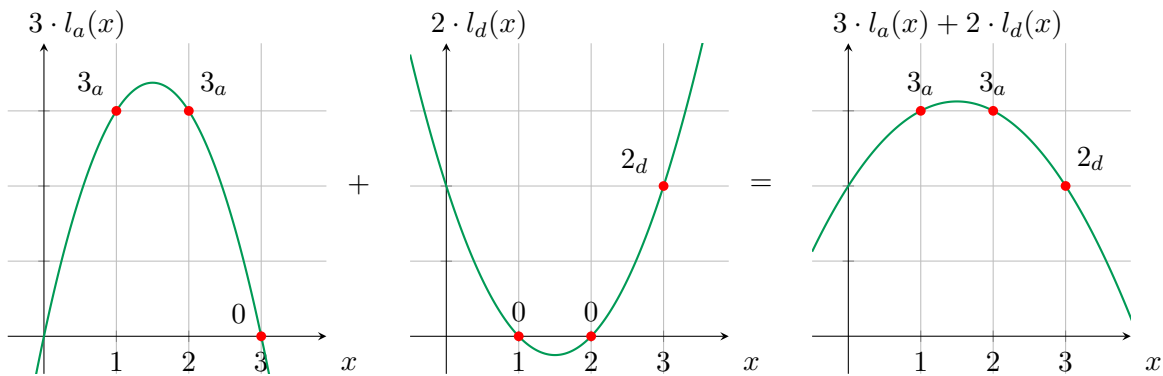
\begin{center}
	\plotgraph{$3\cdot l_a(x)$}{
		\plotpoly{(3*(-x^2 + 3*x))/2}{\lcolor};
		\plotdotangle{1}{3}{100}{$3_a$}
		\plotdotangle{2}{3}{80}{$3_a$}
		\plotdot{3}{0}{$0$}
	}{
		\node at (12em,6.5em) {$+$};
	}{ymax=3.9, xmax=3.9, height=15.5em, width=15em, ytick={}}
	\plotgraph{$2\cdot l_d(x)$}{
		\plotpoly{(2*(x^2 - 3*x + 2))/2}{\lcolor};
		\plotdotangle{1}{0}{90}{$0$}
		\plotdotangle{2}{0}{90}{$0$}
		\plotdot{3}{2}{$2_d$}
	}{
		\node at (12em,6.5em) {$=$};
	}{ymax=3.9, xmax=3.9, height=15.5em, width=15em, ytick={}}
	\plotgraph{$3\cdot l_a(x) + 2\cdot l_d(x)$}{
		\plotpoly{(3*(-x^2 + 3*x) + 2*(x^2 - 3*x + 2))/2}{\lcolor};
		\plotdotangle{1}{3}{90}{$3_a$}
		\plotdotangle{2}{3}{90}{$3_a$}
		\plotdotangle{3}{2}{80}{$2_d$}
	}{}{ymax=3.9, xmax=3.9, height=15.5em, width=15em, ytick={}}
\end{center}

\note{we are using subscript next to a value to indicate which variable it represents, e.g., $3_a$ is a variable $a$ instantiated with value $3$.}

Let us denote such composite \term{operand polynomial} with an upper-case letter from now on, e.g., $L(x) = a\, l_a(x) + d\, l_d(x)$, and its evaluation value as $L$, i.e., $L = L(s)$. This construction will only be effective if each \term{operand variable polynomial} is restricted by the verifier, the interaction concerning left operand shall be altered accordingly:

\zkProtocol{
	\item construct $l_a(x)$, $l_d(x)$ such that it passes through 1 at ``operation $x$" where it is used and through 0 in all other operations
	\item sample random $s$, $\alpha$
	\item evaluate and encrypt \term{unassigned variable polynomials}: \\ 
	$g^{l_a(s)}, g^{l_d(s)}$
	\item calculate \emph{shifts} of these polynomials: \\
	$g^{\alpha l_a(s)}, g^{\alpha l_d(s)}$
	\item set proving key: \\
	$\left( g^{l_a(s)}, g^{l_d(s)}, g^{\alpha l_a(s)}, g^{\alpha l_d(s)} \right)$
	\item set verification key: \\
	$\left( g^\alpha \right)$
}{
	\item \term{assign} values $a$ and $d$ to the variable polynomials: \\
	$\left(g^{l_a(s)}\right)^a, \left( g^{l_d(s)} \right)^d$
	\item \term{assign} same values to the \term{shifted} polynomials: \\
	$\left(g^{\alpha l_a(s)}\right)^a, \left( g^{\alpha l_d(s)} \right)^d$
	\item add all \term{assigned} variable polynomials to form an \term{operand polynomial}: \\
	$g^{L(s)} = g^{a l_a(s)} \cdot g^{d l_d(s)} = g^{a l_a(s) + d l_d(s)}$
	\item add \term{shifted} \term{assigned} \term{variable polynomials} to form a \term{shifted operand polynomial}: \\
	$g^{\alpha L(s)} = g^{a \alpha l_a(s)} \cdot g^{d \alpha l_d(s)} = g^{\alpha \left(a l_a(s) + d l_d(s) \right)}$
	\item provide proof of valid assignment of \term{left operand}: \\
	$\left( g^{L(s)}, g^{\alpha L(s)} \right)$
}{
	\item parse proof as $\left( g^L, g^{L'} \right)$
	\item check that provided polynomials is a sum of multiples of originally provided \term{unassigned variable polynomials}:
	\flalignskip{
		& \prng{ g^{L'}, g } = \prng{ g^{L}, g^\alpha } \textrm{\qquad which checks that} &\\ 
		& \alpha\, a l_a(s) + \alpha\, d l_d(s) = \alpha \times (a l_a(s) + d l_d(s)) &
	}
}
\note{$L(s)$ and $\alpha L(s)$ represent all variable polynomials at once and since $\alpha$ is used only in evaluation of variable polynomials, the prover has no option but to use provided evaluations and assign same coefficients to original and shifted variable polynomials.}

As a consequence the prover:
\begin{itemize}
	\item is not able to modify provided \term{variable polynomials} by changing their coefficients, except ``assigning" values, because prover is presented only with encrypted evaluations of these polynomials, and because necessary encrypted powers of $s$ are unavailable separately with their $\alpha$-shifts
	\item is not able to add another polynomial to the provided ones because the $\alpha$-ratio will be broken
	\item is not able to modify operand polynomials through multiplication by some other polynomial $u(x)$, which could disproportionately modify the values because encrypted multiplication is not possible in pre-pairings space
\end{itemize} 
\note{if we add (or subtract) one polynomial, e.g., $l_a(x)$, to the other, e.g., $l_d'(x) = c_{d} \cdot l_d(x) + c_{a}' \cdot l_a(x)$, that is not really a modification of the polynomial $l_d(x)$, but rather changing of the resulting coefficient of the $l_a(x)$, because they are summed up in the end:
\begin{center}
	$L(x) = c_{a} \cdot l_a(x) + l_d'(x) = \left(c_a + c_a'\right) \cdot l_a(x) + c_d \cdot l_d(x)$
\end{center}
}
While the prover restricts the use of polynomials, there is still some freedoms which are not necessary to counteract:
\begin{itemize}
	\item it is acceptable if the prover decides not to add some of the assigned variable polynomials $l_i(x)$ to form the operand polynomial $L(x)$ because it is the same as to assign the value $0$: $g^{a l_a(x)} = g^{a l_a(x) + 0 l_d(x)}$
	\item it is acceptable if the prover adds same variable polynomials multiple times because it is the same as to assign the multiple of that value once, e.g., $g^{a l_a(x)}\cdot g^{a l_a(x)} \cdot g^{a l_a(x)} = g^{3a l_a(x)}$
\end{itemize}

This approach is applied similarly to the right operand and output polynomials $R(x)$, $O(x)$.

\subsection{Construction Properties}

There are multiple additional useful properties which are acquired as a side-effect of such modification.

\subsubsection{Constant Coefficients}

In the above construction, we have been using evaluations of \term{unassigned variable polynomials} $1$ or $0$ as a means to signify if the variable is \textsl{used} in operation or \textsl{not}. Naturally, there is nothing that stops us from using other coefficients as well, including negative ones, because we can \emph{interpolate} polynomials through any necessary points\footnote{Provided that no two operations occupy same $x$}. Examples of such operations are:

\begin{align*}
	\oper{2a}{1b}{\phantom{-}3r} \\
	\oper{-3a}{1b}{-2r}
\end{align*}

Therefore our program can now use constant coefficients, for example:

\begin{algorithm} 
\caption{Constant coefficients}
\begin{algorithmic} 
    \Function{\textnormal{calc}}{w, a, b}
    		\If{w}
    			\State \Return{\underline{3}a $\times$ b}
    		\Else
    			\State \Return{\underline{5}a $\times$ \underline{2}b}
    		\EndIf
    \EndFunction
\end{algorithmic}
\end{algorithm}

These coefficients will be ``hardwired" during the setup stage and similarly to $1$ or $0$ will be immutable. We can modify the form of operation accordingly:
$$ \begin{aligned} \oper{c_a \cdot a}{c_b \cdot b}{c_r \cdot r} \end{aligned} $$

Or more formally, for variables $v_i \in \{v_1, v_2, ..., v_n\}$:
$$ \begin{aligned} \oper{c_l \cdot v_l}{c_r \cdot v_r}{c_o \cdot v_o} \end{aligned} $$
where $l, r, o$ are indices of a variable used in operation.

\note{constant coefficient for the same variable can be different in different operations and operands/outputs.}

\subsubsection{Addition for Free}

Considering the updated construction, it is apparent that in polynomial representation every \term{operand} expressed by some distinct $x$ is a sum of all \term{operand variable polynomials} such that only single \emph{used} variable can have a non-zero value and all others are zero. The graph demonstrates it best:

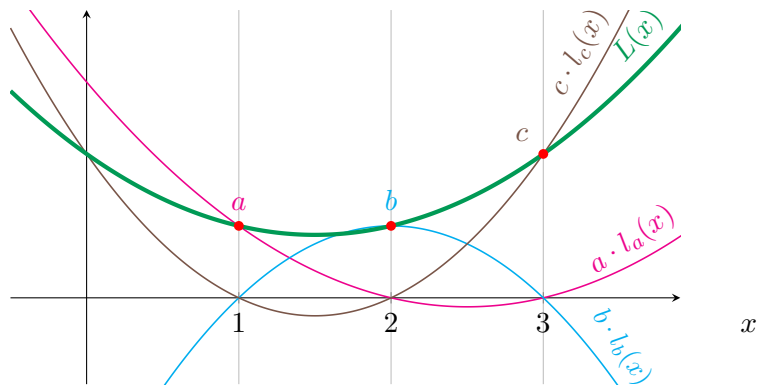
\definecolor{ccolor}{RGB}{121,85,72}
\definecolor{bcolor}{RGB}{0,188,212}
\newcommand{\acolor}{Magenta}
\newcommand{\bcolor}{Cyan}
\newcommand{\ccolor}{ccolor}
\begin{center}
	\plotgraph{$ $}{
 
		\plotpolylabeled{(x-2)*(x-3)/2}{\acolor,style={semithick}}{$a \cdot l_a(x)$}{sloped,at end};
		\plotpolylabeled{-(x-1)*(x-3)}{\bcolor,style={semithick}}{\small{$b \cdot l_b(x)$}}{sloped,pos=0.865}
		\plotpolylabeled{(x-1)*(x-2)}{\ccolor,style={semithick}}{$c \cdot l_c(x)$}{sloped,pos=0.82};
		
		\plotpolylabeled{(x-2)*(x-3)/2 +  -(x-1)*(x-3) + (x-1)*(x-2)}{\lcolor, style={ultra thick}}{$L(x)$}{sloped,at end,xshift=-0.2cm};
		
		\plotdotangle{1}{1}{90}{\color{\acolor} $a$}
		\plotdotangle{2}{1}{90}{\color{\bcolor} $b$}
		\plotdot{3}{2}{\color{\ccolor} $c$}
		 
	}{ 
	}{ymax=4, ymin=-1.2, xmax=3.9, xmin=-0.5, height=17em, width=27em, ytick=\empty,xtick={0,...,3},domain=-0.5:4}
\end{center}

We can take advantage of such construction and allow to add any number of necessary \term{variables} for each operand in operation. For example in the first operation, we can add $a + c$ first and only then multiply it by some other operand, e.g., $\maligned{\opershort{(a+c)}{b}{r}}$, this can be represented as:

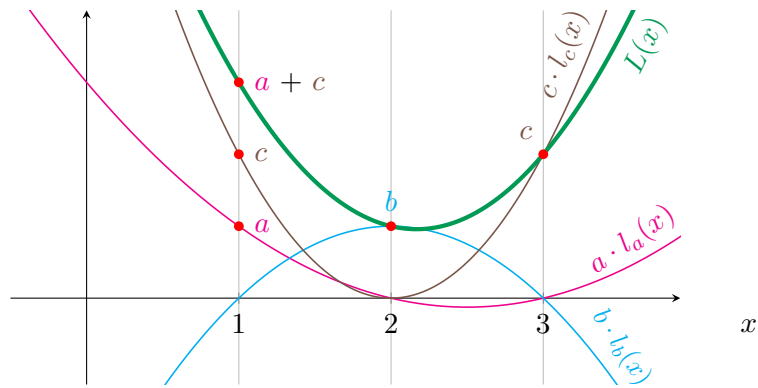
\begin{center}
	\plotgraph{$ $}{
		\plotpolylabeled{(x-2)*(x-3)/2}{\acolor, style={semithick}}{$a \cdot l_a(x)$}{sloped,at end};
		\plotpolylabeled{-(x-1)*(x-3)}{\bcolor, style={semithick}}{\small{$b \cdot l_b(x)$}}{sloped,pos=0.865};
		\plotpolylabeled{(x-2)*(x-3) + (x-1)*(x-2)}{\ccolor, style={semithick}}{$c \cdot l_c(x)$}{sloped,pos=0.811};

		\plotpolylabeled{(x-2)*(x-3)/2 + -(x-1)*(x-3) + (x-2)*(x-3) + (x-1)*(x-2)}{\lcolor, style={ultra thick}}{$L(x)$}{sloped,pos=0.89,below left};

		\plotdotangle{1}{1}{0}{\color{\acolor} $a$}
		\plotdotangle{1}{2}{0}{\color{\ccolor} $c$}
		\plotdotangle{1}{3}{0}{{\color{\acolor} $a$} + {\color{\ccolor} $c$}}
		
		\plotdotangle{2}{1}{90}{\color{\bcolor} $b$}
		\plotdotangle{3}{2}{100}{\color{\ccolor} $c$}
	}{}{ymax=4, ymin=-1.2, xmax=3.9, xmin=-0.5, height=17em, width=27em, ytick=\empty,xtick={0,...,3},domain=-0.5:4}
\end{center}

Therefore it is possible to add any number of present variables in a single \term{operand}, using arbitrary coefficients for each of them, to produce an operand value which will be used in a corresponding operation, as needed in a respective program.
Such property effectively allows changing the operation construction to:
\begin{align*}
	\oper{(c_{\,\textrm{l},a} \cdot a + c_{\,\textrm{l},b} \cdot b + \ldots)}
		{(c_{\,\textrm{r},a} \cdot a + c_{\,\textrm{r},b} \cdot b + \ldots)}
		{(c_{\,\textrm{o},a} \cdot a + c_{\,\textrm{o},b} \cdot b + \ldots)}
\end{align*}

Or more formally, for variables $v_i \in \{v_1, v_2, ..., v_n\}$ and operand variable coefficients $c_{\,\textrm{l},i} \in \{c_{\,\textrm{l},1}, c_{\,\textrm{l},2}, ..., c_{\,\textrm{r},n}\}$, $c_{\,\textrm{r},i} \in \{c_{\,\textrm{r},1}, c_{\,\textrm{r},2}, ..., c_{\,\textrm{r},n}\}$, $c_{\,\textrm{o},i} \in \{c_{\,\textrm{o},1}, c_{\,\textrm{o},2}, ..., c_{\,\textrm{o},n}\}$:
\newcommand{\operFinal}{
	\oper{ \sum_{i=1}^{n} c_{\,\textrm{l},i} \cdot v_i }
	{\sum_{i=1}^{n} c_{\,\textrm{r},i} \cdot v_i}
	{\sum_{i=1}^{n} c_{\,\textrm{o},i} \cdot v_i}
}
\begin{align*}
	\operFinal
\end{align*}

\note{each operation's operand has its own set of coefficients $c$.}

\subsubsection{Addition, Subtraction and Division}

We have been focusing on multiplication operation primarily until now. However, in order to be able to execute general computations, a real-life program will also require addition, division, and subtraction.

\paragraph*{Addition} In previous section we have established that we can add variables in context of a single operand, which is then multiplied by another operand, e.g., $\maligned{ \opershort{(3a + b)}{d}{r} }$, but what if we need just addition without multiplication, for example, if a program needs to compute $\mathrm{a} + \mathrm{b}$, we can express this as:
$$ \maligned{ \oper{(a + b)}{1}{r} } $$
\note{because our construction requires both a constant coefficient and a variable ($\ro{c \cdot v}$) for every operand, the value of $\ro{1}$ is expressed as $c\ped{one} \cdot v\ped{one}$, and while $c\ped{one} = 1$ can be ``hardwired" into a corresponding polynomial, the $v\ped{one}$ is a variable and can be assigned any value, therefore we must enforce the value of $v\ped{one}$ through constraints as explained in section \ref{Constraints}.}

\paragraph*{Subtraction} Subtraction is almost identical to addition, the only difference is a negative coefficient, e.g., for $\mathrm{a - b}$:
$$ \maligned{ \oper{(a + -1 \cdot b)}{1}{r} } $$

\paragraph*{Division} If we examine the division operation $\mathsf{\frac{factor}{divisor} = result}$ we would see that the \textsf{result} of the division is the number we need to multiply \textsf{divisor} by to produce the \textsf{factor}. Therefore we can express the same meaning through multiplication: $\mathsf{divisor \times result = factor}$. Consequently, if we want to prove the division operation $\mathrm{\frac{a}{b} = r}$, it can be expressed as:
$$ \maligned{\oper{b}{r}{a}} $$

\note{the operation's construction is also called ``constraint" because the operation represented by polynomial construction does not compute results per se, but rather checks that the prover already knows variables (including result), and they are \emph{valid} for the operation, i.e., the prover is constrained to provide consistent values no matter what they are.}

\note{all those arithmetic operations were already present; therefore modification of the operation's construction is not needed.}

\subsection{Example Computation}

Having the general operation's construction, we can convert our original algorithm \ref{alg:DependsOnInput} into a set of operations and further into polynomial form. Let us consider the mathematical form of the algorithm (we will use variable $v$ to capture the result of evaluation):
\begin{equation*}
	w \times (a \times b) + (1 - w) \times (a + b) = v
\end{equation*}

It has three multiplications, and because the operation construction supports only one, there will be at least 3 operations. However, we can simplify the equation:
\begin{gather*}
	w \times (a \times b) + a + b - w \times (a + b) = v \\
	w \times (a \times b - a - b) = v - a - b
\end{gather*}

Now it requires two multiplications while maintaining same relationships. In complete form the operations are:
\begin{alignat*}{3}
	1:& & \oper{1\cdot a}{1\cdot b}{1 \cdot m} \\
	2:&\qquad\quad &  \oper{1\cdot w}{1\cdot m\ +\ -1\cdot a\ + \ -1\cdot b}{1 \cdot v\ +\ -1 \cdot a\ +\ -1 \cdot b} \\
\noalign{We can also add a constraint that requires $w$ to be binary, otherwise a prover can use any value for $w$ rendering computation incorrect:}
	3:& &  \oper{1\cdot w}{1\cdot w}{1\cdot w}
\end{alignat*}

To see why $w$ can only be 0 or 1, we can represent the equation as $w^2 - w = 0$ and further as $(w - 0)(w - 1) = 0$\ \ where 0 and 1 are the only solutions.

These totals to 5 variables, with 2 in the left operand, 4 in the right operand and 5 in the output. The operand polynomials are: 
\begin{align*}
	\lo{L(x)} & = \lo{ a \cdot l_a(x) + w \cdot l_w(x) } \\
	\ro{R(x)} & = \ro{ m \cdot r_{m}(x) + a \cdot r_a(x) + b \cdot r_b(x) + w\cdot r_w(x) } \\
	\oo{O(x)} & = \oo{ m \cdot o_{m}(x) + v \cdot o_v(x) + a \cdot o_a(x) + b \cdot o_b(x) + w\cdot o_w(x) }
\end{align*}
\noindent where each \term{variable polynomial} must evaluate to a corresponding coefficient for each of 3 operations or to 0 if the variable isn't present in the operation's operand or output: 
\vspace{-1.5em}
{\scalefont{.955}
\begin{vwcol}[widths={0.31,0.34,0.36},sep=0cm,justify=raggedleft,presep=0cm,postsep=0cm,rule=0pt,indent=0em] 
	\begin{alignat*}{6}
		\lo{l_a(1)} &= 1;\ &	\lo{l_a(2)} &= 0;\ & \lo{l_a(3)} &= 0;\ \\
		\lo{l_w(1)} &= 0;\ &	\lo{l_w(2)} &= 1;\ & \lo{l_w(3)} &= 1;\ \\
		\\
		\\
	\end{alignat*}
	
	\begin{alignat*}{6}
		\ro{r_m(1)} &= 0;\	&	\ro{r_m(	2)} &= \phantom{-}1;\	&	\ro{r_m(3)} &= 0;\ \\
		\ro{r_a(1)} &= 0;\	&	\ro{r_a(2)} &= -1;\			&	\ro{r_a(3)} &= 0;\ \\
		\ro{r_b(1)} &= 1;\	&	\ro{r_b(2)} &= -1;\			&	\ro{r_b(3)} &= 0;\ \\
		\ro{r_w(1)} &= 0;\	&	\ro{r_w(2)} &= \phantom{-}0;\	&	\ro{r_w(3)} &= 1;\ \\
		\\
	\end{alignat*}
	
	\begin{alignat*}{6}
		\oo{o_m(1)} &= 1;\	&	\oo{o_m(2)} &= \phantom{-}0;	\ &	\oo{o_m(3)} &= 0;\\
		\oo{o_v(1)} &= 0;\	&	\oo{o_v(2)} &= \phantom{-}1;	\ &	\oo{o_v(3)} &= 0;\\
		\oo{o_a(1)} &= 0;\	&	\oo{o_a(2)} &= -1;\ 			&	\oo{o_a(3)} &= 0;\\
		\oo{o_b(1)} &= 0;\	&	\oo{o_b(2)} &= -1;\ 			&	\oo{o_b(3)} &= 0;\\
		\oo{o_w(1)} &= 0;\	&	\oo{o_w(2)} &= \phantom{-}0;	\ &	\oo{o_w(3)} &= 1;	
	\end{alignat*}
\end{vwcol}
}

Consequently the cofactor polynomial is $t(x) = (x-1)(x-2)(x-3)$, which will ensure that all three operations are computed correctly. 

\vspace{-1em}Next we leverage polynomial interpolation to find each \term{variable polynomial}:
{\vspace{-5em} 
\setstretch{2.1}
\begin{vwcol}[widths={0.31,0.34,0.36},sep=0cm,justify=raggedleft,presep=0cm,postsep=0cm,rule=0pt,indent=0em] 
	\begin{align*}
		\lo{l_a(x)} &= \phantom{-}\frac{1}{2}x^2-\frac{5}{2}x+3; \\
		\lo{l_w(x)} &= -\frac{1}{2}x^2 + \frac{5}{2}x-2; \\
		\\
		\\
	\end{align*}
	
	\begin{alignat*}{6}
		\ro{r_m(x)} &= -x^2+4 x-3; \\
		\ro{r_a(x)} &= \phantom{-}x^2-4 x+3; \\
		\ro{r_b(x)} &= \phantom{-}\frac{3}{2}x^2 - \frac{13}{2}x+6; \\
		\ro{r_w(x)} &= \phantom{-}\frac{1}{2}x^2-\frac{3}{2}x+1; \\
		\\
	\end{alignat*}
	
	\begin{alignat*}{6}
		\oo{o_m(x)} &= \phantom{-}\frac{1}{2}x^2 -\frac{5}{2}x + 3;\\
		\oo{o_v(x)} &= -x^2+4 x-3;\\
		\oo{o_a(x)} &= \phantom{-}x^2-4 x+3; \\
		\oo{o_b(x)} &= \phantom{-}x^2-4 x+3;\\
		\oo{o_w(x)} &= \phantom{-}\frac{1}{2}x^2-\frac{3}{2}x+1;	
	\end{alignat*}
\end{vwcol}
}
\colorlet{Lcolor1}{\lcolor!100!black!80!}
\colorlet{Rcolor1}{\rcolor!90!\bcolor!90!}
\colorlet{Rcolor2}{\rcolor!65!\bcolor!90!}
\colorlet{Rcolor3}{\rcolor!40!\bcolor!100!}
\colorlet{Ocolor1}{\ocolor!90!Magenta!90!}
\colorlet{Ocolor2}{\ocolor!65!Magenta!90!}
\colorlet{Ocolor3}{\ocolor!40!Magenta!90!}
\colorlet{Ocolor4}{\ocolor!70!black!90!}
Which are plotted as:

\newcommand{\exla}{(1 + (-1 + (-2 + x)/2)*(-1 + x))}
\newcommand{\exlw}{(1 + (2 - x)/2)*(-1 + x)}

\newcommand{\exrm}{(3 - x)*(-1 + x)}
\newcommand{\exra}{(-3 + x)*(-1 + x)}
\newcommand{\exrb}{(1 + (-2 + (3*(-2 + x))/2)*(-1 + x))}
\newcommand{\exrw}{((-2 + x)*(-1 + x))/2}

\newcommand{\exom}{(1 + (-1 + (-2 + x)/2)*(-1 + x))}
\newcommand{\exov}{(3 - x)*(-1 + x)}
\newcommand{\exow}{((-2 + x)*(-1 + x))/2}
\newcommand{\exoa}{(-3 + x)*(-1 + x)}
\newcommand{\exob}{(-3 + x)*(-1 + x)}

\newcommand{\exL}{(3 * \exla + 1 * \exlw)}
\newcommand{\exR}{(6 * \exrm + 3 *\exra + 2 *\exrb + 1 *\exrw)}
\newcommand{\exO}{(6 *\exom + 6 *\exov + 1 *\exow + 3 *\exoa + 2 *\exob)}

\begin{center} \small{
\plotgraph{}{
	\plotpolylabeled{ \exla }{\lcolor}{$l_a(x)$}{sloped,pos=0.3} ;
	\plotdotangle{1}{1}{90}{$a$}
	\plotpolylabeled{ \exlw }{Lcolor1}{$l_w(x)$}{sloped,pos=0.35};
	\plotdotangle{2}{1}{90}{$w$} 
	\plotdotangle{3}{1}{90}{$w$}
}{}{ymin=-1.9, ymax=2.9, xmax=3.9, height=19em, width=17.3em, ytick={-1,...,1}}
\plotgraph{}{
	\plotpolylabeled{ \exrm }{Rcolor1}{$r_m(x)$}{sloped,pos=0.27};
	\plotpolylabeled{ \exra }{Rcolor2}{\small{$r_a(x)$}}{sloped,pos=0.22};
	\plotpolylabeled{ \exrb }{Rcolor3}{$r_b(x)$}{sloped,pos=0.35};
	\plotpolylabeled{ \exrw }{\rcolor}{$r_w(x)$}{sloped,pos=0.1};
	
	\plotdotangle{1}{1}{90}{$b$}
	\plotdotangle{2}{1}{90}{$m$}
	\plotdotangle{2}{-1}{-90}{$-a, -b$}
	\plotdotangle{3}{1}{90}{$w$}
}{}{ymin=-1.9, ymax=2.9, xmax=3.9, height=19em, width=17.3em, ytick={-1,...,1}}
\plotgraph{}{
	\plotpolylabeled{ \exom }{\ocolor,draw opacity=0.8}{$o_m(x)$}{sloped,draw opacity=0.5,pos=0.3};
	\plotpolylabeled{ \exov }{Ocolor2}{$o_v(x)$}{sloped,pos=0.29};
	\plotpolylabeled{ \exow }{Ocolor1}{$o_w(x)$}{sloped,pos=0.1};
	\plotpolylabeled{ \exoa }{white}{$o_a(x), o_b(x)$}
	{sloped,below=-0.07cm,pos=0.5585};
	\plotpolylabeled{ \exob }{white}{$o_a(x), o_b(x)$}
	{sloped,below=-0.07cm,pos=0.5615};
	\plotpolylabeled{ \exob }{Ocolor4}{$o_a(x), o_b(x)$}
	{sloped,below=-0.07cm,pos=0.56};
	
	\plotdotangle{1}{1}{90}{$m$}
	\plotdotangle{2}{1}{90}{$v$}
	\plotdotangle{2}{-1}{-90}{$-a,-b$}
	\plotdotangle{3}{1}{90}{$w$}
}{}{ymin=-1.9, ymax=2.9, xmax=3.99, height=19em, width=17.3em, ytick={-1,...,1}}
}\end{center}

We are ready to prove computation through polynomials. Firstly, let us choose input values for the function, for example $w = 1, a = 3, b = 2$. Secondly, calculate values of intermediary variables from operations:
	\begin{align*}
		m	&=	a \times b = 6  \\
		v	&=	w(m - a - b) + a + b = 6
	\end{align*}

After, we assign all values involved in the computation of the result to the corresponding \term{variable polynomials} and sum them up to form operand and output polynomials:

{\vspace{-5em} 
\setstretch{2}
\begin{align*}
	\lo{L(x)} &= \lo{ 3 \cdot l_a(x) + 1 \cdot l_w(x) } = x^2 -5x + 7 \\
	\ro{R(x)} &= \ro{ 6 \cdot r_{m}(x) + 3 \cdot r_a(x) + 2 \cdot r_b(x) + 1\cdot r_w(x) } = \frac{1}{2} x^2- 2\frac{1}{2}x +4 \\
	\oo{O(x)} &= \oo{ 6 \cdot o_{m}(x) + 6 \cdot o_v(x) + 3 \cdot o_a(x) + 2 \cdot o_b(x) + 1\cdot o_w(x) } = 2\frac{1}{2} x^2 - 12\frac{1}{2}x+16
\end{align*}
}
and in the graph form these are:

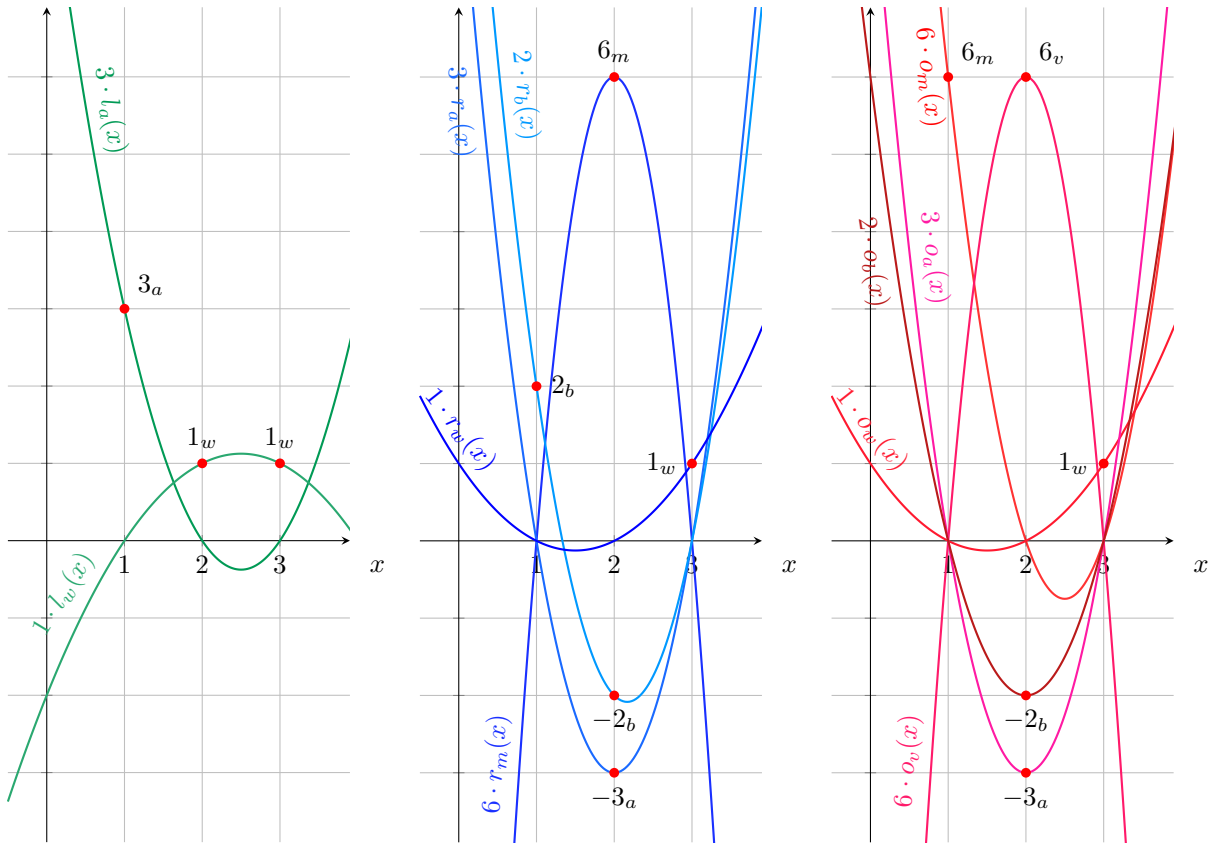
\begin{center} 
\small{
\plotgraph{}{
	\plotpolylabeled{ 3 * \exla }{\lcolor}{$3 \cdot l_a(x)$}{sloped,pos=0.35};
	\plotpolylabeled{ 1 * \exlw }{Lcolor1}{$1 \cdot l_w(x)$}{sloped,pos=0.35};
	
	
	\plotdotangle{1}{3}{45}{$3_a$}
	\plotdotangle{2}{1}{90}{$1_w $} 
	\plotdotangle{3}{1}{90}{$1_w$}
}{}{ymin=-3.9, ymax=6.9, xmax=3.9, height=36em, width=17.3em, ytick={-3,...,6}}
\plotgraph{}{
	\plotpolylabeled{ 6 * \exrm }{Rcolor1}{$6 \cdot r_m(x)$}{sloped,pos=0.32};
	\plotpolylabeled{ 3 *\exra }{Rcolor2}{\small{$3 \cdot r_a(x)$}}{sloped,pos=0.22,below};
	\plotpolylabeled{ 2 *\exrb }{Rcolor3}{$2 \cdot r_b(x)$}{sloped,pos=0.31};
	\plotpolylabeled{ 1 *\exrw }{\rcolor}{$1 \cdot r_w(x)$}{sloped,pos=0.14};
	
	
	\plotdotangle{1}{2}{0}{$2_b$}
	\plotdotangle{2}{6}{90}{$6_m$}
	\plotdotangle{2}{-3}{-90}{$-3_a$}
	\plotdotangle{2}{-2}{-90}{$-2_b$}
	\plotdotangle{3}{1}{180}{$1_w$}
	
}{}{ymin=-3.9, ymax=6.9, xmax=3.9, height=36em, width=17.3em, ytick={-3,...,6}}
\plotgraph{}{
	\plotpolylabeled{ 6 *\exom }{\ocolor,draw opacity=0.8}{$6 \cdot o_m(x)$}{sloped,draw opacity=0.5,pos=0.437,below=-0.05cm};
	\plotpolylabeled{ 6 *\exov }{Ocolor2}{$6 \cdot o_v(x)$}{sloped,pos=0.32};
	\plotpolylabeled{ 1 *\exow }{Ocolor1}{$1 \cdot o_w(x)$}{sloped,pos=0.14};
	\plotpolylabeled{ 3 *\exoa }{Ocolor3}{$3 \cdot o_a(x)$}{sloped,pos=0.278};
	\plotpolylabeled{ 2 *\exob }{Ocolor4}{$2 \cdot o_b(x)$}{sloped,below,pos=0.22};
	
	
	\plotdotangle{1}{6}{45}{$6_m$}
	\plotdotangle{2}{6}{45}{$6_v$}
	\plotdotangle{2}{-3}{-90}{$-3_a$}
	\plotdotangle{2}{-2}{-90}{$-2_b$}
	\plotdotangle{3}{1}{180}{$1_w$}
}{}{ymin=-3.9, ymax=6.9, xmax=3.9, height=36em, width=17.3em, ytick={-3,...,6}}
}\end{center}

Summed up to represent operand and output values in corresponding operations:

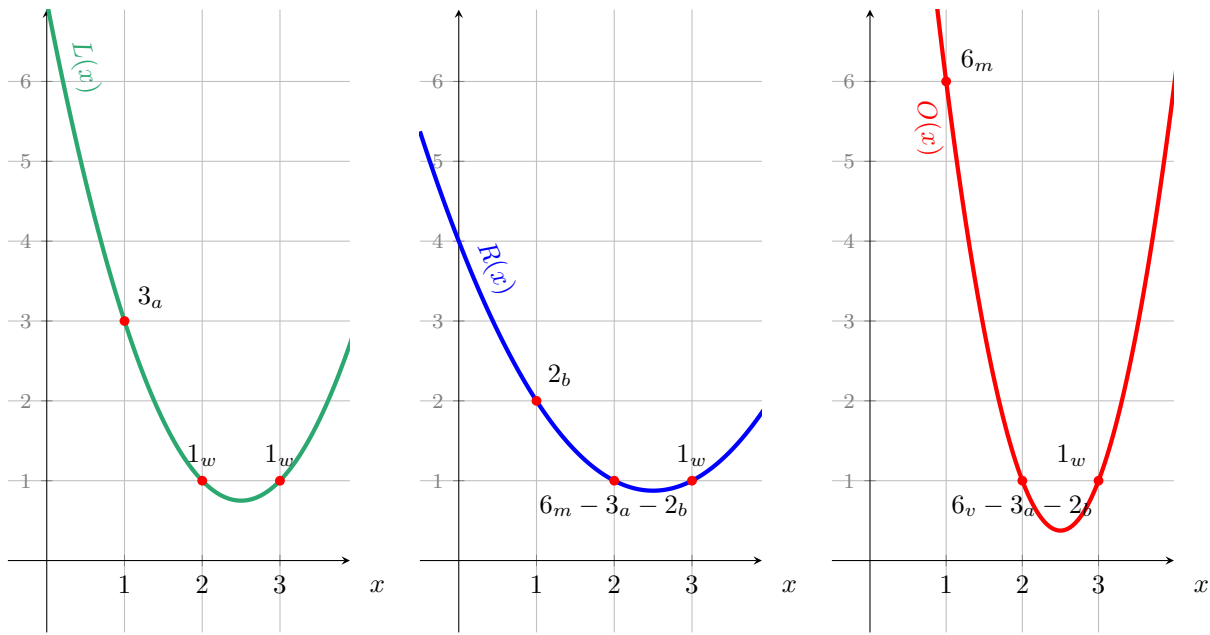
\begin{center} 
\pgfplotsset{every y tick label/.append style={color=gray, font=\scriptsize}}
\small{
\plotgraph{}{
	\plotpolylabeled{ \exL }{Lcolor1, ultra thick}{$L(x)$}{sloped,pos=0.25};
	
	\plotdotangle{1}{3}{45}{$3_a$}
	\plotdotangle{2}{1}{90}{$1_w$} 
	\plotdotangle{3}{1}{90}{$1_w$}
}{}{ymin=-0.9, ymax=6.9, xmax=3.9, height=28em, width=17.3em, ytick={-3,...,6}, yticklabel={} }
\plotgraph{}{
	
	\plotpolylabeled{ \exR }{\rcolor,ultra thick}{$R(x)$}{sloped,pos=0.25};
	
	\plotdotangle{1}{2}{70}{$2_b$}
	\plotdotangle{2}{1}{-90}{$6_m -3_a -2_b$}
	\plotdotangle{3}{1}{90}{$1_w$}
	
}{}{ymin=-0.9, ymax=6.9, xmax=3.9, height=28em, width=17.3em, ytick={-3,...,6}, yticklabel={} }
\plotgraph{}{	
	\plotpolylabeled{ \exO }{\ocolor, ultra thick}{$O(x)$}{sloped,pos=0.45,below};
	
	\plotdotangle{1}{6}{15}{$6_m$}
	\plotdotangle{2}{1}{-90}{$6_v -3_a -2_b$}
	\plotdotangle{3}{1}{110}{$1_w$}
}{}{ymin=-0.9, ymax=6.9, xmax=3.99, height=28em, width=17.3em, ytick={-3,...,6}, yticklabel={} }
}\end{center}

We need to prove that $L(x) \times R(x) - O(x) = t(x)\, h(x)$, therefore we find $h(x)$:
\begin{gather*}
	h(x) = \frac{L(x) \times R(x) - O(x)}{t(x)} = \frac{\frac{1}{2}x^4 - 5x^3 + \frac{35}{2} x^2 - 25x + 12}{(x-1)(x-2)(x-3)} = \frac{1}{2}x - 2
\end{gather*}

In a graph form it is represented as:
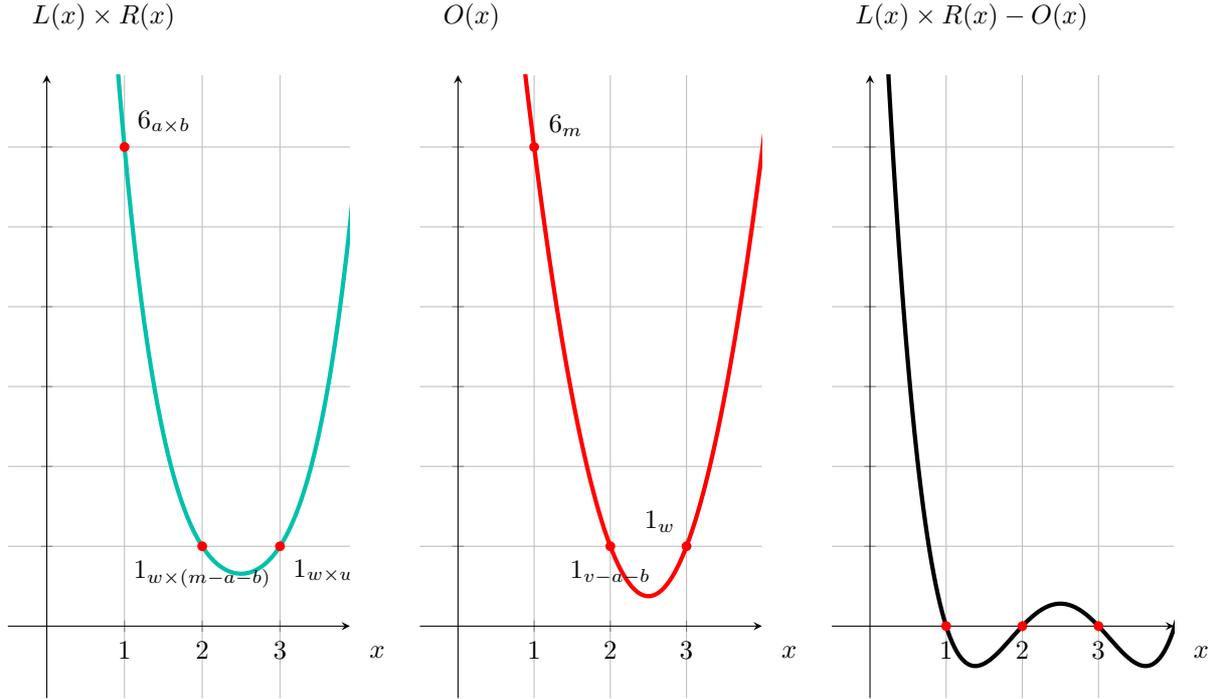
\begin{center} \small{
\plotgraph{$L(x) \times R(x)$}{
	\plotpoly{ \exL * \exR }{LRcolor, ultra thick} ;
	
	\plotdotangle{1}{6}{45}{$6_{a\times b}$}
	\plotdotangle{2}{1}{-90}{$1_{w \times (m-a-b)}$} 
	\plotdotangle{3}{1}{-45}{$1_{w\times w}$}
}{}{ymin=-0.9, ymax=6.9, xmax=3.9, height=28em, width=17.3em, ytick={-3,...,6}}
\plotgraph{$O(x)$}{	
	\plotpoly{ \exO }{\ocolor, ultra thick};
	
	\plotdotangle{1}{6}{15}{$6_m$}
	\plotdotangle{2}{1}{-90}{$1_{v -a -b}$}
	\plotdotangle{3}{1}{110}{$1_w$}
}{}{ymin=-0.9, ymax=6.9, xmax=3.99, height=28em, width=17.3em, ytick={-3,...,6}}
\plotgraph{$L(x) \times R(x) - O(x)$}{	
	\plotpoly{ \exL * \exR - \exO }{ultra thick};
	
	\plotdotangle{1}{0}{15}{}
	
	
	\plotdotangle{2}{0}{-90}{}
	\plotdotangle{3}{0}{110}{}
}{}{ymin=-0.9, ymax=6.9, xmax=3.99, height=28em, width=17.3em, ytick={-3,...,6}}
}\end{center}

Where it's visible that polynomial $L(x)\times R(x) - O(x)$ has solutions $x=1$, $x=2$ and $x=3$, and therefore $t(x)$ is its cofactor, which would not be the case if we used inconsistent values of variables.

That is how the knowledge of variable values for a correct computation execution is proven on the level of polynomials. A prover is then proceeding with a cryptographic portion of the protocol.

\subsection{Verifiable Computation Protocol}

We went through many important modifications of the knowledge of polynomial protocol (section \ref{section:zkSNARKOP}) to make it general-purpose, so let us see how it is defined now. Assuming agreed upon function $f(*)$ the result of computation of which is the subject of the proof, with the number of operations $d$, the number of variables $n$ and corresponding to them coefficients $\left\{ \coef{l}{i,j}, \coef{r}{i,j}, \coef{o}{i,j} \right\}_{i \in \{1, \ldots, n\}, j \in \{1, \ldots, d\}}$:

\zkProtocol{
	\item construct \term{variable polynomials} for left operand $\left\{ l_i(x) \right\}_{i \in \{1, \ldots, n\}}$ such that for all operations $j \in \{1, \ldots, d\}$ they evaluate to corresponding coefficients, i.e., $l_i(j) = \coef{l}{i,j}$, and similarly for right operand and output
	\item sample random $s, \alpha$
	\item calculate $t(x) = (x-1)(x-2)\ldots(x-d)$ and its evaluation $g^{t(s)}$
	\item compute proving key: $ \left( \
		\left\{ g^{s^k} \right\}_{k \in [d]},
		\left\{ g^{l_i(s)}, g^{r_i(s)}, g^{o_i(s)}, g^{\alpha l_i(s)}, g^{\alpha r_i(s)}, g^{\alpha o_i(s)} \right\}_{i \in \{1,\ldots,n\}}
	\right)$
	\item compute verification key: $\left(g^{t(s)}, g^{\alpha}\right)$
}{
	\item compute function $f(*)$ and therefore corresponding variables	 values $\left\{ v_i \right\}_{i \in \{1,\ldots,n\}}$
	\item calculate $h(x) = \frac{L(x) \times R(x) - O(x)}{t(x)}$, where $L(x) = \sum_{i=1}^n v_i \cdot l_i(x)$, and similarly $R(x), O(x)$
	\item assign variable values and \emph{sum up} to get operand polynomials:
	\[ g^{L(s)} = 
		  \left(g^{l_1(s)}\right)^{v_1} \cdots \left(g^{l_n(s)}\right)^{v_n},\ \ 
		  g^{R(s)} = \prod_{i=1}^{n} \left(g^{r_i(s)}\right)^{v_i}, \ \
		  g^{O(s)} = \prod_{i=1}^{n} \left(g^{o_i(s)}\right)^{v_i} 
	\]
	\item assign variable values to the shifted polynomials:
	\[
		g^{\alpha L(s)} = \prod_{i=1}^{n} \left(g^{\alpha l_i(s)}\right)^{v_i},\ \ 
		g^{\alpha R(s)} = \prod_{i=1}^{n} \left(g^{\alpha r_i(s)}\right)^{v_i}, \ \
		g^{\alpha O(s)} = \prod_{i=1}^{n} \left(g^{\alpha o_i(s)}\right)^{v_i}
	\]
	\item calculate encrypted evaluation $g^{h(s)}$ using provided powers of $s$: $\left\{ g^{s^k} \right\}_{k \in [d]}$
	\item set proof: $\left( g^{L(s)}, g^{R(s)}, g^{O(s)}, g^{\alpha L(s)}, g^{\alpha R(s)}, g^{\alpha O(s)}, g^{h(s)} \right)$
}{
	\item parse proof as $\left( g^L, g^R, g^O, g^{L'}, g^{R'}, g^{O'}, g^h \right)$
	\item \checkVarPolynomials\ check: \\ 
	$e(g^{L}, g^\alpha) = e(g^{L'}, g),\quad e(g^{R}, g^\alpha) = e(g^{R'}, g), \quad e(g^{O}, g^\alpha) = e(g^{O'}, g)$
	\item \checkOperations\ check: \\
	$e(g^{L}, g^{R}) = e(g^{t}, g^{h}) \cdot e(g^{O}, g)$
}

\note{using symbol $\prod$ allows for a concise way to express product of multiple elements, i.e., \\$\prod_{i=1}^n v_i = v_1 \cdot v_2 \cdot \ldots \cdot v_n$\,.}

The set of all the variable polynomials $\{l_i(x), r_i(x), o_i(x)\}_{i \in \{1, \ldots, n\}}$ and the target polynomial $t(x)$ is called a \term{quadratic arithmetic program} (QAP\footcite{cryptoeprint:2012:215}).  

While the protocol is sufficiently robust to allow a general computation verification, there are two security considerations that must be addressed.

\subsubsection{Non-Interchangeability of Operands and Output}

Because we use the same $\alpha$ for all the operands of \term{\checkVarPolynomials} there is nothing that prevents prover from:
\begin{itemize}
	\item using variable polynomials from other operands, e.g., $L'(s) = o_1(s) + r_1(s) + r_5(s) + \ldots$
	\item swapping \term{operand polynomials} completely, e.g., $O(s)$ with $L(s)$ will result in operation \(\maligned{\operspace{O(s)}{R(s)}{L(s)}{\ }}\)
	\item re-using same operand polynomials e.g., $\maligned{\operspace{L(s)}{L(s)}{O(s)}{\ }}$
\end{itemize} 

This interchangeability means that the prover can alter the execution and effectively prove some other computation.
The obvious way to prevent such behavior is to use different $\alpha$-s for the different operands, concretely we modify:

\zkProtocol{
		\item[] \ldots
		\item sample random $\alpha_{l}, \alpha_{r}, \alpha_{o}$ instead of $\alpha $
		\item calculate corresponding ``shifts" $\left\{g^{\alpha_{l} l_i(s)}, g^{\alpha_{r} r_i(s)}, g^{\alpha_{o} o_i(s)}\right\}_{i \in \{1 \ldots n\}}$
		\item proving key: $
		\left( 
			\left\{ g^{s^k} \right\}_{k \in [d]},
			\left\{
				g^{l_i(s)}, g^{r_i(s)}, g^{o_i(s)}, g^{\alpha_{l} l_i(s)}, g^{\alpha_{r} r_i(s)}, g^{\alpha_{o} o_i(s)}
			\right\}_{i \in \{1 \ldots n\}}
		\right)$
		\item verification key: $\left( g^{t(s)}, g^{\alpha_{l}}, g^{\alpha_{r}}, g^{\alpha_{o}} \right)$
}{
		\item[] \ldots
		\item assign variables to the ``shifted'' polynomials
	  	\begin{flalign*}
		  	& g^{\alpha_{l} L(s)} = \prod_{i=1}^{n} \left( g^{\alpha_{l} l_i(s)} \right)^{v_i},\
			g^{\alpha_{r} R(s)} = \prod_{i=1}^{n} \left(g^{\alpha_{r} r_i(s)} \right)^{v_i},\ 
			g^{\alpha_{o} O(s)} = \prod_{i=1}^{n} \left(g^{\alpha_{o} o_i(s)} \right)^{v_i} &
		\end{flalign*}
		\item set proof: $\left(
			g^{L(s)}, g^{R(s)}, g^{O(s)}, 
			g^{\alpha_{l} L(s)}, g^{\alpha_{r} R(s)}, g^{\alpha_{o} O(s)}, 
			g^{h(s)} 
		\right)$
}{
		\item[] \ldots
		\item \checkVarPolynomials\ check: \\ 
		$\prng{g^{L}, g^{\alpha_{l}}} = \prng{g^{L'}, g}, \quad
		 \prng{g^{R}, g^{\alpha_{r}}} = \prng{g^{R'}, g}, \quad		
		 \prng{g^{O}, g^{\alpha_{o}}} = \prng{g^{O'}, g}$
}

It is now not possible to use variable polynomials from other operands since $\alpha_{l}, \alpha_{r}, \alpha_{o}$ are not known to the prover.

\subsubsection{Variable Consistency Across Operands}

For any variable $v_i$ we have to \term{assign} its value to a \term{variable polynomial} for each corresponding operand, i.e., $\left(g^{l_i(s)}\right)^{v_i}, \left(g^{r_i(s)}\right)^{v_i}, \left(g^{o_i(s)}\right)^{v_i}$.
Because the validity of each of the \term{operand polynomials} is checked separately,  no enforcement requires to use same variable values in the corresponding \term{variable polynomials}. This means that the value of variable $v_1$ in left operand can differ from variable $v_1$ in the right operand or the output.

We can enforce equality of a variable value across operands through already familiar approach of restricting a polynomial (as we did with variable polynomials). If we can create a ``shifted checksum'' variable polynomial across all operands, that would restrain prover such that he can assign only same value. A verifier can combine polynomials for each variable into one, e.g., $g^{l_i(s) + r_i(s) + o_i(s)}$, and shift it by some other random value $\beta$, i.e., $g^{\beta\left(l_i(s) + r_i(s) + o_i(s)\right)}$. This shifted polynomials are provided to the prover to assign values of the variables alongside with variable polynomials:
	\[
		\left(g^{l_i(s)}\right)^{v_{\textsc{l},i}}, 
		\left(g^{r_i(s)}\right)^{v_{\textsc{r},i}}, 
		\left(g^{o_i(s)}\right)^{v_{\textsc{o},i}}, 
		\left(g^{\beta(l_i(s) + r_i(s) + o_i(s))}\right)^{v_{\upbeta,i}}
	\]
	
And the $\beta$ is encrypted and added to the verification key $g^\beta$. Now, if the values of all $v_i$ were the same (i.e., $v_{\textsc{l},i} = v_{\textsc{r},i} = v_{\textsc{o},i} = v_{\upbeta,i}$\ \ for $i \in \{1, \ldots, n\}$), the equation shall hold:
	\[ 
		\prng{ g^{v_{\textsc{l},i}\, \cdot\, l_i(s)} \cdot g^{v_{\textsc{r},i}\, \cdot\, r_i(s)} \cdot g^{v_{\textsc{o},i}\, \cdot\, o_i(s)}, g^\beta} = 
		\prng{ g^{v_{\upbeta,i}\, \cdot\, \beta(l_i(s) + r_i(s) + o_i(s))}, g }
	\]

While this is a useful consistency 	check, due to the non-negligible probability that at least two of $l(s), r(s), o(s)$ could either have same evaluation value or one polynomial is divisible by another etc., this would allow the prover to factor values $v_{\textsc{l},i}, v_{\textsc{r},i}, v_{\textsc{o},i}, v_{\upbeta,i}$ such that at least two of them are non-equal but the equation holds, rendering the check ineffective:
\[\left(v_{\textsc{l},i} \cdot l_i(s) + v_{\textsc{r},i} \cdot r_i(s) + v_{\textsc{o},i} \cdot o_i(s)\right) \cdot \beta = v_{\upbeta,i} \cdot \beta\cdot\left(l_i(s) + r_i(s) + o_i(s)\right)\]

For example, let us consider a single operation, where it is the case that $l(x) = r(x)$. We will denote evaluation of those two as $w = l(s) = r(s)$ and $y = o(x)$.
The equation then will look as:
	\[
		\beta (v_{\textsc{l}}\, w + v_{\textsc{r}} \,w + v_{\textsc{o}}\, y) =
		v_{\upbeta} \cdot \beta ( w + w + y)
	\]

Such form allows, for some arbitrary $v_{\textsc{r}}$ and $v_{\textsc{o}}$, to set $v_{\upbeta} = v_{\textsc{o}}$, $v_{\textsc{l}} = 2 v_{\textsc{o}} - v_{\textsc{r}}$, which will translate into:
	\[
		\beta (2 v_\textsc{o}\, w - v_{\textsc{r}}\, w + v_{\textsc{r}} \,w + v_{\textsc{o}}\, y) =
		v_\textsc{o} \cdot \beta (2w + y)
	\]

Hence such consistency strategy is not effective. A way to mitigate this is to use different $\beta$ for each operand, ensuring that operand's \term{variable polynomials} will have unpredictable values. Following are the protocol modifications:
\zkProtocol{
	\item \ldots\ sample random $\beta_{l}, \beta_{r}, \beta_{o}$
	\item calculate, encrypt and add to the proving key the \term{variable consistency polynomials}: \\ 
		$\left\{ g^{\beta_{l} l_i(s) + \beta_{r} r_i(s) + \beta_{o} o_i(s)} \right\}_{i \in \{1,\ldots,n\}}$ 
	\item encrypt $\beta$-s and add to the verification key: $\left( g^{\beta_{l}}, g^{\beta_{r}}, g^{\beta_{o}} \right)$
}{
	\item \ldots\ assign variable values to the \term{variable consistency polynomials}:\\ 
		$g^{z_i(s)} = \left( g^{\beta_{l} l_i(s) + \beta_{r} r_i(s) + \beta_{o} o_i(s)} \right)^{v_i}$\quad for $i \in \{1,\ldots,n\}$
	\item add assigned polynomials in encrypted space:
	\begin{flalign*}
		& g^{Z(s)} = \prod_{i=1}^n g^{z_i(s)} =
		g^{\beta_{l} L(s) + \beta_{r} R(s) + \beta_{o} O(s)} &
	\end{flalign*}
	\item add to the proof: $g^{Z(s)}$
}{
	\item \ldots\ check the consistency between provided \term{operand polynomials} and the ``checksum" polynomial: 
	\begin{flalign*}
		& \prng{g^{L}, g^{\beta_{l}}} \cdot \prng{g^{R}, g^{\beta_{r}}} \cdot \prng{g^{O}, g^{\beta_{o}}} = \prng{g^{Z}, g} &
	\end{flalign*}
	
	which is equivalent to:
	\begin{flalign*}
		& \prng{g,g}^{\beta_{l} L + \beta_{r} R + \beta_{o} O} = \prng{g,g}^{Z} &
	\end{flalign*}
}

Same variable values tempering technique will fail in such construction because different $\beta$-s makes the same polynomials incompatible for manipulation. There is however a flaw similar to the one in remark \ref{remark:polyModification}, concretely because the terms $g^{\beta_l}, g^{\beta_r}, g^{\beta_o}$ are publicly available an adversary can modify the zero-index coefficient of any of the variable polynomials since it does not rely on $s$, i.e., $g^{\beta_l s^0} = g^{\beta_l}$\,.

\subsubsection{Non-malleability of Variable and Variable Consistency Polynomials} \label{subsection:poly-non-malleability}

\subparagraph{Malleability of Variable Polynomials} \mbox{}

Let us exemplify remark \ref{remark:polyModification} with the following two operations:
	\begin{align*}
		\oper{a}{1}{b} \\
		\oper{3a}{1}{c}
	\end{align*}
	
The expected result is $b = a$ and $c = 3a$, with clear relationship $c = 3b$.
This implies that the \term{left operand's variable} polynomial has evaluations $l_a(1) = 1$ and $l_a(2) = 3$. Regardless of the form of $l_a(x)$, a prover can unproportionately assign the value of $a$, by providing modified polynomial $l'_a(x) = a l_a(x) + 1$. Therefore evaluations will be $l'_a(1) = a + 1$ and $l'_a(2) = 3a + 1$, hence the results $b = a + 1$ and $c = 3a + 1$ where $c \neq 3b$, effectively meaning that the value of $a$ is different for different operations.

Because the prover has access to $g^{\alpha_l}$ and $g^{\beta_l}$ he can satisfy both the \term{correct operand polynomials} and \term{\checkConsistency}\ checks:
	\begin{itemize}
		\item \ldots proving:
		\begin{itemize}
			\item form left operand polynomial by unproportionately assigning variable $a$: \\ $L(x) = a \cdot l_a(x) + 1$
			\item form right operand and output polynomials as usual: \\
			$R(x) = r_1(x)$, $O(x) = b \cdot o_b(x) + c \cdot o_c(x)$
			\item calculate the remainder $h(x) = \frac{L(x)\cdot R(x) - O(x)}{t(x)}$
			\item compute encryption: $g^{L(s)} = \left( g^{l_a(s)} \right)^a \cdot g^1$\ \  and as usual for $g^{R(s)}, g^{O(s)}$
			\item compute $\alpha$-shifts: $g^{\alpha L(s)} = \left( g^{\alpha l_a(s)} \right)^a \cdot g^\alpha$\ \ and as usual for $g^{\alpha R(s)}, g^{\alpha O(s)}$
			\item compute variable consistency polynomials:
				\begin{flalign*}
					& g^{Z(s)} = 
					\prod_{i \in \{1, a, b, c\}} \left( g^{\beta_l l_i(s) + \beta_r r_i(s) + \beta_o o_i(s)} \right)^i \ \cdot g^{\beta_l} = 
					g^{\beta_l (L(s) + 1) + \beta_r R(s) + \beta_o O(s)}
					&
				\end{flalign*}
				
				{\footnotesize where the subscript $_i$ represents symbol of the corresponding variable while the exponent $^i$ represents the value of variable; moreover undefined \term{variable polynomials} are equal to zero.}
			\item set proof: \( \left(
				g^{L(s)}, g^{R(s)}, g^{O(s)}, 
				g^{\alpha_l L(s)}, g^{\alpha_r R(s)}, g^{\alpha_o O(s)}, 
				g^{Z(s)}
				g^{h(s)} 
			\right) \)
		\end{itemize}
		\item verification:
		\begin{itemize}
			\item \checkVarPolynomials\ check:
			\begin{flalign*}
				& \prng{g^{L'}, g} = \prng{g^{L}, g^\alpha} \ \ \Rightarrow \ \ \prng{g^{\alpha a \cdot l_a(s) + \alpha}, g} = \prng{g^{a l_a(s) + 1}, g^\alpha}&
			\end{flalign*}
			and as usually for $g^{R'}, g^{O'}$
			\item \checkConsistency\ check 
			\begin{flalign*}
				& \prng{g^{L}, g^{\beta_l}} \cdot \prng{g^{R}, g^{\beta_r}} \cdot \prng{g^{O}, g^{\beta_o}} = \prng{g^{Z}, g} \Rightarrow & \\
				& \prng{g, g}^{(a\cdot l_a + 1) \beta_l + R\beta_r + O\beta_o} = 
				  \prng{g, g}^{\beta_l (L + 1) + \beta_r R + \beta_o O} &
			\end{flalign*}
			\item \checkOperations\ check $e(g^{L}, g^{R}) = e(g^{t}, g^{h}) \cdot e(g^{O}, g)$
		\end{itemize}
	\end{itemize}

\subparagraph{Malleability of Variable Consistency Polynomials} \mbox{}

Moreover the availability of $g^{\beta_l}, g^{\beta_r}, g^{\beta_o}$ allows to use different values of same variable in different operands. For example, if we have an operation:
\begin{align*}
	\oper{a}{a}{b}
\end{align*}
Which can be represented by the variable polynomials: 
\begin{align*}
	{\color{\lcolor} l_a(x) = x},\quad {\color{\rcolor}r_a(x) = x},\quad {\color{\ocolor}o_a(x) = 0} \\
	{\color{\lcolor} l_b(x) = 0},\quad {\color{\rcolor}r_b(x) = 0},\quad {\color{\ocolor}o_b(x) = x} 
\end{align*}

While the expected output is $b = a^2$, we can set different values of $a$, for example ${\color{\lcolor}a = 2}$, ${\color{\rcolor}a = 5}$ as following:

\begin{itemize}
	\item proving:
	\begin{itemize}
		\item \ldots form left operand polynomial with $a = 2$: $L(x) = 2 l_a(x) + 10 l_b(x)$
		\item form right operand polynomial with $a = 5$: $R(x) = 2 r_a(x) + 3 + 10 r_b(x)$
		\item form output polynomial with $b = 10$: $O(x) = 2 o_a(x) + 10 o_b(x)$
		\item \ldots compute encryptions: 
			\begin{flalign*}
				& g^{L(s)} = \left( g^{l_a(s)} \right)^2 \cdot \left( g^{l_b(s)} \right)^{10} = g^{2l_a(s) + 10 l_b(s)} &\\
				& g^{R(s)} = \left( g^{r_a(s)} \right)^2 \cdot (g)^3 \cdot \left( g^{r_b(s)} \right)^{10} = g^{2r_a(s) + 3 + 10 r_b(s)} &\\
				& g^{O(s)} = \left( g^{o_a(s)} \right)^2 \cdot \left( g^{o_b(s)} \right)^{10} = g^{2 o_a(s) + 10 o_b(s)} &
			\end{flalign*}
		\item compute variable consistency polynomial:
			\begin{flalign*}
				& g^{Z(s)} = \left( g^{\beta_l l_a(s) + \beta_r r_a(s) + \beta_o o_a(s)} \right)^{2} \ \cdot 
					\left( g^{\beta_r} \right)^{3} \cdot 
					\left( g^{\beta_l l_b(s) + \beta_r r_b(s) + \beta_o o_b(s)} \right)^{10} = &\\
				& g^{\beta_l \left(2 l_a(s) + 10 l_b(s) \right)\ +\ \beta_r (2 r_a(s) + 3 + 10 r_b(s))\ +\ \beta_o (2 o_a(s) + 10 o_b(s))} &
			\end{flalign*}
	\end{itemize}
	\item verification
	\begin{itemize}
		\item \ldots \checkConsistency\ check, should hold:
			\begin{flalign*}
				& \prng{g^{L}, g^{\beta_l}} \cdot \prng{g^{R}, g^{\beta_r}} \cdot \prng{g^{O}, g^{\beta_o}} = \prng{g^{Z}, g} &
			\end{flalign*}
	\end{itemize}
\end{itemize}	

\note{polynomials $o_a(x), l_b(x), r_b(x)$ can actually be disregarded since they are evaluating to 0 for any $x$, however we preserve those for completeness.}

Such ability  sabotages the \term{soundness} of proof. It is clear that encrypted $\beta$-s
should not be available to a prover. 

\paragraph{Non-Malleability} \mbox{}

\newcommand{\highlightscript}[1]{%
  \colorbox{yellow!40}{\scriptsize $#1$}}
One way to address malleability is to make $g^{\beta_l},\ g^{\beta_r},\ g^{\beta_o}$ from verification key incompatible with $g^{Z(s)}$ by multiplying them in encrypted space by a random secret $\gamma$ (gamma) during setup stage: $g^{\beta_l \gamma},\ g^{\beta_r \gamma},\ g^{\beta_o \gamma}$. Consecutively such masked encryptions does not allow feasibility to modify $g^{Z(s)}$ in a meaningful way since $Z(s)$ is not a multiple of $\gamma$, e.g., $g^{Z(s)} \cdot g^{v'\cdot\beta_l \gamma} = g^{\beta_l (L(s) + \highlightscript{v' \gamma} ) + \beta_r R(s) + \beta_o O(s)}$. Because a prover does not know the $\gamma$ the alteration will be random. The modification requires us to balance the \checkConsistency\ check equation in the protocol multiplying $Z(s)$ by $\gamma$:

\begin{itemize}	
	\item setup
	\begin{itemize}
		\item \ldots sample random $\beta_l, \beta_r, \beta_o, \gamma$
		\item \ldots set verification key: $\left(\ldots, g^{\beta_l \gamma}, g^{\beta_r \gamma}, g^{\beta_o \gamma}, g^{\gamma} \right)$
	\end{itemize}
	\item proving \ldots
	\item verification
	\begin{itemize}
		\item \ldots \checkConsistency\ check should hold:
			\begin{flalign*}
				& \prng{g^{L}, g^{\beta_l \gamma}} \cdot \prng{g^{R}, g^{\beta_r \gamma}} \cdot \prng{g^{O}, g^{\beta_o \gamma}} = \prng{g^{Z}, g^\gamma } &
			\end{flalign*}
	\end{itemize}
\end{itemize}

It is important to note that we exclude the case when variable polynomials are of 0-degree (e.g., $l_1(x) = 1x^0$), which otherwise would allow to expose encryptions of $\beta$ in variable consistency polynomials of proving key $\left\{ g^{\beta_l l_i(s) + \beta_r r_i(s) + \beta_o o_i(s)} \right\}_{i \in \{1,\ldots,n\}}$ in case when any two of operands / output is zero, e.g., for\ \ $l_1(x) = 1$, $r_1(s) = 0$, $o_1(s) = 0$\ \ this will result in\ \ $g^{\beta_l l_1(s) + \beta_r r_1(s) + \beta_o o_1(s)} = g^{\beta_l}$\,.

We could also similarly \emph{mask} the $\alpha$-s to address the malleability of \term{variable polynomials}. However it is not necessary since any modification of a \term{variable polynomial} needs to be reflected in \term{variable consistency polynomials} \todonot{predefined check} which are not possible to modify.

\subsubsection{Optimization of \titlecap{\checkConsistency}\ Check} 
\label{optimization} The \term{\checkConsistency} check is effective now, but it adds 4 expensive pairing operations and 4 new terms to the verification key. The Pinocchio protocol \cite{cryptoeprint:2013:279} uses a clever selection of the generators $g$ for each operand \emph{ingraining} the ``shifts'':

\zkProtocol{
	\item \ldots sample random $\beta, \gamma, \rho_l, \rho_r$ and set $\rho_o = \rho_l \cdot \rho_r$
	\item set generators $g_l = g^{\rho_l}, g_r = g^{\rho_r}, g_o = g^{\rho_o}$
	\item set proving key: \\ $\left(\left\{ g^{s^k} \right\}_{k \in [d]}, \left\{g_l^{l_i(s)}, g_r^{r_i(s)}, g_o^{o_i(s)}, g_l^{\alpha_l l_i(s)}, g_r^{\alpha_r r_i(s)}, g_o^{\alpha_o o_i(s)}, g_l^{\beta l_i(s)} \cdot g_r^{\beta r_i(s)} \cdot g_o^{\beta o_i(s)} \right\}\right)$
	\item set verification key: $\left(g_o^{t(s)}, g^{\alpha_l}, g^{\alpha_r}, g^{\alpha_o}, g^{\beta \gamma}, g^\gamma \right)$
}{
	\item \ldots assign variable values 
	\begin{flalign*}
		& g^{Z(s)} = \prod_{i=1}^n \left( g_l^{\beta l_i(s)} \cdot g_r^{\beta r_i(s)} \cdot g_o^{\beta o_i(s)} \right)^{v_i} &
	\end{flalign*}
}{
	\item \ldots \checkVarPolynomials\ check:
	\begin{flalign*}
		& \prng{g_l^{L'}, g} = \prng{g_l^{L}, g^{\alpha_l}} \textrm{,\ \ and similarly for \ \ } g_r^{R}, g_o^{O} &
	\end{flalign*}
	\item \checkConsistency\ check:
	\begin{flalign*}
		& \prng{g_l^{L} \cdot g_r^{R} \cdot g_o^{O}, g^{\beta \gamma}} = \prng{g^{Z}, g^\gamma} &
	\end{flalign*}
	\item \checkOperations\ check:
	\begin{flalign*}
		& \prng{g_l^{L} \cdot g_r^{R}} = \prng{g_o^t, g^h} \prng{g_o^O, g} \Rightarrow & \\
		& \prng{g, g}^{\rho_l \rho_r L R} = \prng{g, g}^{\rho_l \rho_r th + \rho_l \rho_r O} &
	\end{flalign*}
}

Such randomization of the generators further adds to the security making \term{variable polynomials} malleability, described in remark \ref{remark:polyModification}, ineffective because for intended change it must be a multiple of either $\rho_l, \rho_r$ or $\rho_o$, raw or encrypted versions of which are not available (assuming, as stated previously that we're not dealing with 0-degree variable polynomials which could expose encrypted versions).

The optimization makes verification key two elements smaller and eliminates two pairing operations from the verification step.

\note{there are further protocol improvements in the Jens Groth's 2016 paper \cite{cryptoeprint:2016:260}.}

\subsection{Constraints} \label{Constraints}

Our analysis has been primarily focusing on the notion of operation. However, the protocol is not actually ``computing" but rather is checking that the output value is the correct result of an operation for the operand's values. That is why it is called a constraint, i.e., a verifier is constraining a prover to provide valid values for the predefined ``program" no matter what are they. A multitude of constraints is called a \term{constraint system} (in our case it is a rank 1 constraint system or R1CS).

\note{This implies that one way to find all correct solutions is to perform a brute-force of all possible combinations of values and select only ``valid'' ones, or use more sophisticated techniques of constraint satisfaction \cite{wiki:constraint-satisfaction}. }

Therefore we can also use constraints to ensure other relationships. For example, if we want to make sure that the value of the variable $a$ can only be $0$ or $1$ (i.e., binary), we can do it with the simple constraint:
\begin{align*}
	\oper{a}{a}{a} \\
	\noalign{We can also constrain $a$ to only be $2$:}	
	\oper{(a - 2)}{1}{0}
\end{align*}

A more complex example is ensuring that number $a$ is a 4-bit number\footnote{Also called nibble}, in other words it is possible to represent $a$ with 4 bits. We can also call it ``ensuring number range'' since a 4-bit number can represent $2^4$ combinations, therefore 16  numbers in the range from 0 to 15. In the decimal number system any number can be represented as a sum of powers of the base 10 (as the number of fingers on our hands) with corresponding coefficients, for example, $123 = 1\cdot 10^2 + 2\cdot 10^1 + 3\cdot 10^0$. Similarly a binary number can be represented as a sum of powers of base 2 with corresponding coefficients, for example, $1011\ \text{(binary)} = 1\cdot 2^3 + 0\cdot 2^2 + 1\cdot 2^1 + 1\cdot 2^0 = 11\ \text{(decimal)}$.

Therefore if $a$ is a 4-bit number, then $a = b_3\cdot 2^3 + b_2\cdot 2^2 + b_1\cdot 2^1 + b_0\cdot 2^0$ for some boolean $b_0, b_1, b_2, b_3$. The constraint can be following:
\begin{align*}
	1:&& \oper{a}{1}{8\cdot b_3 + 4\cdot b_2 + 2\cdot b_1 + 1\cdot b_0} \\
\noalign{and to ensure that $b_0, b_1, b_2, b_3$ can only be binary we need to add:}
	2:&& \oper{b_0}{b_0}{b_0} \\
	3:&& \oper{b_1}{b_1}{b_1} \\
	4:&& \oper{b_2}{b_2}{b_2} \\
	5:&& \oper{b_3}{b_3}{b_3} 
\end{align*}

Quite sophisticated constraints can be applied this way, ensuring that the values used are complying with the rules. It is important to note that the above constraint 1 is not possible in the current operation's construction:
\begin{align*}
	\operFinal
\end{align*}

Because the value $\ro{1}$ (and $\lo{2}$ from the previous constraint) has to be expressed through $\ro{c \cdot v\ped{one}}$, where $\ro{c}$ can be ingrained into the proving key, but the $\ro{v\ped{one}}$ may have any value because the prover supplies it. While we can enforce the $c \cdot v$ to be $0$ by setting $c = 0$, it is hard to find a constraint to enforce $v\ped{one}$ to be $1$ in the construction we are limited by. Therefore there should be a way for a verifier to set the value of $v\ped{one}$.

\subsection{Public Inputs and One}

The proofs would have limited usability if it were not possible to check them against the verifier's inputs, e.g., knowing that the prover has multiplied two values without knowing what was the result and/or values. While it is possible to ``hardwire" the values to check against (e.g., the result of multiplication must always be 12) in the proving key, this would require to generate separate pair of keys for each desired ``verifier's input."

Therefore it would be universal if the verifier could specify some of the values (inputs or/and outputs) for the computation, including the $v\ped{one}$, instead of the prover.

First, let us consider the proof values $\lo{g^{L(s)}}, \ro{g^{R(s)}}, \oo{g^{O(s)}}$. Because we are using the homomorphic encryption it is possible to augment these values, for example, we can add another encrypted polynomial evaluation $g^{L(s)} \cdot g^{l_v(s)} = g^{L(s) + l_v(s)}$, which means that the verifier could add other variable polynomials to the already provided ones. Therefore if we could exclude necessary variable polynomials from the ones available to the prover, the verifier would be able to set his values on those variables, while the computation check should still match.

It is easy to achieve since the verifier is already constraining the prover in the choice of polynomials he can use empolying the $\alpha$-shift. Therefore those variable polynomials can be moved from the proving key to the verification key while eliminating its $\alpha$-s and $\beta$ checksum counterparts.

The necessary protocol update:
{
\setlist[itemize]{leftmargin=6mm}
\zkProtocol{
	\item \ldots separate all $n$ variable polynomials into two groups:
	\begin{itemize}
		\item verifier's $m + 1$: \\
		$L_v(x) = l_0(x) + l_1(x) + \ldots + l_m$, and alike for $R_v(x)$ and $O_v(x)$, \\
		where index $0$  is reserved for the value of $v\ped{one} = 1$
		\item prover's $n - m$: \\
		$L_p(x) = l_{m+1}(x) + \ldots+ l_n(x)$, and alike for $R_p(x)$ and $O_p(x)$
	\end{itemize}
	\item set proving key: 
	\begin{flalign*}
		& \left( \left\{ g^{s^k} \right\}_{k \in [d]},
			\left\{
				g_l^{l_i(s)}, g_r^{r_i(s)}, g_o^{o_i(s)}, 
				g_l^{\alpha_l l_i(s)}, g_r^{\alpha_r r_i(s)}, g_o^{\alpha_o o_i(s)}, 
				g_l^{\beta l_i(s)} \cdot g_r^{\beta r_i(s)} \cdot g_o^{\beta o_i(s)} 
			\right\}_{i \in \{m + 1, \ldots, n\}} 
		\right) &
	\end{flalign*}
 
	\item add to the verification key: \\
	$ \left( \ldots, \left\{ g_l^{l_i(s)}, g_r^{r_i(s)}, g_o^{o_i(s)} \right\}_{i \in \{0, \ldots, m\}} \right) $
}{
	\item \ldots calculate $h(x)$ accounting for the verifier's polynomials: $\displaystyle h(x) = \frac{L(x)\cdot R(x) - O(x)}{t(x)}$, where $L(x) = L_v(x) + L_p(x)$, and similarly for $R(x), O(x)$
	\item provide the proof: \\
	$\left( g_l^{L_p(s)}, g_r^{R_p(s)}, g_o^{O_p(s)}, g_l^{\alpha_l L_p(s)}, g_r^{\alpha_r R_p(s)}, g_o^{\alpha_o O_p(s)}, g^{Z(s)}, g^{h(s)} \right)$
	
}{

	\item assign verifier's variable polynomial values and add to $1$:
	\begin{flalign*}
		& g_l^{L_v(s)} = g_l^{l_0(s)} \cdot 
		\prod_{i=1}^m \left( g_l^{l_i(s)} \right)^{v_i} &
	\end{flalign*}
	
	and similarly for $g_r^{R_v(s)}$ and $g_o^{O_v(s)}$
	\item \checkVarPolynomials\ check: \\
	$ \prng{g_l^{L_p}, g^{\alpha_l}} = \prng{g_l^{L'_p}, g} $ and similarly for $g_r^{R_p}$ and $g_o^{O_p}$
	\item \checkConsistency\ check: \\
	$ \prng{g_l^{L_p} g_r^{R_p} g_o^{O_p}, g^{\beta \gamma}} = \prng{g^{Z}, g^{\gamma}} $
	\item \checkOperations\ check: \\
	$ \prng{g_l^{L_v(s)} g_l^{L_p}, g_r^{R_v(s)} g_r^{R_p}} = 
	\prng{g_o^{t}, g^{h}} \cdot \prng{g_o^{O_v(s)} g_o^{O_p}, g} $ \\
}
}

\note{following from the protocol properties (section \ref{section:single-var-oper-poly}) the $1$ represented by polynomials $l_0(x), r_0(x), o_0(x)$ already have appropriate values at the corresponding operations and therefore needs no assignment.}

\note{verifier will have to do extra work on the verification step, which is proportionate to the number of variables he assigns.}

Effectively this is taking some variables from the prover into the hands of verifier while still preserving the balance of the equation. Therefore the \term{\checkOperations} check should still hold, but only if the prover has used the same values that the verifier used for his input.

The value of $1$ is essential and allows to derive any number\footnote{In the chosen finite field} through multiplication by a constant term, for example, to multiply $a$ by $123$:
$$\maligned{\oper{1 \cdot a}{123 \cdot v\ped{one}}{1 \cdot r}}$$

\subsection{Zero-Knowledge Proof of Computation}

Since the introduction of the general-purpose computation protocol (\hyperref[sec:proofOfOperation]{section \ref*{sec:proofOfOperation} proof of operation}) we had to let go of the \zeroKnowledge{} property, to make the transition simpler. Until this point, we have constructed a verifiable computation protocol.

Previously to make a proof of polynomial \zeroKnowledge{} we have used the random $\delta$-shift, which makes the proof indistinguishable from random (section \ref{section:polyzk}):
$$ \delta p(s) = t(s) \cdot \delta h(s) $$

With the computation we are proving instead that:
$$ L(s) \cdot R(s) - O(s) = t(s)h(s)$$

While we could just adapt this approach to the multiple polynomials using same $\delta$, i.e., supplying randomized values $\delta L(s), \delta R(s), \delta^2 O(s), \delta^2 h(s) $, which would satisfy the \checkOperations\ check through pairings:
\begin{equation*}
	e\left(g, g \right)^{\delta^2 L(s) R(s)} = e(g,g)^{\delta^2 \left(t(s) h(s) + O(s) \right)}
\end{equation*}

The issue is that having same $\delta$ hinders security, because we provide those values separately in the proof:
\begin{itemize}
	\item one could easily identify if two different polynomial evaluations have same value (e.g., $g^{\delta L(s)} = g^{\delta R(s)}$, etc.), i.e., learning some knowledge
	\item potential insignificance of differences of values between $L(s)$ and $R(s)$ could allow factoring of those differences through brute-force, for example if $L(s) = 5R(s)$, iterating check $g^{L(s)} = \left( g^{R(s)} \right)^i $, for $i \in \{1 ... N\}$ would reveal the $5\times$ difference in just 5 steps. Same brute-force can be performed on encrypted addition operation, e.g., $g^{L(s)} = g^{R(s) + 5}$
	\item other correlations between elements of the proof may be discovered, e.g., if\ \  $e(g^{\delta L(s)}, g^{\delta R(s)}) = e(g^{\delta^2 O(s)}, g)$\ \ then\ \ $L(x) \cdot R(x) = O(x)$, etc.
\end{itemize}

\note{the optimization \ref{optimization} makes such data mining harder but still allows to discover relationships, apart from the fact that verifier can choose $\rho_l, \rho_r$ in a particular way that can facilitate revealing of knowledge\footnote{As long as it is not a diversified setup}.}

\newcommand*\circled[1]{\tikz[baseline=(char.base)]{
   \node[shape=circle,draw,inner sep=1pt] (char) {#1};}}

\newcommand{\anyop}{\raisebox{1pt}{\circled{\tiny{?}}}}   

Consequently, we need to have different randomness ($\delta$-s) for each polynomial evaluation, e.g.:
$$ \delta_l L(s) \cdot \delta_r R(s) - \delta_o O(s) = t(s) \cdot (\Delta\ \anyop\ h(s)) $$

To resolve inequality on the right side, we can only modify the proof's value $h(s)$, without alteration of the protocol which would be preferable. Delta ($\Delta$) here represents the difference we need to apply to $h(s)$ in order to counterbalance the randomness on the other side of the equation and $\anyop$ represents either multiplication or addition operation (which in turn accommodates division and subtraction). If we chose to apply $\Delta$ through multiplication ($\anyop = \times$) this would mean that it is impossible to find $\Delta$ with overwhelming probability, because of randomization:
\vspace{-2mm}\begin{equation*}
	\Delta = \frac{\delta_l L(s) \cdot \delta_r R(s) - \delta_o O(s)}{t(s) h(s)}
\end{equation*}

We could set $\delta_o = \delta_l \cdot \delta_r$, which transforms into:
\vspace{2mm}\begin{equation*}
	\Delta = \frac{\delta_l \delta_r( L(s) \cdot R(s) - O(s))}{t(s) h(s)} = \delta_l \delta_r
\end{equation*}

However, as noted previously this hinders the zero-knowledge property, and even more importantly such construction will not accommodate the verifier's input polynomials since they must be multiples of the corresponding $\delta$-s, which would require an interaction.

We can try adding randomness to the evaluations:
\begin{gather*}
	(L(s) + \delta_l) \cdot (R(s) + \delta_r) - (O(s) + \delta_o) = t(s) \cdot (\Delta \times h(s)) \\
	\Delta = \frac{ \overbrace{L(s) R(s) - O(s)}^{t(s) h(s)} + \delta_r L(s) + \delta_l R(s) + \delta_l \delta_r - \delta_o }{t(s) h(s)} = 
	1 + \frac{\delta_r L(s) + \delta_l R(s) + \delta_l \delta_r - \delta_o}{t(s) h(s)} 
\end{gather*}

However due to randomness it is non-divisible. Even if we address this by multiplying each $\delta$ with $t(s)h(s)$, because we apply $\Delta$ through multiplication of $h(s)$, and $\Delta$ will consist of encrypted evaluations (i.e., $g^{L(s)}$, etc.) it will not be possible to compute $g^{\Delta h(s)}$ without use of pairings (result of which is in another number space). Likewise computation is not possible through encrypted evaluation of $\Delta h(x)$ using encrypted powers $\left\{ g^{s^i} \right\}_{i \in [d]}$, because the degree of $h(x)$ and $\Delta$ is $d$, hence the degree of $\Delta h(x)$ is up to $2d$. 
Moreover, it is not possible to compute such randomized operand polynomial evaluation $g^{L(s) + \delta_l t(s)h(s)}$ for the same reason.

Therefore we should try applying $\Delta$ through addition ($\anyop = +$), since it is available for homomorphically encrypted values. 
\begin{gather*}
	(L(s) + \delta_l) \cdot (R(s) + \delta_r) - (O(s) + \delta_o) = t(s) \cdot (\Delta + h(s)) \\[2mm]
	\Delta = \frac{ L(s) R(s) - O(s) + \delta_r L(s) + \delta_l R(s) + \delta_l \delta_r - \delta_o - t(s) h(s)}{t(s)} \Rightarrow \\[3mm]
	\Delta = \frac{ \delta_r L(s) + \delta_l R(s) + \delta_l \delta_r - \delta_o}{t(s)}
\end{gather*}

Every term in the numerator is a multiple of a $\delta$, therefore we can make it divisible by multiplying each $\delta$ with $t(s)$:
\begin{gather*}
	(L(s) + \delta_l t(s)) \cdot (R(s) + \delta_r t(s)) - (O(s) + \delta_o t(s)) = t(s) \cdot (\Delta + h(s)) \\
	\cancel{L(s) R(s) - O(s)} + t(s) (\delta_r L(s) + \delta_l R(s) + \delta_l \delta_r t(s) - \delta_o)  = t(s) \Delta + \cancel{t(s) h(s)} \\
	\Delta = \delta_r L(s) + \delta_l R(s) + \delta_l \delta_r t(s) - \delta_o
\end{gather*}

Which we can efficiently compute in the encrypted space:
\begin{align*}
	g^{L(s) + \delta_l t(s)} &= g^{L(s)} \cdot \left( g^{t(s)} \right)^{\delta_l} \text{\ ,\quad etc.} \\
	g^\Delta &= \left( g^{L(s)} \right)^{\delta_r} \cdot \left( g^{R(s)} \right)^{\delta_l} \cdot \left( g^{t(s)} \right)^{\delta_l \delta_r} g^{-\delta_o}
\end{align*}

This leads to passing of \term{\checkOperations} check while concealing the encrypted values.
$$ {L\cdot R - O} +
	{\color{magenta}t(\delta_r L + \delta_l R + \delta_l \delta_r t - \delta_o)} = 
	t(s) h + 
		 {\color{magenta}t(s) (\delta_r L + \delta_l R + \delta_l \delta_r t - \delta_o)}
$$

The construction is statistically \term{zero-knowledge} due to addition of uniformly random multiples of $\delta_l, \delta_r, \delta_o$ (see theorem 13 of \cite{cryptoeprint:2012:215}).

\note{this approach is also consistent with the verifier's operands, e.g., $g_l^{L_p + \delta_l t} \cdot g_l^{L_v} = g_l^{L_p + L_v + \delta_l t}$, therefore the \term{\checkOperations} check holds but still only if the prover have used verifier's values to construct the proof (i.e., $\Delta = \delta_r (L_p + L_v) + \delta_l (R_p + R_v) + \delta_l \delta_r t - \delta_o$), see next section for more details.}

To make the ``\checkVarPolynomials" and ``\checkConsistency" checks coherent with the \zeroKnowledge\ alterations, it is necessary to add the following parameters to the proving key:
$$ g_l^{t(s)}, g_r^{t(s)}, g_o^{t(s)}, g_l^{\alpha_l t(s)}, g_r^{\alpha_r t(s)}, g_o^{\alpha_o t(s)}, g_l^{\beta t(s)}, g_r^{\beta t(s)}, g_o^{\beta t(s)} $$

It is quite curious that the original Pinocchio protocol \cite{cryptoeprint:2013:279} was concerned primarily with the verifiable computation and less with the \zeroKnowledge\ property, which is a minor modification and comes almost \emph{for free}.

\newcommand{\zkc}[1] {{\color{Magenta}#1}}

\subsection{\zkSNARKraw\ Protocol}

Considering all the gradual improvements the final zero-knowledge succinct non-interactive arguments of knowledge protocol is (the \zeroKnowledge\ components are optional and highlighted with a \zkc{different color}):
{
\setlist[itemize]{leftmargin=6mm}
\zkProtocol{
	\item select a generator $g$ and a cryptographic pairing $e$
	\item for a function $f(u) = y$ with $n$ total variables of which $m$ are input/output variables, convert into the polynomial form\footnote{A quadratic arithmetic program} $\left( \{l_i(x), r_i(x), o_i(x)\}_{i \in \{0, \ldots, n\}}, t(x) \right)$ of degree $d$ (equal to the number of operations) and size $n + 1$ 
	\item sample random $ s, \rho_l, \rho_r, \alpha_l, \alpha_r, \alpha_o, \beta, \gamma $
	\item set $\rho_o = \rho_l \cdot \rho_r$ and the operand generators $g_l = g^{\rho_l}, g_r = g^{\rho_r}, g_o = g^{\rho_o}$
	\item set the proving key: 
		\begin{flalign*}
			\bigg(\ \ 
				& \left\{ g^{s^k} \right\}_{k \in [d]},
				\left\{ g_l^{l_i(s)}, g_r^{r_i(s)}, g_o^{o_i(s)} \right\}_{i \in \{0, \ldots, n\}}, &\\[1.5mm]
				& \left\{  
					g_l^{\alpha_l l_i(s)}, g_r^{\alpha_r r_i(s)}, g_o^{\alpha_o o_i(s)}, 
					g_l^{\beta l_i(s)} g_r^{\beta r_i(s)} g_o^{\beta o_i(s)} 
				\right\}_{i \in \{m + 1, \ldots, n\}}, &\\[1mm]
				& \ \ \zkc{
					g_l^{t(s)}, g_r^{t(s)}, g_o^{t(s)},
					g_l^{\alpha_l t(s)}, g_r^{\alpha_r t(s)}, g_o^{\alpha_o t(s)}, g_l^{\beta t(s)}, g_r^{\beta t(s)}, g_o^{\beta t(s)} 
				}
			\ \  \bigg) &
		\end{flalign*}
	
	\item set the verification key:
	\begin{flalign*}
	 & \left( 
		g^1, g_o^{t(s)}, 
		\left\{ g_l^{l_i(s)}, g_r^{r_i(s)}, g_o^{o_i(s)} \right\}_{i \in \{0, \ldots, m\}},
		g^{\alpha_l}, g^{\alpha_r}, g^{\alpha_o}, g^{\gamma}, g^{\beta \gamma}
	  \right) &
	\end{flalign*}
	
} {
	\item for the input $u$, execute the computation of $f(u)$ obtaining values $\{v_i\}_{i \in \{m+1, \dots, n\}}$ for all the itermediary variables
	\item assign all values to the unencrypted variable polynomials $L(x) = l_0(x) + \sum_{i = 1}^n v_i \cdot l_i(x)$ and similarly $R(x), O(x)$
	\item \zkc{sample random $\delta_l, \delta_r$ and $\delta_o$}
	\item find\ \ $\displaystyle h(x) = \frac{L(x) R(x) - O(x)}{t(x)} 
	\zkc{\ +\ \delta_r L(x) + \delta_l R(x) + \delta_l \delta_r t(x) - \delta_o }$
	\item assign the prover's variable values to the encrypted variable polynomials \zkc{and apply \zeroKnowledge\ $\delta$-shift} $\displaystyle g_l^{L_p(s)} = 
	\zkc{ \left(g_l^{t(s)}\right)^{\delta_l} }
	\cdot \prod_{i = m+1}^n 
	\left( g_l^{l_i(s)} \right)^{v_i}$ and similarly $g_r^{R_p(s)}$, $g_o^{O_p(s)}$
	\item assign its $\alpha$-shifted pairs $\displaystyle g_l^{L'_p(s)} = 
	\zkc{ \left(g_l^{\alpha_l t(s)}\right)^{\delta_l} } \cdot 
	\prod_{i = m+1}^n 
	\left( g_l^{\alpha_l l_i(s)} \right)^{v_i}$ and similarly $g_r^{R'_p(s)}$, $g_o^{O'_p(s)}$
	\item assign the \checkConsistency\ polynomials
	\flalignskip{
		& g^{Z(s)} = 
			\zkc{
				\left(g_l^{\beta t(s)}\right)^{\delta_l} 
				\left(g_r^{\beta t(s)}\right)^{\delta_r} 
				\left(g_o^{\beta t(s)}\right)^{\delta_o} 
			}
			\cdot
			\prod_{i=m+1}^n \left( g_l^{\beta l_i(s)} g_r^{\beta r_i(s)} g_o^{\beta o_i(s)} \right)^{v_i} &
	}
	\item compute the proof\ \ $\left(
		g_l^{L_p(s)}, g_r^{R_p(s)}, g_o^{O_p(s)}, g^{h(s)}, g_l^{L'_p(s)}, g_r^{R'_p(s)}, g_o^{O'_p(s)}, g^{Z(s)}
	\right)$
} {
	\item parse a provided proof as\ \ $\left( 
		g_l^{L_p}, g_r^{R_p}, g_o^{O_p}, g^{h}, g_l^{L'_p}, g_r^{R'_p}, g_o^{O'_p}, g^{Z}	
	\right)$
	\item assign input/output values to verifier's encrypted polynomials and add to $1$:\\ 
		$\displaystyle g_l^{L_v(s)} = g_l^{l_0(s)} \cdot 
	\prod_{i=1}^m \left( g_l^{l_i(s)} \right)^{v_i}$
	and similarly for $g_r^{R_v(s)}$ and $g_o^{O_v(s)}$
	\item \checkVarPolynomials\ check : \\
	$ \prng{g_l^{L_p}, g^{\alpha_l}} = \prng{g_l^{L'_p}, g} $ and similarly for $g_r^{R_p}$ and $g_o^{O_p}$
	\item \checkConsistency\ check: \\
	$ \prng{g_l^{L_p} g_r^{R_p} g_o^{O_p}, g^{\beta \gamma}} = \prng{g^{Z}, g^{\gamma}} $
	\item \checkOperations\ check: \\
	$\prng{g_l^{L_p} g_l^{L_v(s)}, g_r^{R_p} g_r^{R_v(s)}} = 
	\prng{g_o^{t(s)}, g^{h}} \cdot \prng{g_o^{O_p} g_o^{O_v(s)}, g} $
}
}

\section{Conclusions}

We ended up with an effective protocol which allows proving computation: 
\begin{itemize}
	\item succinctly --- independently from the amount of computation the proof is of constant, small size
	\item non-interactively --- as soon as the proof is computed it can be used to convince any number of verifiers without direct interaction with the prover
	\item with argumented knowledge --- the statement is correct with non-negligible probability, i.e., fake proofs are infeasible to construct; moreover prover \emph{knows} the corresponding values\footnote{A witness} for the true statement, e.g., if the statement is ``$B$ is a result of $\mathrm{sha256}(a)$'' then the prover knows some $a$ such that $B = \mathrm{sha256}(a)$ which is useful since $B$ could only be computed with the knowledge of $a$ as well as it's infeasible to compute $a$ from $B$ only\footnote{Assuming $a$ has enough entropy}
	\item in \zeroKnowledge\ --- it is infeasible to extract any knowledge from the proof, i.e., it is indistinguishable from random
\end{itemize}

It was possible to achieve primary due to unique properties of polynomials, modular arithmetic, homomorphic encryption, elliptic curve cryptography, cryptographic pairings and ingenuity of the inventors.

This protocol proves correctness of computation of a unique finite execution machine which in one operation can add together almost any number of variables but may only perform one multiplication. Therefore there is an opportunity to both optimize programs to leverage this specificity efficiently as well as use constructions which minimize the number of operations. 

It is essential that verifier does not have to know any secret data in order to verify a proof so that properly constructed verification key can be published and used by anyone in a non-interactive manner. 
Which is contrary to the ``designated verifier" schemes where the proof will convince only one party, therefore it is non-transferable. In \zkSNARK\ context, we can achieve this property if untrustworthy or a single party generates the keypair.

The field of zero-knowledge proof constructions is continuously evolving, introducing optimizations (\cite{BCTV13, cryptoeprint:2016:260, GM17}), improvements such as updatable proving and verification keys (\cite{cryptoeprint:2018:280}), and new constructions (Bulletproofs \cite{cryptoeprint:2017:1066}, ZK-STARK \cite{cryptoeprint:2018:046}, Sonic \cite{cryptoeprint:2019:099}).

\section*{Acknowledgments}

We are grateful to Mary Maller and Andrew Miller for their valuable comments on this work.

\begin{otherlanguage}{english}
\newpage
\section{References} \label{References}
\begingroup
\hfuzz=2pt
\renewcommand{\section}[2]{}
\printbibliography
\endgroup
\end{otherlanguage}

\end{document}